\documentclass[11pt]{cernrep}
\usepackage{graphicx}
\usepackage{here}
\usepackage{epsfig}
\usepackage{amssymb}
\usepackage{amsmath}

\usepackage{pstricks}
\usepackage[active]{srcltx}

\def\half{{1 \over 2}}
\def\({\left(}
\def\){\right)}
\def\[{\left[}
\def\]{\right]}
\def\k{{\bf k}}
\def\x{{\bf x}}

\def\psp{\hspace{0.1cm}}
\def\nsp{\hspace{-0.1cm}}

\def\ltap{\raisebox{-.55ex}{\rlap{$\sim$}} \raisebox{.4ex}{$<$}}
\def\gtap{\raisebox{-.55ex}{\rlap{$\sim$}} \raisebox{.4ex}{$>$}}
\def\gsim{\mathrel{\gtap}}

\def\agt{\mathrel{\gtap}}
\def\alt{\mathrel{\ltap}}

\newcommand{\e}{\mathop{\rm e}\nolimits}
\newcommand{\be}{\begin{equation}}
\newcommand{\ee}{\end{equation}}
\newcommand{\ba}{\begin{eqnarray}}
\newcommand{\ea}{\end{eqnarray}}

\begin{document}
\bibliographystyle{/home/tkachev/TeX/local/h-physrev3}

 \title{ASTROPARTICLE PHYSICS}
\author{I. Tkachev}
\institute{Theory Division, CERN, CH-1211 Geneva 23, Switzerland \\
      and \\
      Institute for Nuclear Research of the Russian Academy of
      Sciences, 117312, Moscow, Russia
}
\maketitle
\begin{abstract}
In this astro-particle lecture course I shall try to emphasize evidence of the
new physics which we have in cosmological and astrophysical data. This
includes support of the inflationary model, necessity of dark energy and of
non-baryonic dark matter, the Grizen-Zatsepin-Kuzmin puzzle of the ultra-high
energy cosmic rays.
\end{abstract}

\section{INTRODUCTION}

The purpose of these CERN school lectures is to review the evidence for the
new physics in cosmological and astrophysical data, and to give the minimal
theoretical frameworks needed to understand and appreciate the evidence.
Beyond any reasonable doubt, we have solid evidence for the new physics in
these data. The strongest is the case for non-baryonic dark matter, followed
by the case for dark energy. The possibility (though very speculative, since a
consistent and working model has not been constructed yet) that the law of
gravity should be changed instead is not excluded, but that would mean a new
physics anyway. I will not engage in discussion of relevant particle physics
model building, instead the reader is refereed to lectures by G. Gabadadze at
this school \cite{Gabadadze:2003}.  Another solid evidence for the new physics
beyond the standard model is given by neutrino oscillations. I will not
discuss this topic, it is covered in lectures by S. Petcov at this
school \cite{Petcov:2003}. The physics of cosmic rays is partially covered in
lectures by A. Chilingaryan \cite{Chilingaryan:2003},
therefore, I restrict myself to the highest-energy part of the spectrum, which
is related to Grizen-Zatsepin-Kuzmin puzzle and, possibly, to a new physics.

There are many excellent reviews on the subject of Cosmology and Astroparticle
physics, including lectures at previous CERN schools, for a recent one see
\cite{Rubakov:2001,Shaposhnikov:2000,Garcia-Bellido:1999ys}. I've tried to be
complimentary to these lectures to the extent it is possible, so many
additional details can be found there. Proceedings of these schools can be
found at http://physicschool.web.cern.ch/PhysicSchool.  Due to space and time
limitations, I omit several very important traditional topics, most notably
Big Bang Nucleosynthesis (see e.g. the review \cite{Olive:1999ij}) and
Baryogenesis (see e.g. the review \cite{Rubakov:1996vz}).  In the area
covered, I've updated experimental results and resulting constraints, where
applicable. The most important developments since the time of the previous
school were: release of the first year observations of Cosmic Microwave
Background Radiation (CMBR) by the Wilkinson Microwave Anisotropy Probe (WMAP)
\cite{Bennett:2003bz}, first data release by the Sloan Digital Sky Survey
(SDSS) of three-dimensional distribution of galaxies \cite{Tegmark:2003ud},
and the release\cite{Riess:2004nr} of a statistically significant dataset of
Supernovae Ia at large cosmological redshifts, $z>1$, which provide the first
conclusive evidence for cosmic deceleration that preceded the current epoch of
cosmic acceleration. These are long awaited cosmological data of unprecedented
quality, and with their appearance cosmology has truly entered the golden era
and became a precision science.

The plan of the lectures is as follows. In Section~\ref{sec:CosmologyBasics},
I review the basics of cosmology: Friedman equations, Hubble expansion,
cosmography.  In Section~\ref{sec:CMBR}, the Cosmic Microwave Background
Radiation (CMBR) is discussed. In Section~\ref{sec:LSS}, I briefly review
resent results on another cosmological probe - the large-scale distribution of
galaxies. In Section~\ref{sec:darkEnergy}, the evidence for the existence of
dark energy is presented.  Sections~\ref{sec:DM-motivation} and
\ref{sec:DM-models} review the evidence for a dark matter, and particle
physics models of non-baryonic matter are briefly considered. In Section
\ref{sec:inflation}, I review the basics of inflationary cosmology and discuss
support of the inflationary model by the CMBR data. In Section
\ref{sec:UHECR}, the physics of the Ultra-high energy cosmic rays is reviewed.

\section{BASICS OF COSMOLOGY}
\label{sec:CosmologyBasics}

\subsection{Note on units and scales}

\paragraph{Length.} Astronomers are measuring distances in parsecs, which is 
about $3.1 \times 10^{16}$ m or about 3.26 light years, and is comparable with
the distance to the closest star. PARSEC is an abbreviation for the distance
to a star with a semi-annual PARallax of 1 arc SECond. The distance from our
Sun to the Galactic center is $8$ kiloparsecs, so the kpc is an appropriate
unit when discussing galactic structure. The appropriate unit of extragalactic
distance, however, is the megaparsec, or Mpc. The nearest large cluster of
galaxies, the Virgo cluster, is at a distance of 20 Mpc. The size of the
visible Universe is 4200 Mpc or 13.7 billions of light years.
\paragraph{Energy.} Usually astronomers are measuring energy in ergs. 
E.g. the luminosity of our Sun is $4\times 10^{33} {\rm ~erg~ s^{-1}}$, while
luminosity of bright quasars reaches $ 10^{46} {\rm ~erg~ s^{-1}}$.  Galaxy
like our Milky Way contains $10^{11}$ stars, and there are $10^{11}$ galaxies
in the visible part of the Universe. Particle physicists are measuring energy
in electron-volts, 1 erg $ = 6.2\times 10^{11}$ eV, and usually are choosing
units where the velocity of light and the Plank constant are set to unity,
$c=1,~~\hbar =1$, which I am using too, when convenient. In these units, for
example, 1 Mpc $= 1.6\times 10^{29}\; {\rm eV}^{-1}$.

\subsection{Dynamical Frameworks}

Dynamics is provided by General Relativity - the Einstein field equations
\be
R_{\mu \nu} -\half g_{\mu \nu} R = 8\pi G \; T_{\mu \nu}\; ,
\label{Einstein-equations}
\ee 
where $T_{\mu \nu}$ is a stress energy tensor describing the distribution of
mass in space, $G$ is Newton's gravitational constant, and the curvature
$R_{\mu \nu}$ is a complicated function of the metric and its first and second
derivatives.  Clearly, finding a general solution to a set of equations as
complex as the Einstein field equations is a hopeless task. The problem is
simplified greatly considering mass distributions with special symmetries.
The basic assumption underlying the construction of cosmological models is
that of spatial homogeneity and isotropy.  The most general space-time metric
consistent with these symmetries is the Robertson-Walker metric:
\be
ds^{2} = dt^{2} - a^{2}(t)\; {\bf dl}^{2}\; ,
\label{FRW-metric}
\ee 
where $a(t)$ is the dimensionless scale factor by which all distances vary as
a function of cosmic time $t$.  The scale factor contains all the dynamics of
the Universe, while the vector product ${\bf dl}^{2}$ describes the geometry
of the space,
$$
{\bf dl}^{2} = \frac{dr^{2}}{1-k\,r^{2}} + r^{2}\,(d\theta^{2} +
\sin^{2}\theta \, d\phi^{2})\; ,
$$
which can be either Euclidian, or positively or negatively curved.
For the spatial 3-dimensional curvature we find, explicitly
\be
{}^{(3)}\!R = {6k\over a^2(t)}\; \; \; 
\left\{\begin{array}{ll}
k=-1&\hspace{0.5cm}{\rm   Open} \\
k=0&\hspace{0.5cm}{\rm Flat}\\
  k=+1&\hspace{0.5cm}{\rm   Closed} 
\end{array}\right.
\label{Spatial-curvature} 
\ee 
E.g., the space with $k=+1$ can be thought of as a 3-dimensional sphere
with a curvature being inversely proportional to the square of its radius.  In
this Section we will model the matter content of the Universe as a perfect
fluid with energy density $\rho$ and pressure $p$, for which the stress-energy
tensor in the rest frame of the fluid is
\begin{eqnarray}
T_{\mu}^{~\nu}=\left( \begin{array}{cccc}
{\rho} & 0 & 0 & 0\\
0 & {-p} & 0 & 0 \\
0 & 0 & {-p} & 0 \\
0 & 0 & 0 &{-p}
\end{array} \right) 
\label{Tmunu-ideal}
\end{eqnarray} 
With these assumptions the Einstein equations simplify to the Friedmann
equations, which form the dynamical basis of cosmology 
\ba \label{Friedmann-Eq1}
\frac{\dot{a}^{2}}{a^{2}} &=& \frac{8\pi G}{3}\, \rho ~-~
\frac{k}{a^{2}} \; , \\ 
\frac{\ddot{a}}{a}~ &=& -~ \frac{4\pi G}{3}\; (\rho + 3\, p) \; .
 \label{Friedmann-Eq2}
\ea 
Let us have a look at the basic physics behind of these equations.

\paragraph{1.} Differentiating Eq.~(\ref{Friedmann-Eq1})
and subtracting  Eq.~(\ref{Friedmann-Eq2}) we obtain
\be
\frac{d{\rho}}{dt} ~+~ 3\,\frac{\dot{a}}{a}\, ({\rho}+{p})=0 \; ,
\label{internal-energy-conservation}
\ee
which is nothing but energy-momentum conservation,
\be
T^{~\nu}_{\mu\,\,\,\,;\nu}=0 \; .  
\label{relativ-energy-conservation}
\ee
On the other hand, the result is nothing but the First Law of thermodynamics
\be
dE ~+~ p\, dV ~=~ T\, dS \; ,
\label{first-law}
\ee 
with $ dS = 0$.  Here $ E = \rho V = \rho a^{3} V_0$ is energy, $T$ is
temperature and $ S$ is entropy of some (fixed) comoving volume $V_0$.
Therefore, Friedmann expansion driven by an ideal fluid is isentropic, $ dS =
0$. This is not unexpected, and relaxing the assumption of a perfect fluid
will lead to entropy production. However, the dissipation is negligible in
cosmological frameworks (except of special moments, like the initial matter
creation and possible phase transitions, which will be considered separately)
and isentropic expansion is a very good approximation.  This gives a useful
integral of the motion, $S =$ const.  On dimensional grounds, $S \propto T^3
a^{3} V_0$ and we obtain frequently used relation between the scale factor and
temperature in an expanding Universe 
\be
\label{a-Temperature} a \propto \frac{1}{T}\; .  
\ee To be precise, 
\be
\label{Entropy-reativistic}
S = {2\pi^2\over45}\, g_*\; T^3\, a^3 = {\rm\footnotesize const}\; , 
\ee 
where the factor $ g_*$ counts the effective number of relativistic degrees of
freedom 
\be 
g_* = {\footnotesize \sum_{i={\rm bosons}} g_i +
\frac{7}{8}\sum_{j={\rm fermions}} g_j } \equiv (g_B + \frac{7}{8}g_F)\; .
\ee 

\begin{figure}
\begin{center}
\includegraphics[width=12.5cm]{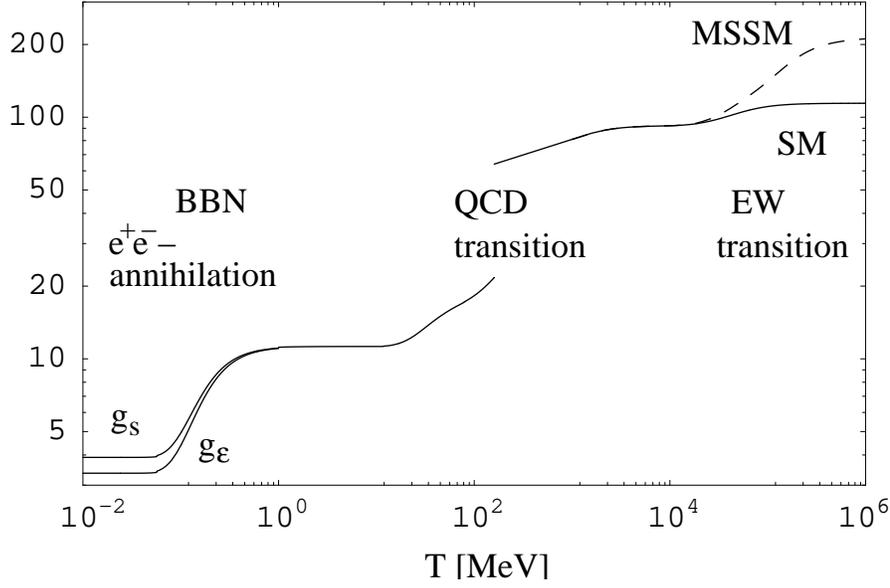} 
\caption{Number of relativistic degrees of freedom as a 
function of temperature. From Ref. \cite{Schwarz:2003du}.}
\label{fig:Dominik}
\end{center}
\end{figure}

At any given temperature, only particles with $m \ll T$ should be counted,
i. e.  $g_*$ is a function of temperature, which is shown in
Fig.~\ref{fig:Dominik}. For a gas of photons, $g_* = 2$.  Considering the
current epoch of the Universe evolution, we have to add the neutrino
contribution, which will be discussed later and leads to a different account
of effective degrees of freedom at temperatures below $e^+e^-$ annihilation in
entropy, $ g_S$, and in energy density, $g_\rho$. Namely, $g_S(T_0) = 3.909$
and $g_\rho(T_0) = 3.363$. At temperatures above the electroweak scale $g_*
\sim 100$ in the Standard Model.

Let us give here also other thermodynamical relations, similar to
Eq.~(\ref{Entropy-reativistic}), but for the energy density, $\rho$, and
particle number density, $n$
\begin{eqnarray}\label{EnergyDensity-reativistic}
\rho &=& \frac{\pi^2}{30}\; g_* T^4 \;,\\
n &=& \frac{\zeta(3)}{\pi^2}\;(g_B + \frac{3}{4}g_F)\; T^3\; ,
\label{numberDensity-reativistic}
\end{eqnarray}
where $\zeta(3) = 1.202$. These relations are a simple consequence of the
integration of Bose-Einstein or Fermi-Dirac distributions 
\be
\frac{g}{(2\pi)^3} \int \frac{d^3 q}{\e^{q/T}_{~} \pm 1}\;  q^a \; ,
\ee
where $q$ is particle momentum, the plus (minus) sign corresponds to fermions
(bosons), and $a=1$ in calculation of $\rho$, while $a=0$ in calculation of
$n$ (in the latter case the integral cannot be evaluated in terms of
elementary functions and the Riemann $\zeta$-function appears).  With the use
of Eq.~(\ref{first-law}), the entropy density,
Eq.~(\ref{Entropy-reativistic}), can be found as $s = 4\rho/3T$, since for the
relativistic particles $p=\rho/3$ regardless of spin.

\begin{figure}
\begin{center}
\includegraphics[width=7cm]{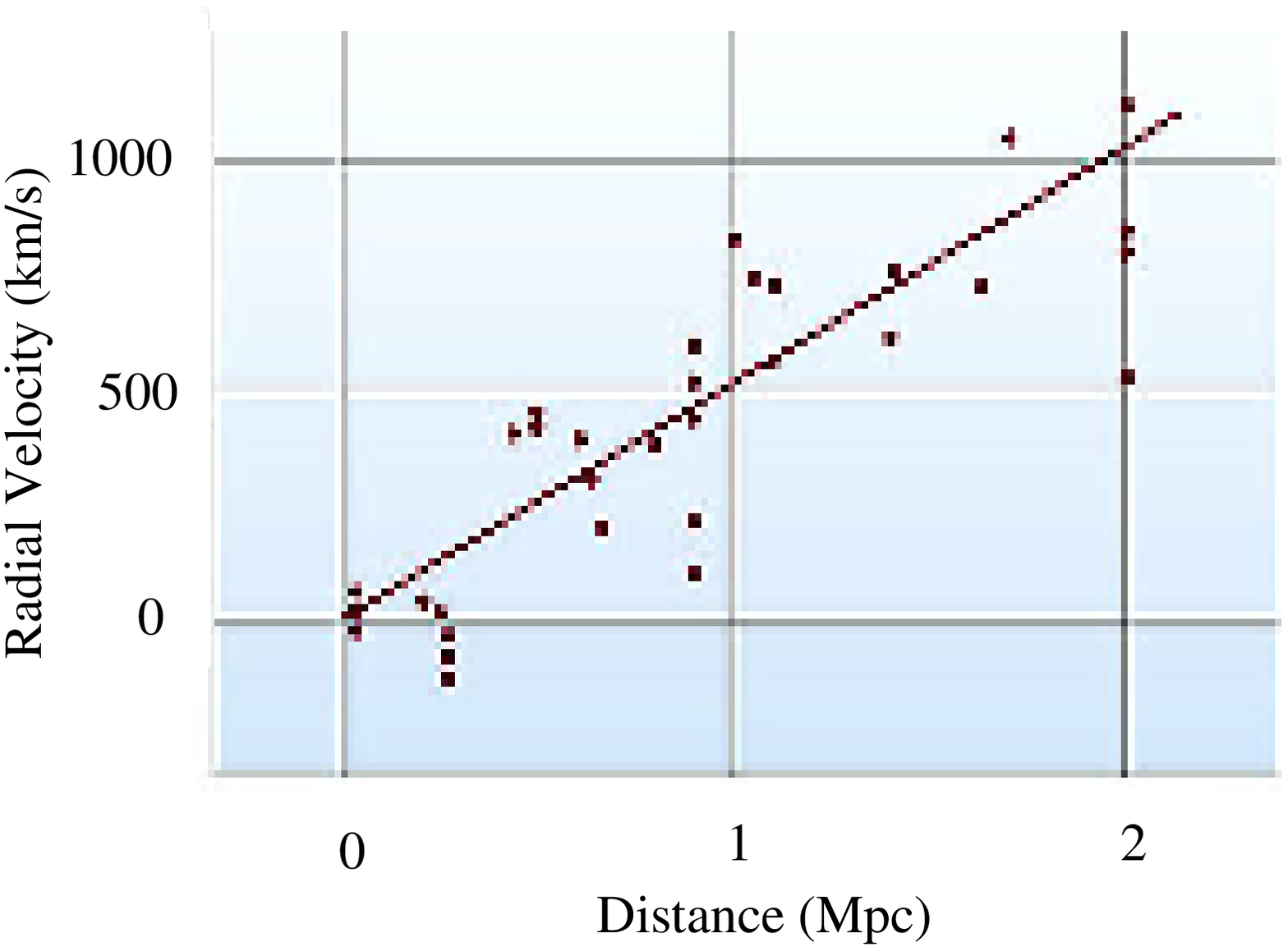}
\hspace{1cm}
\includegraphics[width=7cm]{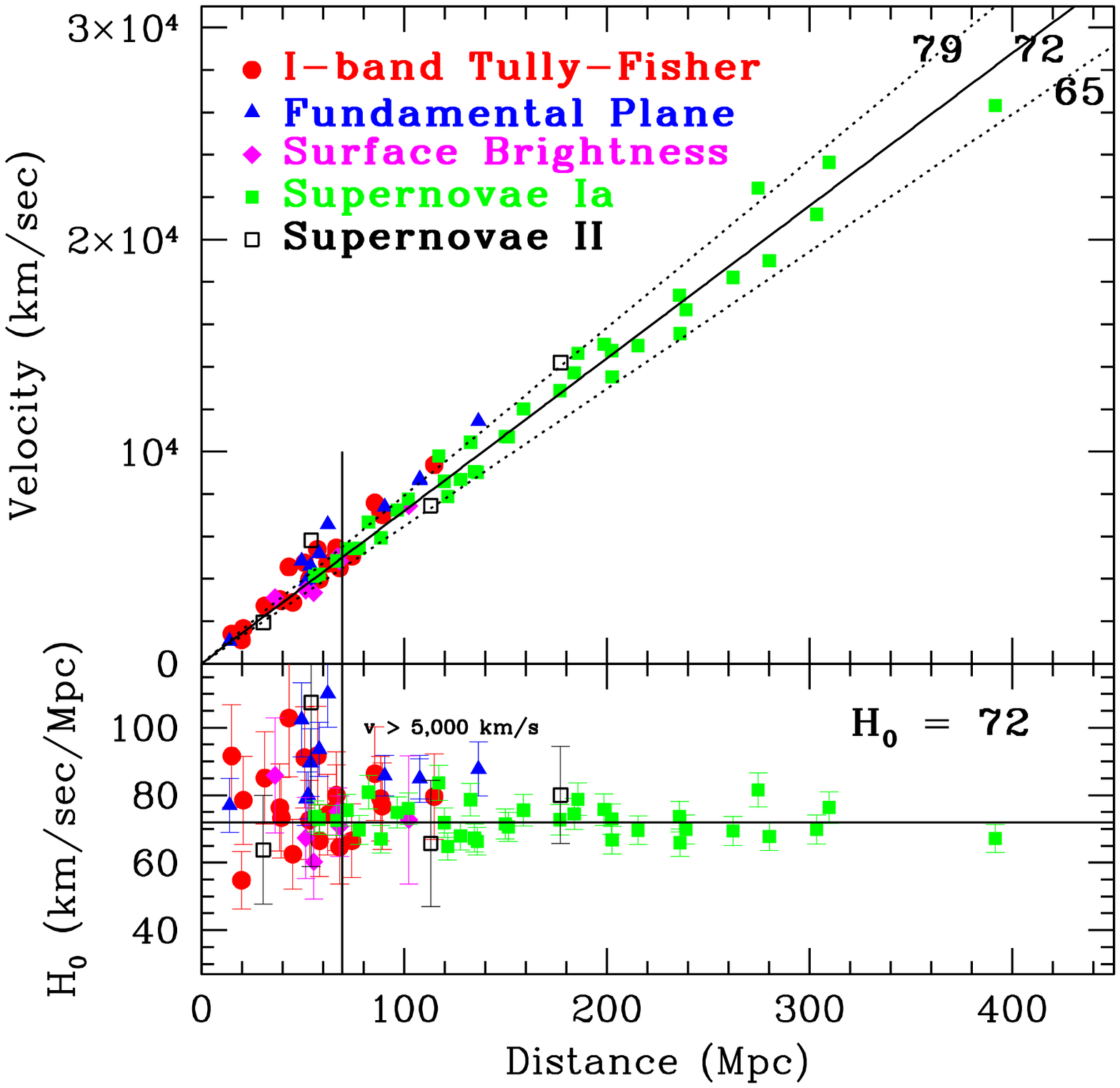}
\caption{Hubble Diagram. From Ref.~\cite{Freedman:2000cf}.}
\label{fig:HubbleDiagram}
\end{center}
\end{figure}

\paragraph{2.} Friedmann equation,  Eq.~(\ref{Friedmann-Eq1}), can be
interpreted within Newtonian mechanics. Indeed, let us first re-arrange it as
\be 
\half\, {\dot{a}}^{2} - \frac{4\pi G}{3}\, \rho \, a^{2} = - \frac{k}{2}\; .
\ee
Now, it is easy to see that for $ r = a\, r_0$, the Friedmann equation takes
the form of energy conservation for test particles bounded in the
gravitational potential created by mass $M = \frac{4\pi}{3}\, \rho \, r^{3}$,
\be
\half\, {\dot{r}}^{2} - \frac{G\, M}{r} = - \frac{k\, r_0^{2}}{2}\; .
\label{energy-conserv-Newton}
\ee
We see that the constant $k$, which determines the sign of spacial curvature in
the language of General relativity, also determines the sign of the binding 
energy
\begin{center}\begin{tabular}{ll}
${ k = + 1}$ &  \hspace{0.2cm} Binding energy is negative, \\
 & \hspace{0.2cm}  the Universe will recollapse\\
${ k = - 1}$ &   \hspace{0.2cm} Binding energy is positive, \\
& \hspace{0.2cm}  the Universe will expand forever\\
\end{tabular}\end{center}
Therefore, the case of zero spatial curvature, or zero binding energy, ${k =
0}$, is special and corresponds to fine tuning between initial kinetic and
potential energies. Setting $k=0$ in Eq.~(\ref{Friedmann-Eq1}), this fine
tuning can be expressed as $\rho=\rho_c$, where the critical density is defined
as
\be
\rho_c \equiv \frac{3}{8\pi G}\; \(\frac{\dot{a}}{a}\)^{2} \; .
\label{critical-density}
\ee
The critical density is proportional to the square of another fundamental
parameter, 
\be
H \equiv \frac{\dot{a}}{\huge a} \; .
\label{Hubble-def}
\ee
The present value of this parameter is called the Hubble constant. It
describes the rate of the Universe expansion, and can be related to
observations in the following way. Consider two points with a fixed comoving
distance $r_0$ between them (this means that points do not feel any other
forces and do not participate in any other motion beyond general expansion of
the Universe). The physical distance between points increases as
$r(t) = a(t) r_0$, and we can find the relative velocity as
\be
v \equiv \dot{a}r_0 = \frac{\dot{a}}{a}\; a r_0 = H r \; .
\label{Hubble-law}
\ee 
The relation $v = H r $ is called the Hubble law. This is shown in
Fig.~\ref{fig:HubbleDiagram}. The left panel is original data used by
E. Hubble, the right panel presents recent data from
Ref.~\cite{2002ApJ...574L.155H}. We will discuss it in more
detail later on, especially in relation to observations. But the first thing
we may notice is that according to Hubble law, $v \sim 1$ at $r \sim
H^{-1}$. Separations (or wavelengths) of this order are therefore special in
cosmology and mean {\it super-horizon} length scale.  At smaller separations,
Newtonian gravity should be valid.  Einstein equations tell us that energy
conservation in Newtonian mechanics, Eq.~(\ref{Friedmann-Eq1}), and the first
law of thermodynamics, Eq.~(\ref{internal-energy-conservation}), applied to
the Universe as a whole, can be extended beyond horizons without any
change. The second Friedmann equation, Eq.~(\ref{Friedmann-Eq2}), can be
derived as a consequence of these two equations. However, hardly would Newton
do it, even if he had known the first law of thermodynamics. Indeed, in
Eq.~(\ref{Friedmann-Eq2}) we recognize the Newton's second law, $F = m \dot{v}$
with $F$ being the gravity force, and energy conservation is derived from
the equations of motion, not vice versa.

We see that, according to Einstein's theory, the force law is modified. Not
only mass gravitates, but the pressure, too, gives its contribution into
gravitational force. This is a very important modification, since pressure can
be negative, leading to anti-gravity and to accelerated expansion. As we will
see, this stage of expansion may have lead to creation of the Universe in our
classical understanding. At present, the Universe expansion seems to be
dominated by anti-gravity as well. This has an interesting history.  Newton
did not know that one should worry about horizons, but he worried why the
Universe does not collapse under the pull of gravity.  Einstein was worried
about this too. He added (1917) a cosmological constant to the equations of
motion, thinking that it will make the Universe static. (As we will see, the
cosmological constant corresponds to a vacuum with non-zero energy density and
negative pressure, $p=-\rho$.) However, Friedmann had shown (1922) that the
Universe will not be static anyway.  After some debate, Einstein admitted his
mistake and called the introduction of a cosmological constant ``the greatest
blunder of my life''.

So, why did the Universe not collapse under the pull of gravity? Resolution is
in awkward initial conditions called the Big Bang, where velocity in
Eq.~(\ref{energy-conserv-Newton}) is highly tuned to potential energy, leading
to practically zero spatial curvature and to $\rho = \rho_c$.  This implies
enormous fine-tuning for the Universe to survive till present. Such
fine-tuning is hard to accept, and a modification of classical cosmology was
called for.  We will see how modern inflationary cosmology solves the problem
of initial conditions. Again, the resolution is in anti-gravity caused by
negative pressure.

\subsection{Matter content in the Universe}

To solve Friedmann equations, Eq.~(\ref{Friedmann-Eq1})-(\ref{Friedmann-Eq2}),
one has to specify the matter content of the Universe and the equation of
state for each of the constituents.  To fit current observations we need at
least four components
\begin{itemize}
\item[--] {\it Radiation} (relativistic degrees of freedom). Today this
  component consists of the photons and neutrino and gives negligible
  contribution into total energy density. However, it was major fraction at
  early times.
\item[--] {\it Baryonic matter}. Makes up the observable world today.
\item[--] {\it Dark matter}. Was not directly detected yet, but should be
  there. Constitutes major matter fraction today. Has rather long
  observational history and can be fitted within frameworks of modern particle
  physics nicely, at the price of ``moderate'' tuning of parameters to
  provide required fraction of matter.
\item[--] {\it Dark energy}. Looks like it also should be there. It provides
  the major fraction of the total energy density today. Was not anticipated
  and appears as a biggest surprise and challenge for particle physics, though
  conceptually it can be very simple, being just a ``cosmological constant''
  or vacuum energy.
\end{itemize}
\paragraph{Equations of state.} Each of these components have very simple 
equation of state, parameterized by a single constant $w$
\be
w \equiv \frac{p}{\rho} \; .
\label{w-def}
\ee
With a constant $w$, solutions of the first law of thermodynamics,
Eq.~(\ref{internal-energy-conservation}), are readily found
\be
\rho(t) = a(t)^{-3(1+w)}\; \rho_0 \; ,
\label{solutions-for-w-const}
\ee 
where $\rho_0$ stands for the present day density. For example, for ordinary
forms of matter we have
\begin{itemize}
  \item[--] Radiation: $w = \frac{1}{3}$ ~and $\rho = a^{-4}\; \rho_0 $.
Result can be understood as a simple consequence of entropy conservation, $aT
=$ const, since for radiation $\rho \propto T^4$.
  \item[--] Matter: $w = 0$ and $\rho = a^{-3}\; \rho_0 $.  Result can be
understood as a simple consequence of particle number conservation, $na^3 =
{\rm const}$ and $\rho = m n$, where $m$ is particle mass.
\end{itemize}
For hypothetical matter, which may play the role of dark energy, $w$ is
negative
\begin{itemize}
  \item[--] Network of cosmic strings: $w = - \frac{1}{3}$ ~and 
    $\rho = a^{-2}\; \rho_0$.
  \item[--] Network of domain walls: $w = - \frac{2}{3}$ ~and
    $\rho = a^{-1}\; \rho_0$.
  \item[--] Cosmological constant: $w = -1$ ~and $\rho = \rho_0$.  Result can
be understood as a consequence of the Lorentz invariance of a vacuum, which
restricts $T_{\mu}^{\,\nu}$ to be proportional to the Kronecker tensor,
$T_{\mu}^{\,\nu}=V\, \delta_{\mu}^{\,\nu}$. Comparing with
Eq.~(\ref{Tmunu-ideal}) we find $p=-\rho$, or $w=-1$.
\end{itemize}

\paragraph{Law of expansion.} Friedmann equation (\ref{Friedmann-Eq1}) in 
a spatially flat Universe and with a single matter component, which energy
density evolves according to Eq.~(\ref{solutions-for-w-const}), has the
solution
\be
a = \( {t}/{t_0} \) ^{\frac{2}{3(1+w)}} \; .
\label{solutions-Friedmann}
\ee
In particular
\begin{itemize}
  \item[--] Radiation:~~~ $w = \frac{1}{3}$,~~~~~~~ $a = (t/t_0)^{1/2} $
  \item[--] Matter:~~~~~~~~ $w = 0$,~~~~~~~ $a = (t/t_0)^{2/3}$
  \item[--] Cosmological constant:~ $w = -1$. This case is special, and $a =
 \e^{H_0 t}\;$.
\end{itemize}

\subsection{Cosmological Parameters.}

These are used to parameterize the Fiedmann equation and its solution
$a(t)$. Let me first summarize the current knowledge of numerical values of
those parameters which were introduced already; later in the course we
will discuss how these values were deduced.
\begin{center}
Cosmological parameters
\vskip0.2cm
\begin{tabular}{|l|l|l|}\hline
Symbol \& Definition & Description & Present value\\\hline
$t$                 & {Age of the Universe} & $ t_0 =(13.7 \pm 0.2)$ ~Gyr\\ 
$H = \dot{a}/a$     & {Hubble constant} & $H_0 = 71\;
\rm {~km~ s^{-1}~ Mpc^{-1}}$\\
$\rho_c = 3H^2/8\pi G$  & {Critical density} & $\rho_c = 10\; h^{2} \rm  ~~GeV~ m^{-3} $ \\
$\Omega = \rho/\rho_c$  & Omega &  $\Omega_0 = 1.02 \pm 0.02$  \\
$\Omega^{}_{\rm CMB} = \rho^{}_{\rm CMB}/\rho_c$ & Fraction of CMB photons &
$\Omega^{}_{\rm CMB} = 2.4\cdot 10^{-5} h^{-2}$ \\
$\Omega_b = \rho_b/\rho_c$ &  Baryonic fraction & $\Omega_b = 0.044\pm 0.004$ \\
$\Omega_m = \rho_m/\rho_c$ &  Matter fraction & $\Omega_m = 0.27\pm 0.04$ \\
$\Omega_\Lambda = \rho_\Lambda/\rho_c$ &  Dark Energy fraction &
$\Omega_\Lambda = 0.73 \pm 0.04$ \\
\hline
\end{tabular}
\end{center}

\subsection{Cosmography.}

We can define cosmography (this is my custom definition for these lectures) as
a part of cosmology which tries to map observations into reconstruction of the
scale factor. I.e., the goal is to find and tabulate the function $a(t)$. This
is important in many respects: e.g. it allows to determine the matter content
in the Universe (assuming the Friedmann equations are correct). One simple and
straightforward way of tabulating the function is in determining its
coefficients in Taylor expansion. This can be done making a Taylor
decomposition around present time, $t=t_0$. The value of the scale factor at
any moment of time can be fixed arbitrarily, we can use this freedom to choose
$a(t_0) = 1$. The second term in the Tailor decomposition is naturally the
value of the Hubble constant, $H_0$, Eq.~(\ref{Hubble-def}). It gives us the
rate of the Universe expansion at present and can be measured using the Hubble
law, Eq.~(\ref{Hubble-law}).  One can go further in this decomposition and
define the second derivative of the universe at present, the ``deceleration
parameter'', and so on. We will not do it (at least at this point), since
modern observations probe the whole function $a(t)$. Therefore, let us start
with the preparation of the necessary machinery which allows us to deduce
$a(t)$ from observations.

Let me stress now that Eq.~(\ref{Hubble-law}) involves some degree of cheating
since it is not a relation between the observables. (However, it is a valid
relation for small separations.) To apply the Hubble law to observations, we
have to derive its generalization, which would connect quantities we can
measure.

\paragraph{Redshift.}
Looking at distant objects we see only the light they emit. How can physical
quantities like distance and velocity be derived?  Recall how a police officer
determines the speed of a car. A similar principle is used in cosmology to
determine the velocity of distant bodies. The shift of emission lines with
respect to the frequency measurements by the local observer is related to
velocity, and is used as an observable instead of the velocity. Systematic
recession of objects, or cosmological expansion, leads to redshift. Note that
cosmological redshift is not entirely due to the Doppler effect, but, rather,
can be interpreted as a mixture of the Doppler effect and of the gravitational
redshift.

Let us relate now the redshift to cosmological expansion, $\dot{a}/a$. To this
end, we consider photon trajectories in a cosmological background with metric
Eq.~(\ref{FRW-metric}). The trajectory is given by $ds^{2} = 0$.  Since the
overall scale factor does not change the solutions of $ds^{2} = 0$, it is
convenient to introduce the conformal time $\eta$ defined as
\be 
dt = a(\eta)\,d\eta \; . 
\label{time-conformal}
\ee
It is sufficient to consider radial trajectories with the observer at the
center, and I restrict myself to a spatially flat metric
$
ds^{2} = a^{2}(d\eta^{2} - d\chi^{2}) = 0 , 
$
where $\chi$ denotes the radial coordinate. The solution of~ $ d\eta^{2} -
d\chi^{2} = 0$ is $\chi = \pm \eta \;+\; {\rm const}$. Since the comoving
distance between source and observer does not change, the conformal time
interval between two light pulses is the same at the point of emission and at
the point of observation, $\Delta \eta =$ const. Using the definition of
conformal time, $d\eta = dt/ a$, we find
$$
\frac{\Delta t}{a}|_{\rm emission} = \frac{\Delta t}{a}|_{\rm  detection} \; .
$$
Therefore, for a signal frequency we get~ $\omega_d {a_d} = \omega_e {a_e}$.
Defining (measurable) redshift as
\be
z \equiv \frac{\omega_e - \omega_d}{\omega_d} \; 
\label{z-definition}
\ee
we obtain
\be
1+z = \frac{a_d}{a_e} \; .
\label{z-a}
\ee 
It is convenient to normalize the scale factor by the condition $a_d = 1$ at
the point of detection, and to consider the scale factor at the point of
emission as a function of redshift $z$. On the basis of this relation, the
expansion history of the Universe can be parameterized as
\be
a(z) = \frac{1}{1+z} \; .
\label{z-a-ehist}
\ee 
The differential form of Eq.~(\ref{z-a-ehist}) is $da/dz = - a^2$.  For future 
use, let us find now the relation between $d\eta$ and $dz$
\be
d\eta ~=~ \frac{d\eta}{dt}\; \frac{dt}{da}\; \frac{da}{dz}\; dz ~=~ 
-\frac{a}{\dot{a}}\; dz
~=~ -\frac{dz}{H(z)} \; .
\label{deta-dz} 
\ee
Observing that $d\eta = -d\chi$, we obtain the Hubble low for small
separations, $dz = H d \chi$. At this point, we have succeeded in replacing
the velocity by the redshift. Now we aim to relate the distance to some other
quantity, directly measurable in cosmology.

\paragraph{Luminosity distance.}
Looking at distant objects we see only the light they emit. How can physical
quantities like distance and velocity be derived?  There are several ways to
introduce a quantity related to the distance: different definitions are not
equivalent in curved space-time. A definition based on flux measurements is
the appropriate one, if ``standard candles'' can be found and used.
Detected flux $\rm [erg\; s^{-1}\, sm^{-2}]$ is inversely proportional
to the distance from a source, $F \propto D^{-2}$. Namely, if $L$ is intrinsic
luminosity $\rm [erg\; s^{-1}]$, we have
\be
D_L^{2} = \frac{L}{4\pi F} \; .
\label{luminocity-distance-def}
\ee 
$D_L$ is called the ``luminosity distance''.  For this technique to work, one
has to find a set of sources with a known or calibrated luminosity. If such
can be defined, they are called {\it standard candles.}

To see how the luminosity distance enters the Hubble law, let us consider a
space-time with the metric $ds^{2} = a^{2}\;(d\eta^{2} - d\chi^{2} -
\chi^{2}\; d\Omega)$. Now, go through the following list to find out what
happens with the flux emitted into a frequency interval ~$d\nu$~ by a source
located at redshif ~$z$:
\begin{itemize}
\item[--] Surface area at the point of detection  is~ $4\pi \chi^{2}$.~
(Recall our choice $a_d=1$).
\item[--] Energy and arrival rates are redshifted between the points of
  emission and detection. This reduces the flux by~ $(1+z)^{2}$. 
\item[--] Opposing this tendency, the bandwidth $d\nu$ is reduced by~ $(1+z)$. 
\item[--] Photons observed at a frequency $\nu$, were emitted at~ $(1+z)\,\nu$.
\end{itemize}
Therefore, the measured spectral flux density is
\be
S(\nu) = \frac{L((1+z)\,\nu)}{4\pi\, \chi^{2}(1+z)}\;  .
\ee
For the bolometric flux (i.e. integrated over $\nu$) we find 
\be
F = \frac{L}{4\pi\, \chi^{2}(1+z)^{2}}\; .
\ee
Comparing this with the definition, Eq.~(\ref{luminocity-distance-def}), we
find for the luminosity distance
\be 
D_L = (1+z)\, \chi \; ,
\label{luminocity-distance-result}
\ee 
where $\chi$ is the comoving distance between the point of emission and the
point of detection 
\be 
\chi(z) = \int_{\eta_e}^{\eta_d} d\eta = \int_0^{z}
\frac{dz'}{H(z')}\; .  
\label{comovingD-z}
\ee
In the last equality we have used Eq.~(\ref{deta-dz}). Therefore, the
generalization of the Hubble law, which can be used in observational
cosmology, can be written as

\be 
(1+z)\, \chi(z) = \sqrt{\frac{L}{4\pi F}} \;\; .
\label{z-vs-F}
\ee

\paragraph{Parameterization of $\bf H(z)$.}
Let us express now the function $H(z)$ in the r.h.s. of Eq.~(\ref{comovingD-z})
through the cosmological parameters.  First, we define the ratio of the total
energy density to the critical one
\be \Omega \equiv \frac{\rho_{\rm tot}}{\rho_c} \; .
\label{Omega-def}
\ee
The present day value is referred to as $\Omega_0$.  Similarly, for each energy
component we denote its {\it present day} fractional contribution as $\Omega_i
\equiv {\rho_i}/{\rho_c}$. With these definitions, the Friedmann equation
(\ref{Friedmann-Eq1}) for a spatially flat Universe can be re-written as
$$
H^{2} = \frac{8\pi G}{3}\sum_i \rho_i \; ,
$$
or
\be
H^{2} = H_0^{2}\, \sum_i  \Omega_i\, (1+z)^{3(1+w_i)}\; .
\label{Friedmann-parmetrization}
\ee
Here I have used
$$
\rho_i = \rho_{i,0}\; a^{-3(1+w_i)}\;  = 
\rho_c\, \Omega_i\, (1+z)^{3(1+w_i)}\; ,
$$
and expressed the scale factor as a function of redshift according to
Eq.~(\ref{z-a}), and used the definition of the critical density, $H_0^{2} =
{8\pi G}\rho_c/3$.

Parameterization (\ref{Friedmann-parmetrization}) is ready for use in
Eq.~(\ref{comovingD-z}) for the comoving distance.  In particular, this
finalizes expression Eq.~(\ref{z-vs-F}) for the luminosity distance.


\section{Cosmic Microwave Background Radiation}
\label{sec:CMBR}

The Universe is filled with radiation which is left-over from the Big Bang.
The name for this first light is Cosmic Microwave Background Radiation (CMBR).
Measurements of tiny fluctuations (anisotropy) in CMBR temperature give a
wealth of cosmological information and became a most powerful probe of
cosmology.

This radiation was predicted by Georgi Gamov in 1946, who estimated its
temperature to be $\sim 5\; K^{\circ}$. Gamov was trying to understand the
origin of chemical elements and their abundances. Most abundant, after
hydrogen, is helium, with its shear being $\sim 25\%$. One possibility which
Gamov considered was nucleo-synthesis of He out of H in stars. Dividing the
total integrated luminosity of the stars by the energy released in one
reaction, he estimated the number of produced He nuclei. This number was too
small in comparison with observations. Gamov assumed then the oven where the
light elements were cooked-up was the hot early Universe. He calculated
abundances of elements successfully and found that the redshifted relic of
thermal radiation left over from this hot early epoch should correspond 
to  $\sim 5\; K^{\circ}$ at present. In one stroke G. Gamov founded two 
pillars (out of four) on which modern cosmology rests: CMBR and Big Bang
Nucleosynthesis (BBN). Hot Big Bang was born. 

Cosmic microwave background was accidentally discovered by Penzias and Wilson
\cite{Penzias:1965wn} at Bell Labs in 1965 as the excess antenna temperature
which, within the limits of their observations, was isotropic, unpolarized,
and free from seasonal variations. A possible explanation for the observed
excess noise temperature was immediately given by Dicke, Peebles, Roll, and
Wilkinson and was published in a companion letter in the same issue
\cite{Dicke:1965}. They were preparing dedicated search experiment, but were
one month late. Penzias and Wilson measured the excess temperature as $\sim
3.5\pm 1\; K^{\circ}$. It is interesting to note that the first (unrecognized)
direct measurements of the CMB radiation was done by T. Shmaonov at Pulkovo in
1955, also as an excess noise while calibrating the RATAN antenna
\cite{Shmaonov:1957}. He published the temperature as $ (3.7 \pm 3.7)\;
K^{\circ}$. Prior to this, in 1940, Andrew McKellar \cite{McKellar:1940} had
observed the population of excited rotational states of CN molecules in
interstellar absorption lines, concluding that it was consistent with being in
thermal equilibrium with a temperature of $ \approx 2.7 \; K^{\circ}$.  Its
significance was unappreciated and the result essentially forgotten.  Finally,
before the discovery, in 1964 Doroshkevich and Novikov in an unnoticed paper
emphasized \cite{Doroshkevich:1964} the detectability of a microwave blackbody
as a basic test of Gamov's Hot Big Bang model. To me, as to theorist, the
detection of CMBR looks nowadays like an easy problem. Indeed, a few percent of
the "snow" on TV screens is due to CMBR.

The spectrum of CMBR is a perfect blackbody, with a temperature $T=2.725 \pm
0.002 \; K^{\circ}$ as measured by modern instruments. This corresponds to
410.4 photons per cubic centimeter or to the flux of 10 trillion photons per
second per squared centimeter. The temperature is slightly different in
different patches of the sky - to 1 part in 100,000. And this is most
important: the spectrum of this tiny fluctuations tells us a lot about the
fundamental properties of the Universe.

CMBR is the oldest light in the Universe. When registering it, we are looking
directly at the deepest past we can, using photons. These photons had traveled
the longest distances without being affected by scattering, and geometrically
came out almost from the universe Horizon. More precisely, the CMB comes from
the surface of the last scattering.  We cannot see past this surface.  That is
because at early times the Universe was ionized and not transparent for
radiation. With expansion, it cooled down, and when hydrogen recombined, the
universe became transparent.  Therefore the CMBR gives us a snapshot of the
baby Universe at this time, which is called the time of last scattering. Let
us determine when the last scattering had occurred in the early Universe.

\subsection{ Hydrogen recombination}

At temperatures greater than a few thousands $K$, the ionized plasma in the
Universe consisted of mostly protons, electrons, and photons, with a small
fraction of helium nuclei and a tiny trace of some other light elements.  To a
good approximation we can consider only the hydrogen.  Matter is then ionized
at temperatures higher than the hydrogen ionization energy $E_{\rm ion} =
13.6$ eV.  At lower T neutral atoms start to form.  The baryonic matter is in
thermal equilibrium and the equilibrium fraction of ionized hydrogen can be
described by the Saha equation (see Ref.~\cite{Rubakov:2001} for more details)
\be 
\frac{n_e\,
n_p}{n^{~}_H} = \( \frac{m_e T}{2\pi}\)^{3/2}\e ^{-E_{\rm ion}/T}\; ,
\label{Saha}
\ee
here $n_e,~ n_p$ and $n^{~}_H$ are the number densities of electrons, protons,
and neutral hydrogen respectively.
Plasma is electrically neutral, i.e. $n_e = n_p$. To find the closed relation
for the fraction of ionized atoms, $X\equiv n_p/(n_p+n^{~}_H) = n_p/n^{~}_B$,
we need the relation between the baryon number density, $n^{~}_B$, and
temperature. This relation can be parameterized with the help of an important
cosmological parameter called {\it baryon asymmetry} 
\be 
\eta =
\frac{n^{~}_B}{n_\gamma} =\frac{n_p + n^{~}_H}{n_\gamma} = (6.1 \pm 0.3)\times
10^{-10} \; ,
\label{BAU}
\ee
where ${n_\gamma}$ is the number density of photons
\be 
{n_\gamma} = \frac{2\zeta(3)}{\pi^2} T^3  \; ,
\label{n-gamma}
\ee 
and $\zeta(3) = 1.202$, see Eq.~(\ref{numberDensity-reativistic}). Baryon
asymmetry can be estimated by an order of magnitude by simply counting the
number of baryons, $\eta = 2.68\times 10^{-8}\Omega_b h^2$. This is not the
most precise method, though; the value presented in Eq.~(\ref{BAU}) was
obtained from fitting the spectrum of CMBR fluctuations, see below. Nowadays,
this is the most precise baryometer. Prior to this, the best estimates were
obtained comparing BBN predictions of element abundances to
observations. Defining recombination as the temperature when $X=0.1$, we find
$T_{\rm rec} \approx 0.3 {\rm eV}$.

The Universe became transparent for radiation when the mean free path of
photons became comparable to the size of the Universe at that time. Photons
scatter mainly on electrons and we find that the Universe became transparent
when
\be
(\sigma_{\gamma e}\, n_e)^{-1} ~\sim~ t  \; .
\label{TranspCond}
\ee 
Here, $\sigma_{\gamma e} = 8\pi \alpha^{2}/3m_e^{2}$ is the Compton
cross-section.  For the temperature of last scattering we find $T_{\rm ls}
\approx 0.26\; {\rm eV}$. Taking the ratio to the current CMBR temeperature
we find  $z_{\rm ls} \approx 1000$.

\subsection{Spectrum is not distorted by red-shift}

Prior to recombination photons were in thermal equilibrium. Therefore, at last
scattering they have Planck spectrum
$$ n (p) = \frac{1}{\exp(E_{\rm ls}/T_{\rm ls}) - 1}\; .$$ 
Since then, particle momenta are red-shifted, $p = {k}/{a}$.  Since photons are
massless, $E = p$, their energies are red-shifted at the same rate, $E_0 a_0 =
E_{\rm ls} a_{\rm ls}$, and the spectrum becomes
$$ n ~=~ \frac{1}{\exp(E_{0}/a_{\rm ls}T_{\rm ls}) -1} =
\frac{1}{\exp(E_{0}/T_{0}) -1}\; , $$ 
where we have used the notation $T_{0} \equiv a_{\rm ls}T_{\rm ls}$.
Therefore, after decoupling, the shape of the spectrum is not distorted.  This
statement would not be true for massive particles, $E^{2} = \({p}/{a}\)^{2} +
m^{2}$.

Measuring CMBR, we should still see the Planckian spectrum, but with
red-shifted temperature. Clearly, this conclusion is true not only for
cosmological red-shift, but for the gravitational red-shifts as well. E.g.,
fluctuations in the gravitational potential at the last scattering surface
should cause fluctuations in CMBR temperature, but do not distort the spectrum.

\subsection{ Dipole spectrum}

\begin{figure}
\begin{center}
\includegraphics[width= 5.cm,angle=90]{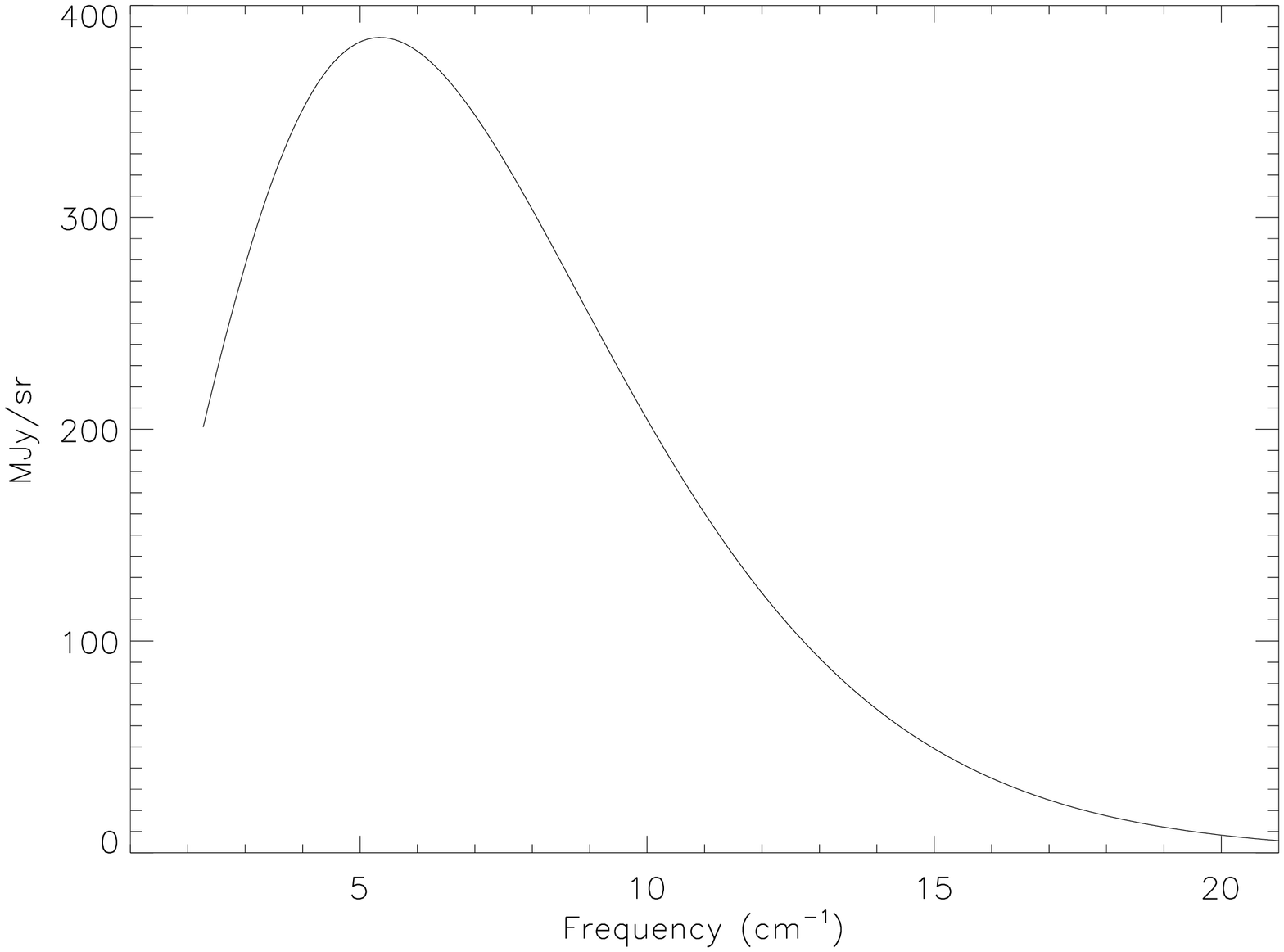}
\includegraphics[width= 5.cm,angle=90]{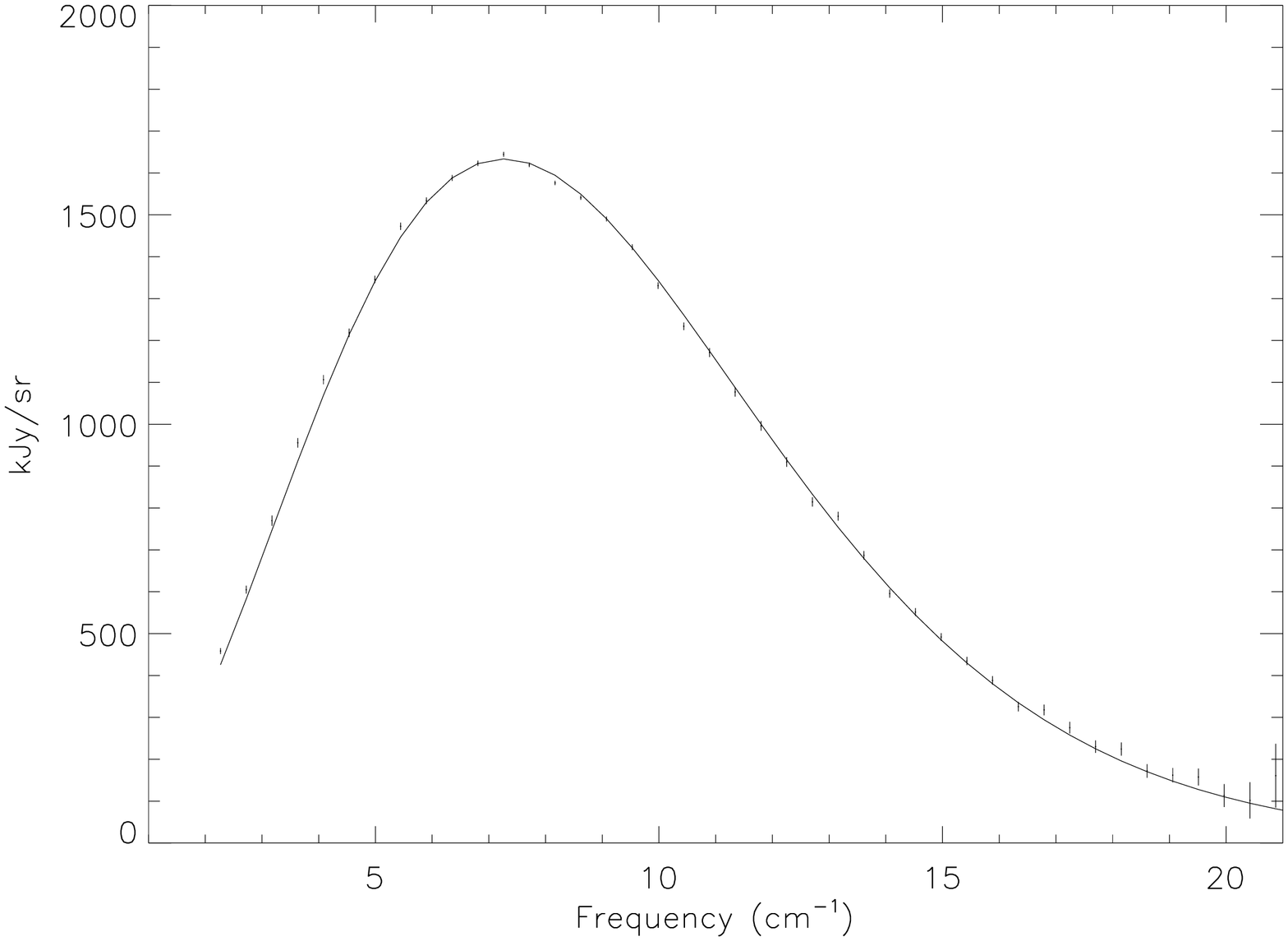}
\caption{Left panel: uniform spectrum; error-bars are a small fraction of the
  line thickness. Right panel: dipole spectrum; vertical lines indicate one
  $\sigma$ uncertainties. From Ref. \cite{Fixsen:1996nj}.}
\label{CMBR-spectrum}
\end{center}
\end{figure}

Intensity of the CMB radiation is a function of the frequency $2\pi\nu=E$ and
the direction on the sky $(l,b)$. As a function of $(l,b)$ it can be
decomposed in spherical harmonics. Coefficients will be the functions of
$\nu$. The first two terms in this decomposition are
\be 
S(\nu, l, b) = I_0(\nu) + D(l,b)\, d(\nu)
+ \dots \; ,
\label{CMB-MonDipole}
\ee 
where $ D(l,b) = \cos \theta$, and $\theta$ is an angle between observation
and the maximum of the dipole $l_0 = 263.85^{\circ}$, $b_0 = 48.25^{\circ}$.
The dipole is caused by our motion with respect to CMBR (which is composed of
the motion of the Sun in the Galaxy and the Galaxy's own motion in the Local
Cluster of galaxies). It gives the direction of this motion, $l_0, b_0$, which
roughly coincides with the direction towards Virgo. The dipole induced by the
velocity $v$ is $vT\cos \theta$. This gives the magnitude of Sun's peculiar
velocity, $\rm (371 \pm 1)\;\, km\; s^{-1}$.

Let $x \equiv E/T$. The monopole term should have the usual black-body
spectrum $I_0(\nu) \propto x^3/(\e^x -1)$. The dipole spectrum is actually
distorted, because the Doppler frequency shift depends upon direction. The
dipole spectrum can be found as a term linear in $v$ in the Taylor
decomposition of $S(\nu, l, b)$, with the result $ d(\nu) \propto
x^4\e^x/(\e^x -1)^2$; for a recent discussion see
Ref.~\cite{Kamionkowski:2002nd}. Functions $I_0(\nu)$ and $d(\nu)$ are shown
in Fig.~\ref{CMBR-spectrum}, left and right panels respectively. Both agree
with theoretical expectation.

\subsection{Multipoles}

\begin{figure}
\begin{center}
\includegraphics[width= 9.cm]{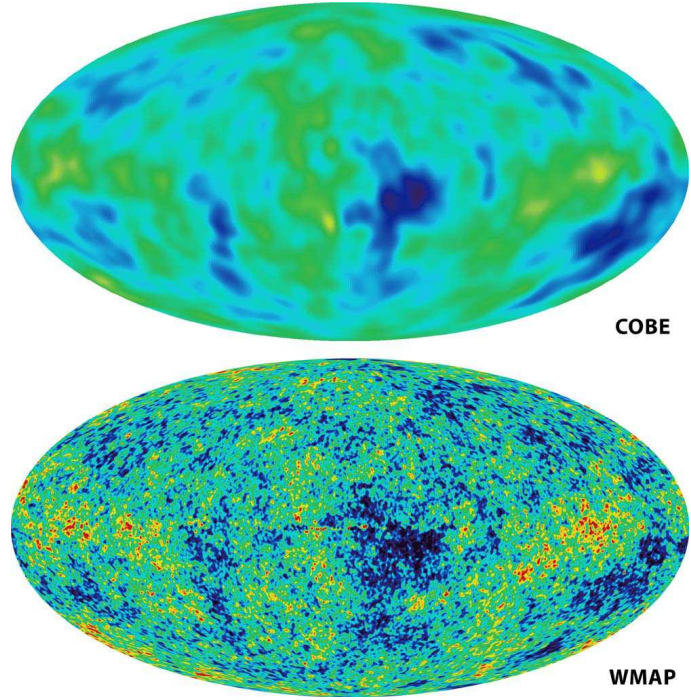}
\caption{Pattern of primordial temperature fluctuations in Galactic
coordinates, from Ref.~\cite{Bennett:2003bz}. The WMAP map has 30 times finer
resolution than the COBE map.}
\label{MAPandCOBE}
\end{center}
\end{figure}

Monopole and dipole contributions to CMBR, Eq.~(\ref{CMB-MonDipole}), can be
subtracted. Emission of our Galaxy and various extragalactic sources can be
subtracted also. This procedure uses the fact that the relic CMBR signal has
a black body spectrum , which allows to distinguish it from other forms of
radiation: measurements of the intensity at different frequencies allow to
subtract contaminating foregrounds. What remains corresponds to the primordial
cosmological pattern of temperature fluctuations, which is shown in
Fig.~\ref{MAPandCOBE}. The upper panel presents the results of early COBE
experiments \cite{Bennett:1996ce}, the lower panel shows the results of a
recent WMAP experiment \cite{Bennett:2003bz}.

\begin{figure}
\begin{center}
~\hspace{-2cm}
\includegraphics[width=9.cm]{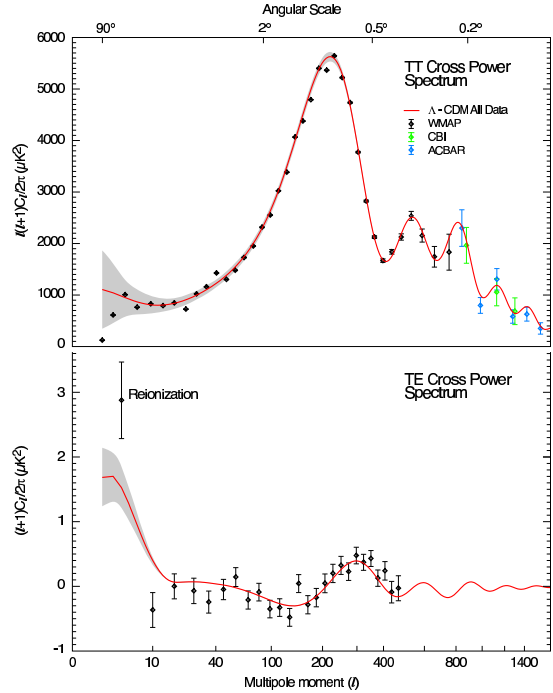}
~\vspace{-2cm}
\caption{The angular power spectrum of primordial CMBR temperature
  fluctuations is shown in the upper panel, from
  Ref.~\cite{Bennett:2003bz}. Black symbols ($l < 700$) are WMAP measurements,
  data points at smaller angular scales represent CBI and ACBAR experiments.
  Lower panel shows the temperature-polarization cross-power spectrum. }
\label{PowerSpectrum-WMAP}
\end{center}
\end{figure}

The temperature anisotropy, $T({\bf n})$, as a function of viewing direction
vector ${\bf n}$, is naturally expanded in a spherical harmonic basis,
$Y_{lm}$
\be
T({\bf n}) = \sum\limits_{l,m} a_{lm} Y_{lm}({\bf n})\; .
\label{T-SpericHarmDec}
\ee
The coefficients in this decomposition, $a_{lm}$, define the angular power
spectrum, $C_l$
\be
C_l={\frac{1}{2l+1}}\sum \limits_m \vert a_{lm} \vert ^2\; .
\label{T-PowSpectr}
\ee
The CMBR angular power spectrum as measured by WMAP experiment is shown in
Fig.~\ref{PowerSpectrum-WMAP}. The harmonic index $l$ is related to the angular
scale $\theta$ as $l \approx \pi/\theta$, so the first peak, at $l \approx
220$, would correspond to an angular scale of about one degree.  Assuming
random phases, the r.m.s. temperature fluctuation assosiated with the angular
scale $l$ can be found as 
\be \Delta T_l = \sqrt{\frac{C_l\; l(l+1)}{ 2\pi}}\; .
\label{DeltaT}
\ee
Another representation of temperature fluctuations is given by the angular
correlation function, which is related to $C_l$ as
\be
C(\theta)  =  {\frac{1}{4\pi}} \sum\limits_l (2l+1) C_l\; P_l(cos\; \theta)\; ,
\label{T-CorrFunct}
\ee
where $P_l$ is the Legendre polynomial of order $l$.


\subsection{Tool of Precision Cosmology}

The functional form of the CMBR power spectrum is very sensitive to both the
various cosmological parameters and to the shape, strength and nature of
primordial fluctuations. Measurements of the power spectrum provide us with a
wealth of cosmological information at an unprecedented level of precision.

Right after the discovery of CMBR, it was realized that fluctuations in its
temperature should have fundamental significance as a reflection of the seed
perturbations which grew into galaxies and clusters. In a pure baryonic
Universe it was expected that the level of fluctuations should be of the order
$\delta T /T \sim 10^{-2}-10^{-3}$. Mesurements of the CMBR anisotropy with
ever-increasing accuracy have begun. Once the temperature fluctuations were
shown to be less than one part in a thousand, it became clear that baryonic
density fluctuations did not have time to evolve into the nonlinear structures
visible today. A gravitationally dominant dark matter component was invoked.
Eventually, fluctuations were detected at the level of $ \delta T / T ~\sim
~10^{-5}$ \cite{Smoot:1992td}, consistent with structure formation in Cold
Dark Matter models with the Harrison-Zel'dovich spectrum of primordial
perturbations motivated by cosmological inflation.  Already this magnitude of
$ \delta T / T$ is very restrictive by itself.  A partial set of best fit
cosmological parameters, as derived from the recent measurments of CMBR
anisotropies, is presented in Table I.

The foundations of the theory of CMBR anisotropy were set out by Sachs \&
Wolfe \cite{Sachs:1967er}, Silk \cite{Silk:1968kq}, Peebles \& Yu
\cite{Peebles:1970ag}, Syunyaev \& Zel'Dovich \cite{Syunyaev:1970}.  The
measured spectrum of CMBR power has a characteristic shape of multiple
peaks. Positions of these peaks and their relative amplitudes are sensitive to
many cosmological parameters in a non-trivial way. Fitting the data to model
predictions gives very accurate values for many of these parameters (though
there are some degeneracies between deferent sets). Numerical calculations for
different models were done already in Ref. \cite{Doroshkevich:1978}, and power
spectra exhibiting acoustic peaks (similar to those in
Fig.~\ref{PowerSpectrum-WMAP}) were presented. It was realized, in particular,
that positions of the peaks are shifted with respect to each other for
adiabatic and isentropic primordial fluctuations.

\subsection{Acoustic oscillations}

Let us give a qualitative picture of why the CMBR power spectrum has a
specific shape of a sequence of peaks, and explain how it depends on the
values of particular cosmological parameters. Insight, sufficient for the
purposes of these lectures, can be gained with the idealization of a perfect
radiation fluid. In complete treatment, one has to follow the evolution of
coupled radiation and metric fluctuations, i.e. to solve the linearized
Einstein equations. However, essential physics of radiation (or matter)
fluctuations can be extracted without going into the tedious algebra of
General Relativity. It is sufficient to consider the energy-momentum
conservation, Eq.~(\ref{relativ-energy-conservation}). To solve for metric
perturbations, full treatment based on Einstein equations,
Eq.~(\ref{Einstein-equations}), is needed of course. We will not do that here,
but simply quote results for the evolution of the gravitational potentials
(coincident in some important cases with the solutions for the Newtonian
potentials).

Perturbations of the ideal radiation fluid, $p = \rho/3$, can be separated
into perturbations of its temperature, velocity and gravitational potential.
In the general-relativistic treatment gravitational potential appears as a
fractional perturbation of the scale factor in the perturbed metric 
\be ds^{2}
= a^{2}(\eta) \, [(1+2\Psi)d\eta^{2} - (1-2\Phi)dx^{i}dx_j] \; .
\label{metric-perturbed}
\ee 
Two equations contained in the energy-momentum conservation,
$T^{\mu\nu}_{\,\,\,\,\,\,\,\,\,;\nu}=0$ (i.e. temporal $\mu=0$ and spatial
$\mu=i$ parts of this equation), written in metric (\ref{metric-perturbed}),
can be combined to exclude the velocity perturbations. The resulting
expression is simple
\be 
\ddot{\theta}_k + \frac{k^{2}}{3} \theta_k = - \frac{k^{2}}{3}
\Phi_k + \ddot{\Phi}_k \; .
\label{eqMotion-radFluid}
\ee
Note that this equation is the exact result for a pure radiation fluid.  Here,
$\theta_k$ are Fourier amplitudes of $\delta T/T$ with wavenumber $k$, and
$\Phi_k$ is a Fourier transform of gravitational potential.  Analysis of
solutions of the Einstein equations for $\Phi$ shows that $\Phi_k = {\rm
const}$ in two important cases:
\begin{enumerate}
\item For superhorizon scales, which are defined as $k\eta \ll 1$.
\item For all scales in the case of matter dominated expansion, $p=0$.
\end{enumerate}
In these situations the last term in the r.h.s. of
Eq.~(\ref{eqMotion-radFluid}), namely, $ \ddot{\Phi}_k$, can be neglected. 
The Einstein equations also restrict the initial conditions for fluctuations.
For the adiabatic mode in the limit  $k\eta \ll 1$ one finds 
\be
\delta_i = - 2\Phi_i  \; ,
\label{GravPot-InCond}
\ee
where $\delta_i \equiv \delta \rho/\rho$. The adiabatic mode is defined as a
perturbation in the total energy density. For the one component fluid, which
we consider here, only the adiabatic mode can exist.  Note that fractional
perturbation of the scale factor in metric (\ref{metric-perturbed}),
$a(\eta,{\bf x}) = a(\eta) +\delta a(\eta,{\bf x}) \equiv a(\eta) (1-\Phi)$,
can be expressed as perturbation of spatial curvature, see
Eq. (\ref{Spatial-curvature}). Therefore, adiabatic perturbations are also
called curvature perturbations.  Let us re-write Eq. (\ref{GravPot-InCond}) for
temperature perturbations:
\begin{itemize}
\item Radiation domination,  $\delta  = 4\, \delta T/T $, and we find 
\be
\theta_i = - \frac{\Phi_i}{2}\; .
\label{theta-phi-rd}
\ee
\item Matter domination,  $\delta  = 3\, \delta T/T $, and we find 
\be
\theta_i = - \frac{2\Phi_i}{3}\; .
\label{theta-phi-md}
\ee
\end{itemize}
Recall now that in the limit $k\eta \ll 1$ the gravitational potentail is
time-independent, $\Phi = {\rm const}$. Therefore,
Eq.~(\ref{eqMotion-radFluid}) has to be supplemented by the following initial
conditions in the case of the adiabatic mode:
\be
\theta_i \neq 0, \hspace{1cm} \dot{\theta}_i = 0\; .
\label{inCond-adiabatic}
\ee

\paragraph{Temperature fluctuations on largest scales.} 
Let us consider the modes which had entered the horizon after matter-radiation
equality, $ k \eta_{\rm eq} < 1$.  For those modes, $\dot{\Phi} = 0 $ all the
way from initial moments till present, and the solutions of
Eq.~(\ref{eqMotion-radFluid}) with adiabatic inital conditions is

\be
\theta(\eta ) + \Phi =
(\theta_i + \Phi)\, \cos \(\frac{k\eta}{\sqrt{3}}\)\; .
\label{radFluid-solution}
\ee
As gravity tries to compress the fluid, radiation pressure resists resulting
in acoustic oscillations. It is important that oscillations are synchronized.
All modes have the same phase regardless of $k$. This is a consequence of
$\dot{\theta}_i = 0$, which is valid for all $k$. At the last scattering, the
universe becomes transparent for the radiation and we see a snapshot of these
oscillations at $\eta = \eta^{}_{\rm ls}$.

To get its way to the observer, the radiation has to climb out of the
gravitational wells, $\Phi$, which are formed at the last scattering surface.
Therefore the observed temperature fluctuations are $\theta_{\rm obs}~=
\theta(\eta^{}_{\rm ls} ) + \Phi$, or
\be
\theta_{\rm obs} = \frac{1}{3}
\Phi_i\, \cos \(\frac{k\eta^{}_{\rm ls}}{\sqrt{3}}\) \; ,
\label{Sachs-Wolfe}
\ee
where we have used Eq.~(\ref{theta-phi-md}), which relates initial values of
$\theta$ and $\Phi$.  Note that overdense regions correspond to cold spots in
the temperature map on the sky, since the gravitational potential is
negative. This is the famous Sachs-Wolfe effect \cite{Sachs:1967er}.

\paragraph{Acoustic peaks in CMBR.}

Modes caught in the extrema of their oscillation, $k_n \eta^{}_{\rm
ls}/\sqrt{3} = n \pi$, will have enhanced fluctuations, yielding a fundamental
scale, or frequency, related to the universe sound horizon, $s_* \equiv
\eta^{}_{\rm ls}/\sqrt{3}$.  By using a simple geometrical projection, this
becomes an angular scale on the observed sky. In a spatially flat Universe,
the position of the first peak corresponds to $l_1 \approx 200$, see below.
Both minima and maxima of the cosine in Eq.~(\ref{Sachs-Wolfe}) give peaks in
the CMBR power spectrum, which follow a harmonic relationship, $k_n = {n
\pi}/s^{}_{*}$,~ see Fig.~\ref{PowerSpectrum-WMAP}.
\begin{figure}
\begin{center}
\includegraphics[width=10.cm]{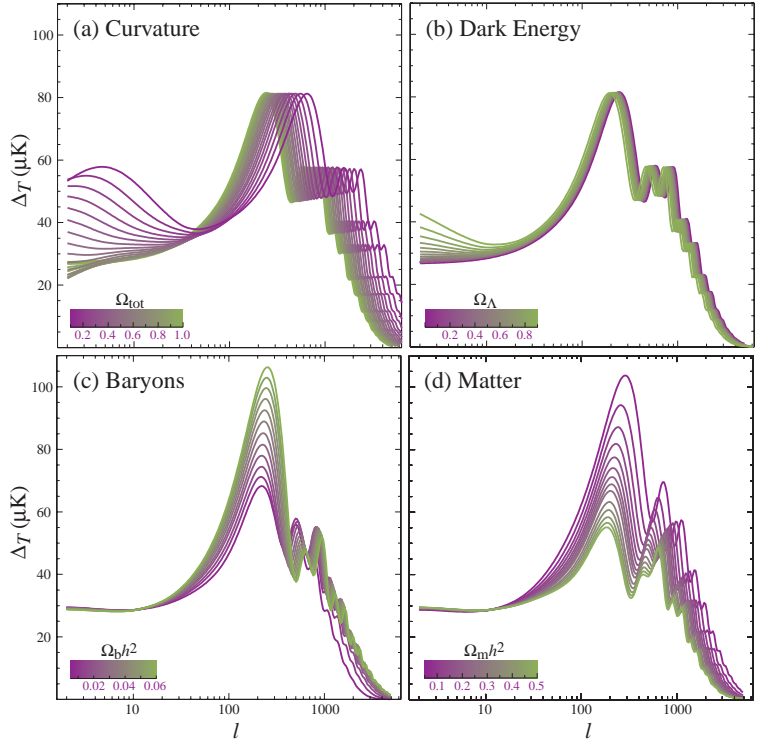}
\caption{Sensitivity of the CMBR angular power spectrum to four fundamental
cosmological parameters (a) the curvature as quantified by $\Omega_{\rm tot}$
(b) the dark energy as quantified by the cosmological constant
$\Omega_\Lambda$ ($w_\Lambda=-1$) (c) the physical baryon density $\Omega_b
h^2$ (d) the physical matter density $\Omega_m h^2$, all varied around a
fiducial model of $\Omega_{\rm tot}=1$, $\Omega_\Lambda=0.65$, $\Omega_b
h^2=0.02$, $\Omega_m h^2=0.147$. ~From Ref.~\cite{Hu:2001bc}. }
\label{fig:CMB-parameter-dependence}
\end{center}
\end{figure}

The amplitudes of the acoustic peaks are recovered correctly after the
following effects are taken into account: 1) baryon loading; 2)
time-dependence of $\Phi$ after horizon crossing in radiation dominated
universe; 3) dissipation.

The effect of baryons is exactly the same for the oscillator equation
Eq.~(\ref{radFluid-solution}), as if we had increased the mass of a load
connected to a spring and oscillating in a constant gravitational field
starting on top of an uncompressed coil at rest. The addition of baryons makes
a deeper compressional phase, and therefore increases every other peak in the
CMBR power spectrum. (First, third, fifth, $\dots$) The CMBR power spectrum is
a precise baryometer.

Gravitational potentials are not constant, but decay inside the horizon during
radiation domination. This decay drives the oscillations: it is timed to leave
compressed fluid with no gravitational force to fight when the fluid turns
around.  Therefore, the amplitudes of the acoustic peaks increase as the cold
dark matter fraction decreases, which allows to measure $\Omega_m$.

Dissipation leads to dumping of higher order peaks in the CMBR power spectrum.

The dependence of the CMBR angular power spectrum on different cosmological
parameters is shown in Fig.~\ref{fig:CMB-parameter-dependence}.

\paragraph{Position of the first peak.}
Position of the first peak is determined by the angular size of the sound
horizon at last scattering. Let us calculate here a similar quantity: the
causal horizon (which is larger by a factor of $\sqrt{3}$ in comparison with
the sound horizon).  The comoving distance traveled by light, $ds^{2} = 0$,
from the ``Big Bang'' to redshift z is determined by a relation similar to
Eq.~(\ref{comovingD-z}), but with different integration limits
\be 
\eta (z) = \int_z^{\infty}
\frac{dz'}{H(z')} \; ,
\label{com-fromInfty-toZ}
\ee 
where $H(z)$ is given by Eq.~(\ref{Friedmann-parmetrization}).  One has to
integrate this relation with a complete set of $\Omega_i$, but for simplicity
let us consider here the Universe dominated by a single component $\rho_j$
$$
\eta (z) = \frac{(1+z)^{-\gamma_j}}{\gamma_j\; H_0} \; ,
$$
where $\gamma_j \equiv (1+3w_j)/2$. From the last scattering to $z \sim 1$,
the Universe was matter dominated.  Therefore, the causal horizon in matter
dominated Universe $(w_j = 0)$
$$
\eta (z) = \frac{2}{H_0\sqrt{1+z}}
$$
should give a reasonable first approximation.

Consider two light rays registered at $z=0$ which were separated by a comoving
distance $\chi = \eta(z)$ at the moment of emission. Since both propagate in
the metric $ ds^{2} = a^{2}(d\eta^{2} - d\chi^{2} - \chi^{2} d\theta^{2}) = 0$,
we find for the angular size of horizon at last scattering
\be
\theta_h =\frac{\eta (z_{\rm ls})}{\eta (0)} =
\frac{1}{\sqrt{1+z_{\rm ls}}} = \sqrt{\frac{T_0}{T_{\rm ls}}} \approx
2^\circ\; .
\label{angSize-hor-ls}
\ee
\begin{itemize}
  \item This gives the position of the first acoustic peak,~ $l \approx
  200$.
  \item Tells us that there were~ $10^{4}$ causally disconnected
  regions at the surface of last scattering.
\end{itemize}

\paragraph{Horizon problem.}

Regions separated by more than $\;>\;2^\circ$ have not been in the causal
contact prior to the last scattering in the standard Friedmann cosmology.  The
microwave sky should not be homogeneous on scales $\;>\;2^\circ$.  Yet, CMB is
isotropic to better than~ $10^{-4}$ on all scales. Observations tell us that
all modes were, indeed, synchronized according to adiabatic initial
conditions, Eq.~(\ref{inCond-adiabatic}), with only small initial
perturbations present, $\Phi_i \ll 1$. This constitutes the so-called
``Horizon problem`` of standard cosmology. In Section~\ref{sec:inflation} we
will see how this problem is solved in frameworks of inflationary cosmology.

\begin{figure}
\begin{center}
\includegraphics[width=6.cm]{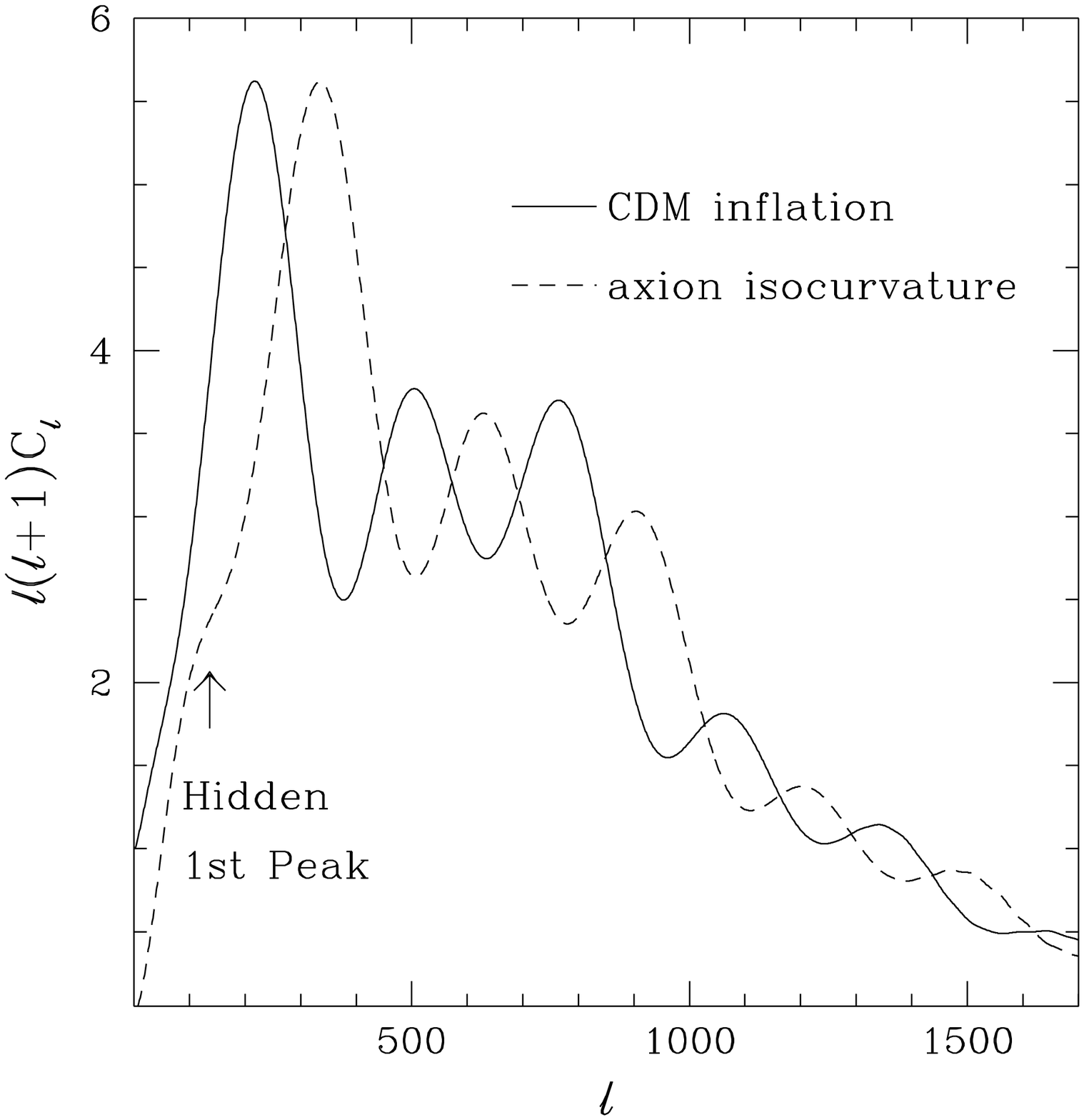}
\hspace{1cm}
\includegraphics[width=8.cm]{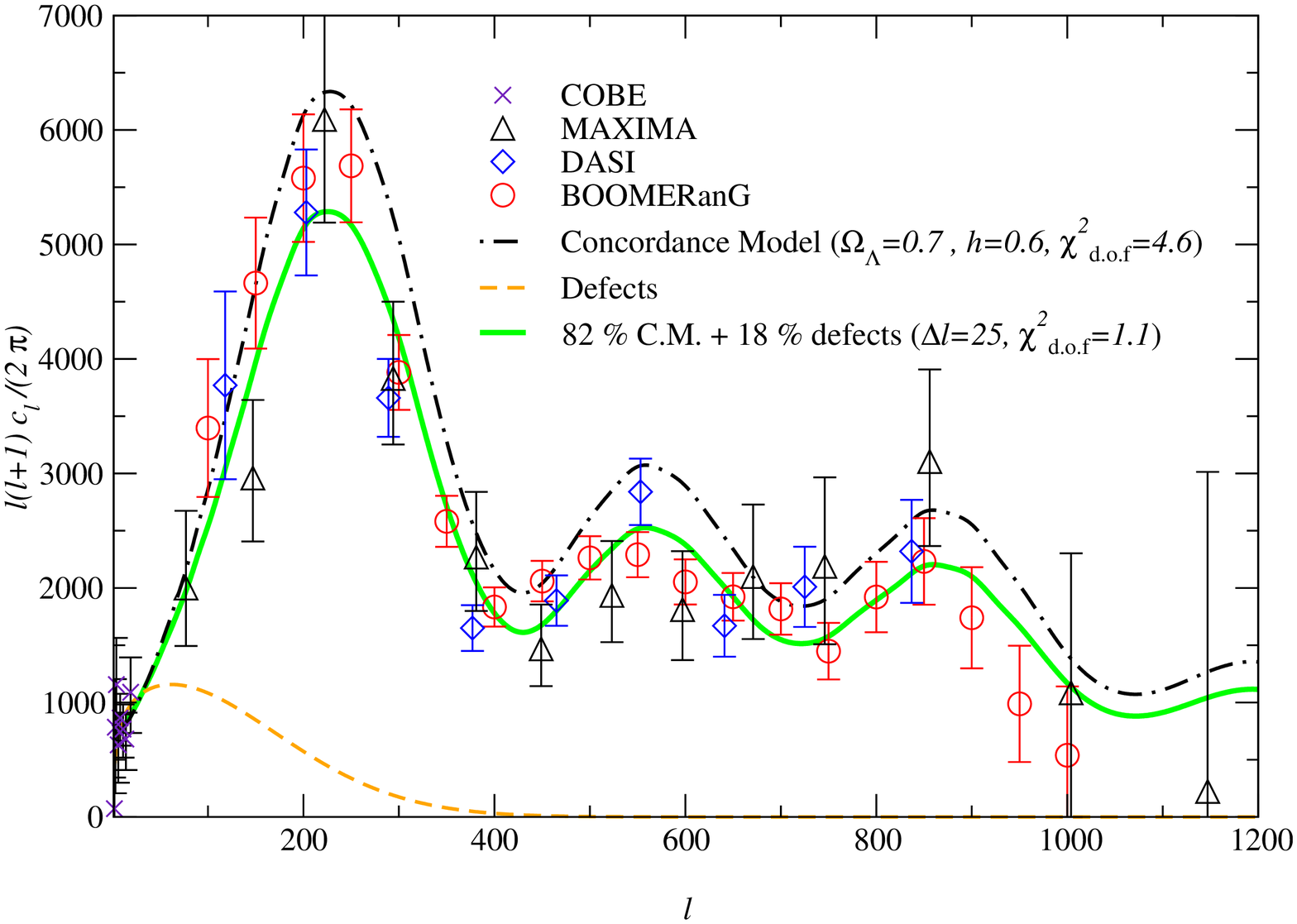}
\caption{Left panel: comparison of CMB power spectra in the models with
  adiabatic and isocurvature initial perturbations, from
  Ref.~\cite{Hu:1996vr}. Right panel: adiabatic power spectra in comparison
  with spectra appearing in models seeded by topological defects, from
  Ref.~\cite{Bouchet:2000hd}. In this panel some older, pre-WMAP, data
  are also shown.}
\label{fig:non-adiabatic}
\end{center}
\end{figure}

\paragraph{Non-adiabatic perturbations.}
For the isocurvature perturbations, instead of Eq.~(\ref{inCond-adiabatic}),
the initial conditions are given by
\be
\delta_i = 0, \hspace{1cm} \dot{\delta}_i \neq 0\; .
\label{inCond-ithocurvature}
\ee
That is because, in this case, perturbation in total density (and therefore in
curvature) are zero initially.  As a consequence, in
Eq.~(\ref{radFluid-solution}) we will have sine instead of cosine. Acoustic
peaks will be shifted by half a period, see
Fig.~\ref{fig:non-adiabatic}. Therefore, isocurvature perturbations are ruled
out by modern CMBR experiments.

If density perturbations would be seeded by topological defects (e.g. cosmic
strings), both sine and cosine will be present in the solution for temperature
fluctuations, Eq.~(\ref{radFluid-solution}). That is because the source for
$\Phi_k$ is active inside the horizon and phases of $\theta_k$ will be random.
Acoustic peaks will be absent, see Fig.~\ref{fig:non-adiabatic}.  Structure
formation seeded primarily by topological defects is ruled out by modern CMBR
experiments.

\section{Large scale distribution of galaxies}
\label{sec:LSS}

\begin{figure}
\begin{center}
\includegraphics[width=7cm]{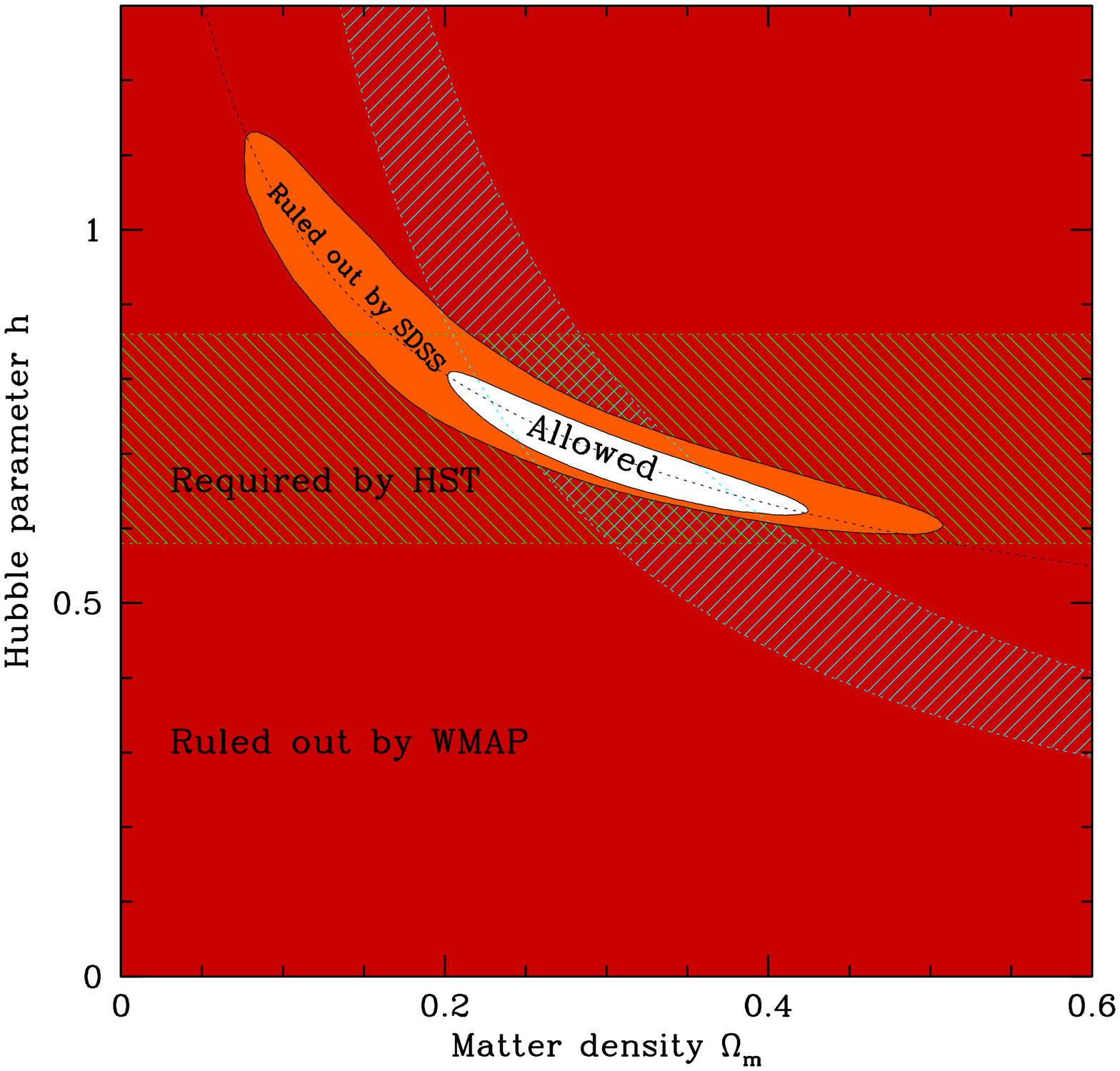}
\hspace{.7cm}\parbox{8cm}{\vspace{-6.7cm}
\includegraphics[width=7.cm]{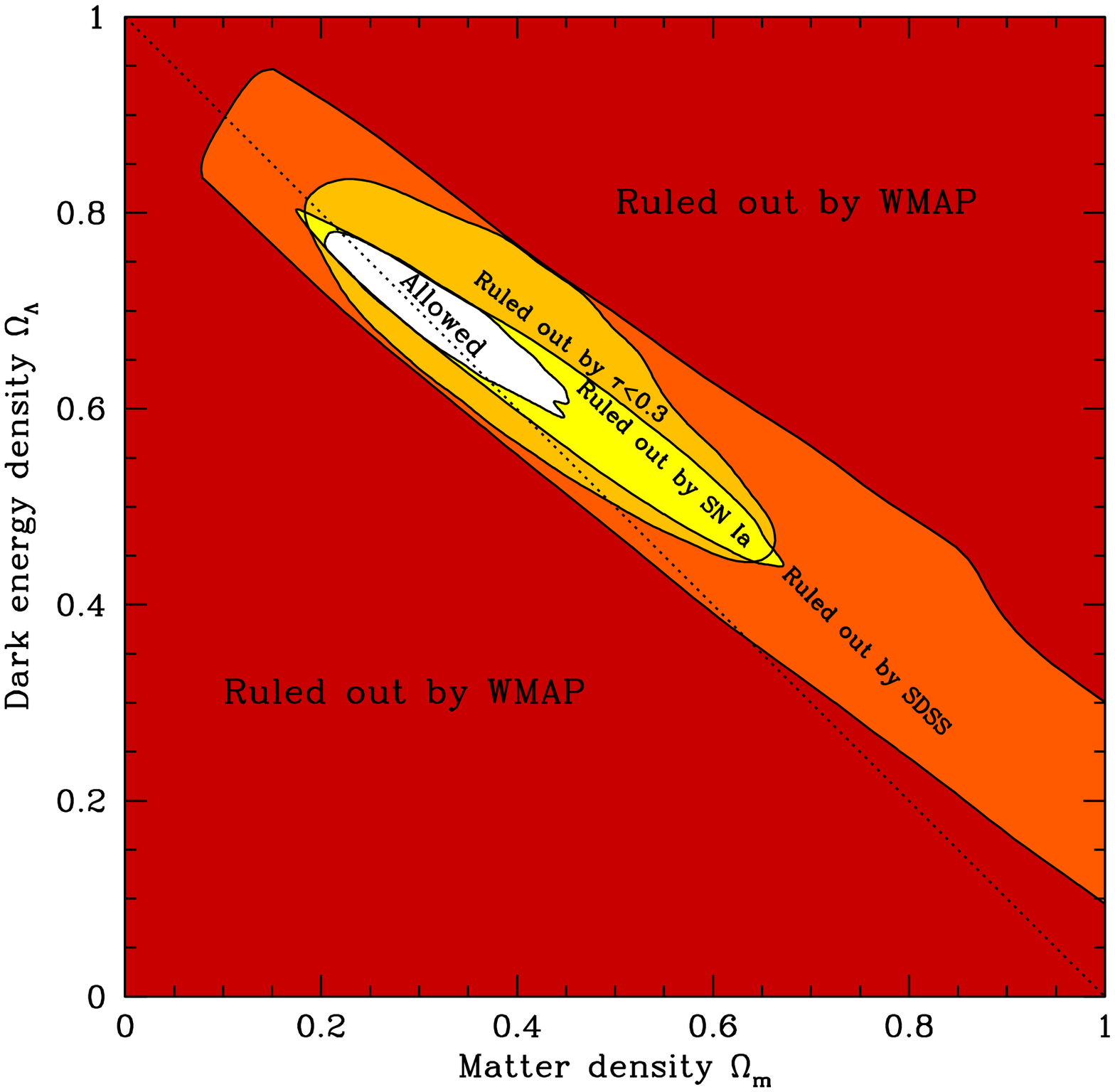}}
\caption{Combined CMBR and large-scale structure constraints.  {\bf Left
panel:} 95\% constraints in the $(\Omega_m,h)$ plane.  {\bf Right panel:} 95\%
constraints in the $(\Omega_m,\Omega_\Lambda)$ plane. From
Ref.~\cite{Tegmark:2003ud}.}
\label{fig:SDSS1}
\end{center}
\end{figure}

Primordial cosmological fluctuations leave their imprint as CMBR anisotropies
(discussed in the previous Section), and as density perturbations which give
rise to galaxies and clusters of galaxies.  CMBR anisotropies are observed on
a two dimensional surface of last scattering, and therefore are measured as a
two dimensional power spectrum. On the other hand, the distribution of
galaxies can be measured in three dimensions.  (Two angular coordinates of the
line of sight to a galaxy and its redshift.)  Different physical processes
influence the initial perturbations until they are transformed into CMBR
fluctuations or fluctuations of the distribution of galaxies. This influence
can be encoded as a function of momenta, the transfer function $T(k)$, which
simply maps the power spectrum of the initial perturbations into the observed
power spectrum, and is a function of cosmological parameters. Therefore, the
distribution of galaxies gives complimentary information with respect to CMBR
anisotropies and helps to break degeneracy between cosmological parameters and
the initial spectrum.

This is illustrated in Fig.~\ref{fig:SDSS1} with CMBR data from WMAP and
large-scale structure data from SDSS. The left panel corresponds to 95\%
constraints in the $(\Omega_m,h)$ plane.  The shaded dark red region is ruled
out by WMAP alone leaving the long banana region.  This shows that these two
basic cosmological parameters are not well constrained by WMAP alone. The
shaded light red region is ruled out when SDSS information is added.  The
small (shown as white) region remain allowed. Note that the allowed region is
in good agreement with a completely independent measurement by HST key project
based on entirely different physics. The combined WMAP + SDSS constraint is
even tighter than HST project measurement.

One should bear in mind that there are caveats here. In deriving the WMAP+SDSS
constraints which are shown in this figure, it was assumed that the universe
is spatially flat, neutrino have negligible masses and the primordial spectrum
is a pure power law. Without these priors the constraints are less tight.

The constraints in the $(\Omega_m,\Omega_\Lambda)$ plane with the assumption
about spatial curvature being relaxed is shown in Fig.~\ref{fig:SDSS1}, right
panel.  The shaded dark red region is ruled out by WMAP alone, illustrating
the well-known geometric degeneracy between models that all have the same
acoustic peak locations.  The shaded light red region is ruled out when adding
SDSS information.  Continuing inwards, the next two regions are ruled out
allowing the assumption that re-ionization optical depth $\tau<0.3$ and when
supernova SN Ia information is included.  $ \Omega_\Lambda > 0$ is required
with high confidence only when CMBR is combined with galaxy clustering
information, or SN Ia information, see the next Section.

\section{Dark energy}
\label{sec:darkEnergy}

Something which is often called ``Dark Energy'' reveals itself in a variety of
cosmological and astrophysical observations.  This form of matter gravitates,
but does not cluster.  Contrary to radiation or dark matter, the dark energy
causes the {\it accelerated} expansion of the universe.  The need for it was
hinted long ago to resolve conflict between the measured Hubble constant and
the lower limits on the age of the Universe. Without the cosmological constant
it was also not possible to obtain the correct growth of large scale
structures in the $\Omega = 1$ Universe.  Recently, the presence of dark
energy was derived from the spectrum CMBR anisotropies and directly detected
in the Hubble diagram of high redshift supernovae.

\paragraph{Age of the Universe.}
If there is no dark energy, the Universe should be matter dominated 
and should expand according to $a = (t/t_0)^{2/3}$. Differentiating this
expansion law we find 
\be H t = 2/3  \; .
\label{Ht-matter}
\ee
The value of the Hubble constant, as derived by the Hubble Key Project
from the Hubble diagram, is $72\pm 8 $~km~s$^{-1}$~Mpc$^{-1}$
\cite{Freedman:2000cf}, see Fig.~\ref{fig:HubbleDiagram}.
On the other hand, the lower bound on the age of the Universe can be
established estimating the ages of various objects it consists of. For example,
the temperature of the coldest white dwarfs in globular clusters yields a
cluster age of $12.7 \pm 0.7$ Gyr \cite{2002ApJ...574L.155H}. This gives
\be H_0 t_0 > 0.93 \pm 0.12  \; ,
\label{Ht-matter}
\ee 
in clear disagreement with Eq.~(\ref{Ht-matter}). In other words, the Universe
appears much younger than the ages of the oldest objects in it. The critical
density Universe, $\Omega=1$, cannot consist of pressureless matter if
measurements of the Hubble constant are correct and Friedmann equations are
valid.

\begin{figure}
\begin{center}
\includegraphics[width=7cm]{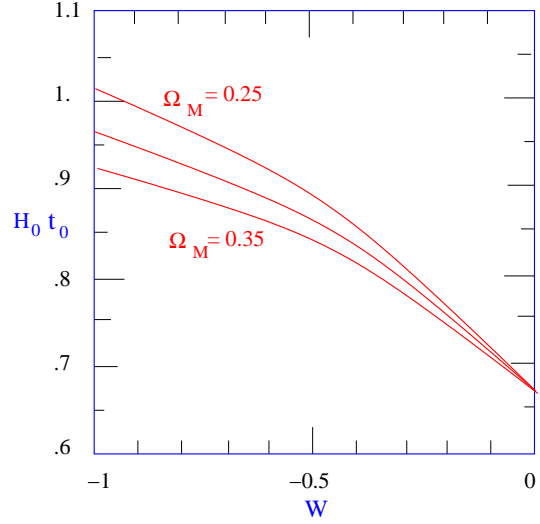}
\caption{Graphical representation of Eq.~(\ref{age-expansion_MandDE}) assuming
critical density, $\Omega_{\rm tot} = 1$.}
\label{fig:Ht-w}
\end{center}
\end{figure}

The simplest cure (but ``embarrassing'' from the point of view of the particle
physicist), is to add a cosmological constant, or dark energy.
It should be stressed that this minimal modification of Friedmann equations 
is consistent with all other current cosmological tests and measurements.
In the general case, the age of the Universe can be related to the expansion 
history as
\be
t_0 = \int_0^{t_0} dt = \int_0^{t_0} a d\eta = 
\int_0^{\infty} \frac{dz}{(1+z)H(z)}\; .
\label{age-expansion}
\ee
Here we have used Eqs.~(\ref{time-conformal}), (\ref{z-a-ehist}),
(\ref{deta-dz}). For two components, pressureless matter and dark energy with
equations of state $w$, this relation can be written as (see
Eq.~(\ref{Friedmann-parmetrization})):  
\be
 H_0 t_0 = \int_0^{\infty} \frac{dz}{(1+z)^{5/2}\sqrt{\Omega_M +
\Omega_{\rm DE} (1+z)^{3w}}} \; .
\label{age-expansion_MandDE}
\ee
The product $H_0 t_0$  as a function of $w$, assuming $\Omega_M
+\Omega_{\rm DE}=1$, is shown in Fig.~\ref{fig:Ht-w}. We see that it matches
the observational constraints when $w \approx -1$ and $\Omega^{~}_M \approx
0.3$. 

A discussion of further evidence for dark energy, e.g., related to the problem
of the growth of density perturbations, can be found in
Ref.~\cite{Huterer:2000mj}.

\paragraph{Redshift - Luminosity Distance relation for Supernovae Ia.}

For the two-component energy content of the Universe, presureless matter and
dark energy, the expression for the luminosity distance,
Eq.~(\ref{comovingD-z}), takes the form
\be
D_L = \frac{1+z}{H_0}\;  \int_0^{z} \,
\frac{dz'}{(1+z')^{3/2}\sqrt{\Omega_M +
\Omega_{\rm DE} (1+z')^{3w}}} \; ,
\label{age-expansion_MandDE}
\ee 
see Eqs.~(\ref{luminocity-distance-result}), (\ref{comovingD-z}),
(\ref{Friedmann-parmetrization}). To use this relation as a cosmological test
in conjunction with Eq. (\ref{z-vs-F}), one has to find a set of standard
candles. This is a big challenge in practice, since we have to find very
bright objects which can be seen from far away. At the same time, all of them
should have the same luminosity, and we have to be sure that they do not
evolve intrinsically.  These requirements rule out galaxies and
quasars. However, supernovae seem to be suitable. They are bright, as bright
as the whole galaxy at the peak of luminosity, and Type Ia supernovae appear
to be standard candles. These type of supernovae are thought to be nuclear
explosions of white dwarfs in binary systems. The white dwarf, a stellar
remnant supported by the degenerate pressure of electrons, accrete matter from
a companion and its mass increases toward the Chandrasekhar limit of about
$1.4\, M_\odot$.  Near this limit, the degenerate electrons become
relativistic, which leads to instability and the white dwarf explodes. This
physics allows the explosions to be calibrated, since instability occur under
the same conditions.

To proceed, I have to remark about the units of flux used by astronomers, which
are magnitudes. The system is ancient, and has its origin in the logarithmic
response of the human eye.  The ratio of the flux of two objects is then given
by a difference in magnitudes; i.e., 
\be 
m_2 - m_1 = - 2.5 \log_{10}(F_2/F_1) \; .
\label{apparent-magnitude}
\ee 
A smaller magnitude means larger flux. 

\begin{figure}
\vspace{-4.5cm}
\begin{center}
\hspace{-3.5cm}
\vspace{-1.5cm}
\includegraphics[width=9cm]{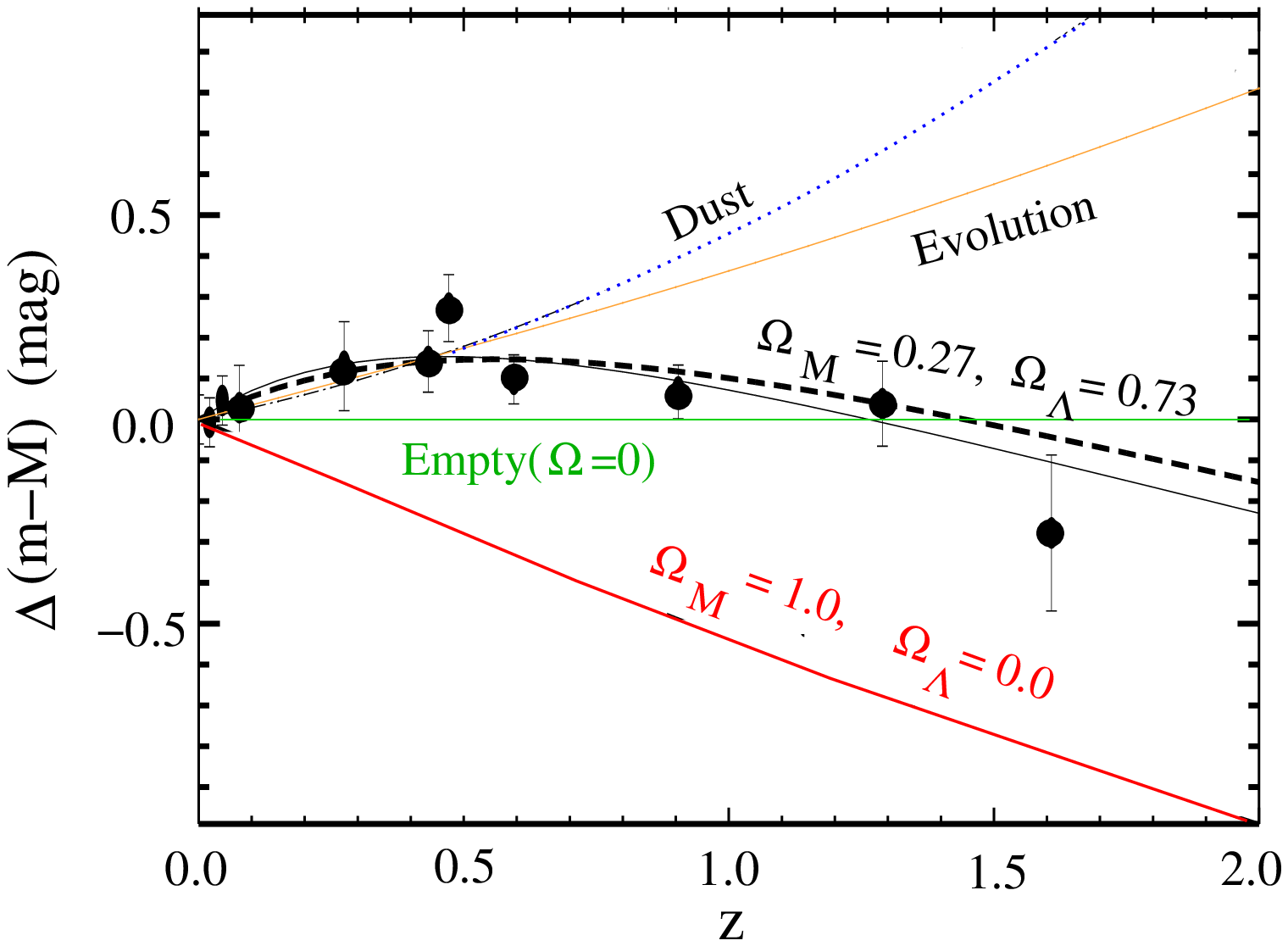}
\hspace{0.5cm}
\parbox{6cm}{\vspace{-7.8cm}
\includegraphics[width=8.cm]{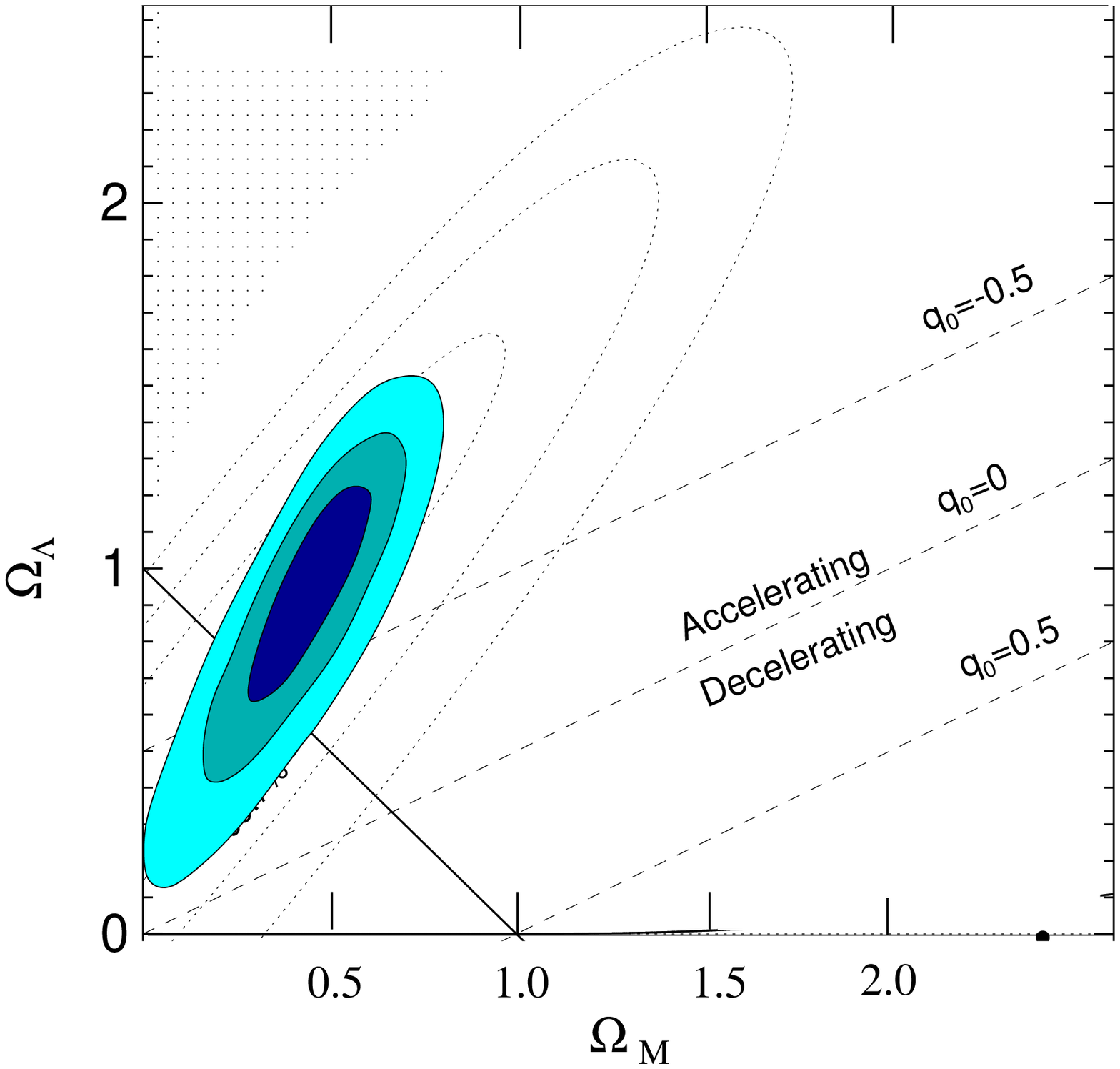}}
\caption{{\bf Left panel:} SN Ia residual Hubble diagram comparing
  astrophysical models and models for astrophysical dimming. Data (weighted
  averages in fixed redshift bins) and models are shown relative to an empty
  Universe model $(\Omega = 0)$, adopted from Ref.~\cite{Riess:2004nr}.  {\bf
  Right panel:} Joint confidence intervals for ($\Omega_M$,$\Omega_\Lambda$)
  from SNe~Ia, Ref~\cite{Riess:2004nr}.  The dotted contours are the results
  from Ref.~\cite{Riess:1998cb}, illustrating the earlier evidence for
  $\Omega_\Lambda > 0$.  The figure is adopted from Ref.~\cite{Riess:2004nr}.}
\label{fig:RDOF}
\end{center}
\end{figure}

Fig.~\ref{fig:RDOF}, left panel, shows the corresponding SNe Ia
redshift-luminosity distance diagram. Data points correspond to magnitudes of
SNe Ia measured at different redshifts.  The case of $\Omega_M=1$ (red curve)
is ruled out. The ``concordance model'' $\Omega_M=0.27$ and
$\Omega_\Lambda=0.73$ (black dashed curve) is within $1\sigma$. For a flat
geometry prior, best fit corresponds to $\Omega_M=0.29\pm^{0.05}_{0.03}$
(correspondingly $\Omega_\Lambda=0.71$).  Data are inconsistent with a simple
model of evolution of SNe Ia, or dimming due to light absorption by dust as an
alternatives to dark energy.  The shaded area in Fig.~\ref{fig:RDOF}, right
panel, corresponds to 68\%, 95\% and 99.7\% confidence levels in the
($\Omega_M$, $\Omega_\Lambda$) plane.

\paragraph{Constraints from CMB.}

\begin{figure}
\begin{center}
\epsfig{file=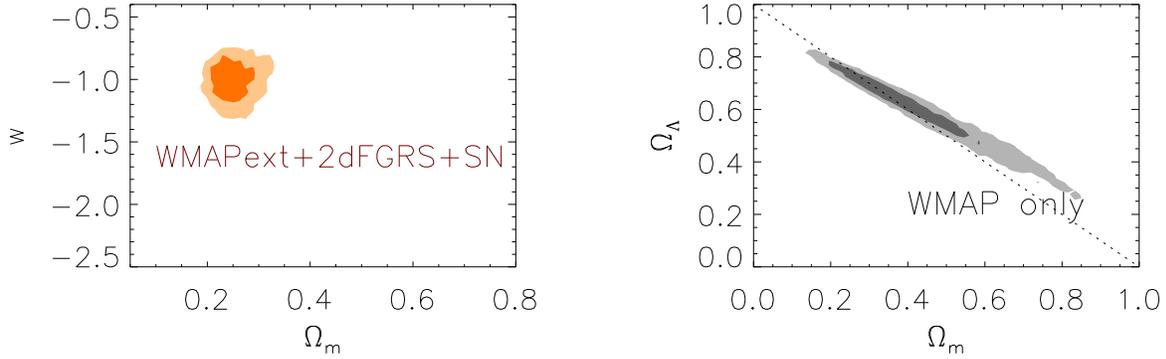,height=5.5cm,%
bbllx=300,bblly=200,%
bburx=860,bbury=390,%
clip=}
\caption{ {\bf Left panel:} Constraints on the equation of state of dark
energy in the ($w$, $\Omega_M$) plane for a combination of the CMBR + 2dF +
SNe Ia data sets.  {\bf Right panel:} Constraints on the geometry of the
universe in the ($\Omega_M$, $\Omega_\Lambda$) plane assuming the prior $h >
0.5$.  From Ref.~\cite{Spergel:2003cb}}.
\label{fig:dark-energy-WMAP}
\end{center}
\end{figure}

Tight constraints on dark energy, and in a direction in parameter space which
is ``orthogonal'' to SNe Ie constraints, are obtained from fitting the power
spectrum of cosmic microwave background anisotropies, see the discussion in
Section~\ref{sec:CMBR} The WMAP data alone rule out the standard $\Omega_M =
1$ CDM model by $7\sigma$ if the prior $h>0.5$ is accepted
\cite{Bennett:2003bz}. The resulting confidence levels in ($\Omega_M$,
$\Omega_\Lambda$) plane are shown in Fig.~\ref{fig:dark-energy-WMAP}.  While
the CMBR data alone are compatible with a wide range of possible properties
for the dark energy, the combination of the WMAP data with either the HST key
project measurement of $H_0$, the 2dFGRS measurements of the galaxy power
spectrum or the Type Ia supernova measurements requires that the dark energy
be $\Omega_\Lambda = 0.73 \pm 0.04$ of the total density of the Universe, and
that the equation of state of the dark energy satisfy $w< -0 78$ (95\%
CL) \cite{Bennett:2003bz}.

\paragraph{Constraints from gravitational lensing.}

Gravitational lensing will be discussed in Section~\ref{sec:DM-motivation}.
Here we just note that the analysis of strong lensing of sources with known
redshift is sensitive to the value of the geometrical cosmological parameters
of the Universe. A recent study \cite{Soucail:2004se} of the lensing
configuration in the cluster Abell 2218 is in agreement with the concordance
model. In particular, assuming the flat Universe, it gives for the equation of
state of dark energy $w< -0.85$. These constraints are consistent with the
current constraints derived from CMB anisotropies or supernovae studies, but
they are completely independent tests, providing nearly orthogonal constraints
in the $(\Omega_M,\Omega_{\Lambda})$ plane, see Fig.~\ref{Abell2218_cosmo}

\paragraph{Biggest Blinder -- Biggest Surprise.}

From the point of view of the particle physicis,t the cosmological constant
just should not be there. Indeed, in quantum field theory, the cosmological
constant corresponds to vacuum energy, which is infinite and has to be
renormalized,
\be
\rho_{\rm vac} =\frac{1}{2\,(2\pi )^{3}}\int_0^{k_{\rm max}}\omega_k\, k^2\, 
dk\;\; .
\label{vacuum-energy}
\ee 

The natural value for the cut-off in this integral is the Plank scale, and
then $\rho_{\rm vac}\sim ~M_{\rm Pl}^{4}~ \approx ~10^{74} \; {\rm GeV}^{4}$.
Exact supersymmetry can make this integral vanish. Indeed, in
Eq.~(\ref{vacuum-energy}), the contribution of one Bosonic degree of freedom
is counted. Fermions contribute with an opposite sign, and if there is an
equal number of Bosons and Fermions with equal masses, the vacuum energy will
be zero.  However, supersymmetry is broken at least at the electroweak scale,
and then $\rho_{\rm vac}$ should not be smaller than $\sim ~M_{\rm W}^{4}~
\approx ~10^{8} \; {\rm GeV}^{4}$. Before dark energy was detected, it was
believed that some yet unknown mechanism reduces the cosmological constant to
zero.  Zero is a natural number.  However, it is hard to understand the
smallness of the observed value $\rho_{\rm vac} \approx 10^{-46} \; {\rm
GeV}^{4}$. Moreover, there is another pressing issue of fine tuning: why the
detected value of $\rho_{\rm vac}$ approximately equals to the energy density
of matter at the {\it present} epoch of cosmological evolution? The ratio of
these two contributions scales as $a^3$ and, say, at recombination the vacuum
energy was only $10^{-9}$ of matter energy...  Detection of dark energy not
only points to a new physics, but hints that we are missing SOMETHING very
fundamental.


\section{DARK MATTER}
\label{sec:DM-motivation}

CMBR observations accurately measure the geometry of the Universe, its present
expansion rate, its composition, and the nature and spectrum of the primordial
fluctuations. Nevertheless, the traditional cosmological tests are still
important. In particular, degeneracies between different parameter sets
exist, which can produce the same CMBR spectra, and the conclusions drawn do
rest upon a number of assumptions. Below we consider cosmological observations
that are independent of the CMB and point to the existence of non-baryonic
dark matter.

\subsection{DARK MATTER: motivation}
The missing mass is seen on all cosmological scales and reveals itself via:
\begin{itemize}
  \item Flat rotational curves in galaxies.
  \item Gravitational potential which confines galaxies and hot gas in
    clusters.
  \item Gravitational lenses in clusters.
  \item Gravitational potential which allows structure formation from tiny
    primeval perturbations.
\end{itemize}

\subsubsection{Dark Matter in Galaxies}

\paragraph{Galactic rotational curves.}

Consider a test particle which is orbiting a body of mass $M$ at a distance
$r$. Within the frameworks of Newtonian dynamics the velocity of a particle is
given by
\be
v_{\rm rot} =\sqrt{G\, M(r) \over r} \; .
\label{rot-velocity}
\ee 
\begin{figure}
\begin{center}
\includegraphics[width=15.cm]{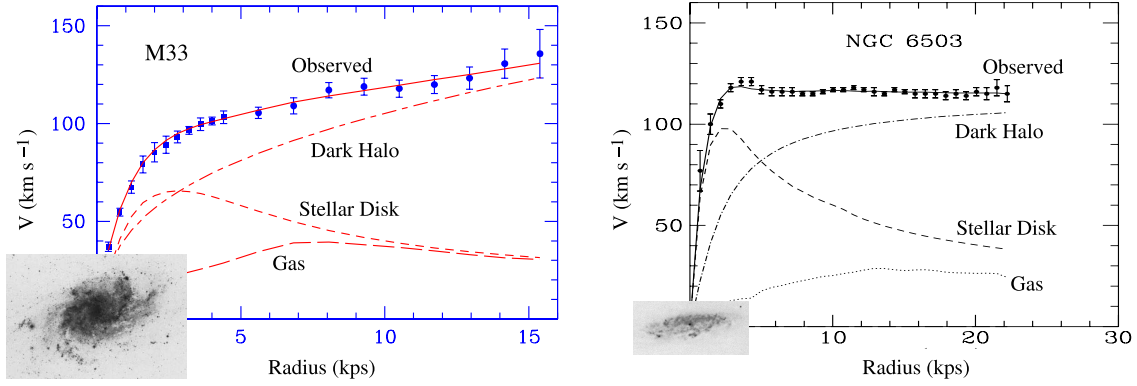}
\vspace{-.7cm}
\caption{The rotational curves of two sample galaxies. Left panel - M33,
  adapted from Ref.~\cite{Corbelli:1999af}. Right panel - NGC6503, adapted
  from Ref.~\cite{1991MNRAS.249..523B}. I superimposed with the rotational
  curves the optical images of corresponding galaxies, approximately to
  scale.}
\label{RotCurves}
\end{center}
\end{figure}

Outside of the body, the mass does not depend on distance, and the rotational
velocity should obey the Kepler law, $v_{\rm rot} \propto r^{-1/2}$.  Planets
of the Solar system obey this law. However, this is not the case for stars or
gas which are orbiting galaxies. Far away from the visible part of a galaxy,
rotational curves are still rising or remain flat. Two examples are shown in
Fig.~\ref{RotCurves}. An optical image of the M33 galaxy is superimposed with
its rotational curve, approximately to correct scale. The contribution of
visible baryons in the form of stars and hot gas can be accounted for, and the
expected rotational curve can be constructed. The corresponding contributions
are shown in Fig.~\ref{RotCurves}. One can see that the data-points are far
above the contribution of visible matter. The contribution of missing dark
mass, which should be added to cope with data, is also shown and is indicated
as Dark halo.  For the rotational velocity to remain flat, the mass in the
halo should grow with the radius as $M(r) \propto r$, i.e., the density of
dark matter in the halo should decrease as $\rho(r) \propto r^{-2}$.

\paragraph{Halo structure.}
\begin{figure}
\begin{center}
~\hspace{-0.5cm}
\includegraphics[width=7.cm]{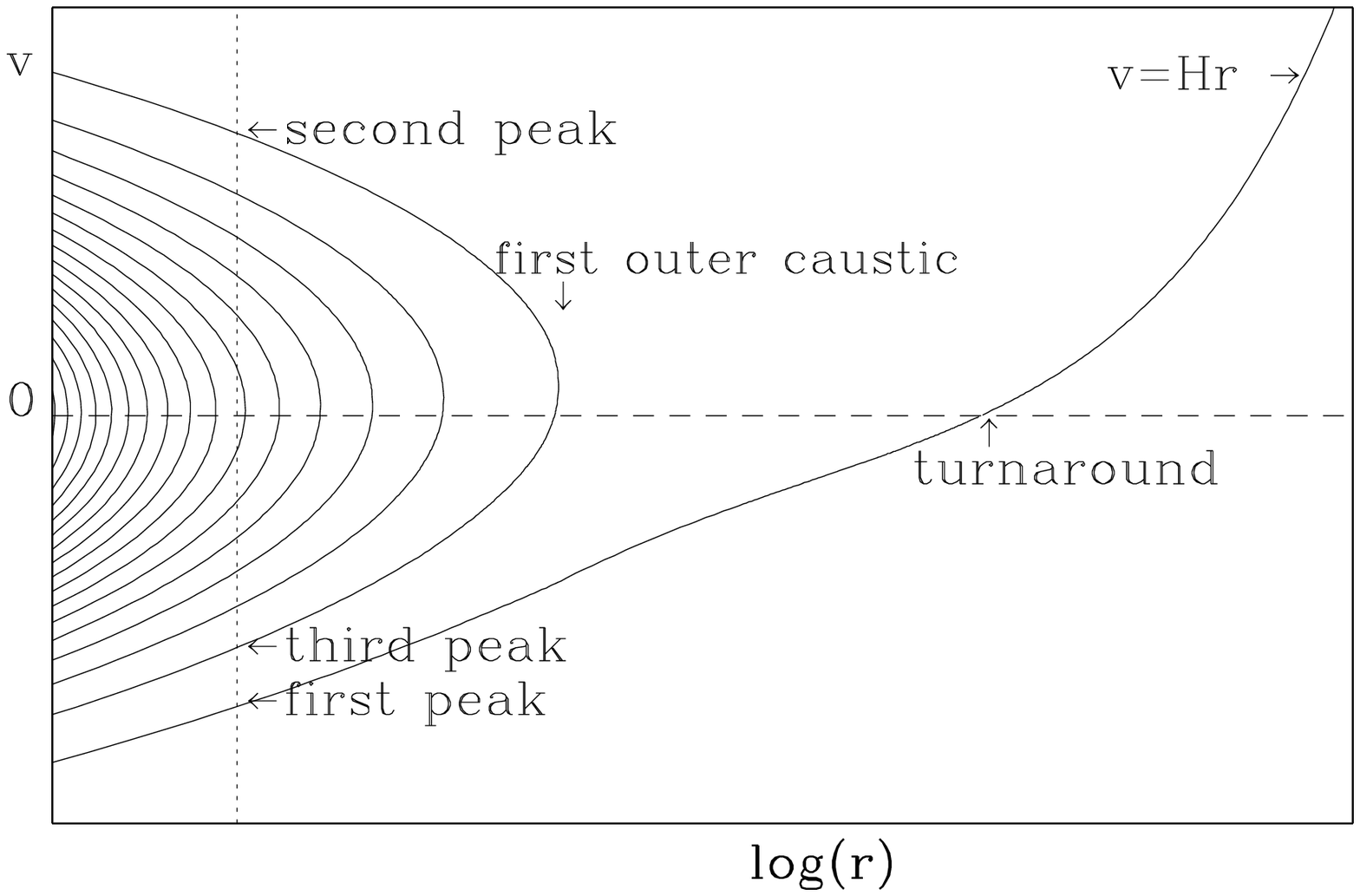}
~\hspace{1.5cm}
\includegraphics[width=7.cm]{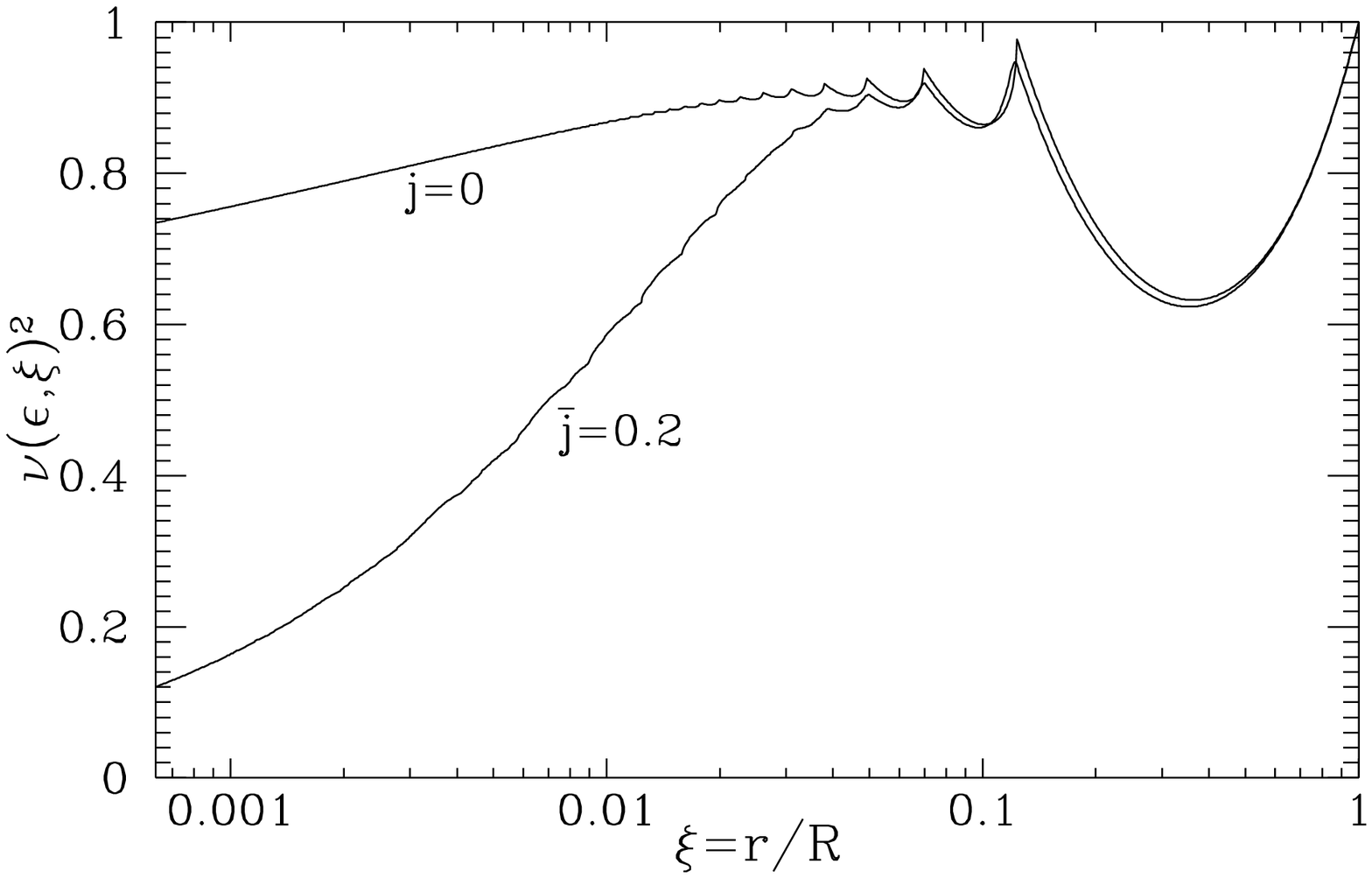}
\caption{Left panel: the phase space structure of an infall model. Right
  panel: rotational curves in an infall models. Two curves which correspond to
  different angular momenta are shown. From Ref. \cite{Sikivie:1997nn}.}
\label{phase-space-infall}
\end{center}
\end{figure}

For direct and indirect dark matter searches it is important to know the
phase-space structure of the dark halo as well. With dark matter particles
that are interactive, a thermal distribution over velocities would eventually
be established. However, in conventional cold dark matter models, particles
are non-interacting, except gravitationally. Binary gravitational interactions
are negligible for elementary particles, and the resulting phase-space
distributions are not unique, even for stationary equilibrium states, and even
if flat rotational curves are reproduced.

1. The simplest self-gravitating stationary solution which gives flat
rotational curves corresponds to an ``isothermal sphere'' with Maxwellian 
distribution of particles over velocities:
\be
n(\vec{r},\vec{v}) = n(r)\; e^{-v^{2}/v^{2}_0}\; .
\label{n-isothermal}
\ee
Solution of the equation of hydrostatic equilibrium can be approximated by the
density profile
\be  \rho (r) = \frac{\rho_0}{(1 + x^{2})},~~~
{\rm where}~~   x \equiv r/r_c \; .
\label{rho-isothermal}
\ee
It should be stressed that the distribution Eq.~(\ref{n-isothermal}), in
contrast to a distribution in real thermal equilibrium, depends on particle
velocities, not on their energies. Such distributions may arise in
time-dependent gravitational potential as a result of collisionless
relaxation.

2. There exist several density profiles which are empirical fits to numerical
simulations, e.g.  Navarro, Frenk \& White (NFW) profile \cite{Navarro:1997he}
and Moore {\it et al.} profile \cite{Ghigna:1999sn}
\begin{eqnarray}
\rho (r) &=& \frac{\rho_0}{x\; (1 + x)^{2}}~~~~~~~~~~~~~~{\rm NFW,}\\
\label{NFW}
\rho (r) &=& \frac{\rho_0}{x^{3/2}\; (1 + x^{3/2})}~~~~~~{\rm  Moore~ 
{\it et~ al.}}
\label{Moore}
\end{eqnarray}

4. In the cold dark matter model, the distribution of particles in the phase
space during initial linear stage prior to structure formation corresponds to
thin hypersurface, ${\bf v} = H {\bf r}$ (or line in the Hubble
diagram). Since during collisionless evolution the phase-space density
conserves, at the non-linear stage the distribution will still be a thin
hypersurface. If can be deformed in a complicated way and wrapped around, but
it cannot tear apart, intersect its own folds, puff up or dissolve. The
corresponding phase-space distribution for the case of spherical symmetry is
shown in Fig.~\ref{phase-space-infall}, left panel. With time, non-linear
structure grows, and the infall of new particles continues. This manifests
itself as a growth of turnaround radius (which is a surface where $v=0$; the
turnaround radius of our Galaxy is at 1 Mpc, see Ref.~\cite{Steigman:1998sb})
and as an increasing number of folds inside turnaround. The energy spectrum of
dark matter particles at a fixed position will be discrete, see
Fig.~\ref{phase-space-infall}, left panel, where several velocity peaks are
indicated at intersections of the vertical dashed line, $r={\rm const}$, with
phase-space sheets.  The overall shape of the spectrum also changes compared
to an isothermal distribution. This may be important for direct dark matter
searches.

The infall model reproduces flat rotational curves, see
Fig.~\ref{phase-space-infall}, right panel. There is one interesting
difference, though; rotational curves of the infall model have several small
ripples in the region where the curve is flat. These ripples appear near the
surfaces where $v=0$. In principle, they may be detectable
\cite{Kinney:1999rk,Charmousis:2003rq} and then it will give a clear, unique
signature of the presence of dark matter in galactic halo, (as opposed to
models which try to explain apparent violation of Kepler's law by modification
of gravity).

Existence of such a folded structure is a topological statement. However, in
the inner halo the number of folds is very large, and limited resolution makes
the distribution indistinguishable from, say, isothermal. It is not clear at
which distances the description of halo in terms of the infall becomes
appropriate.  But for sufficiently isolated galaxy, in regions closer to the
outer rim of the halo, where the number of folds is still small, signatures of
the infall should exist, and they do exist in our Galaxy
\cite{Steigman:1998sb}.

\paragraph{Baryonic Halo Dark Matter ?  No.}
\begin{figure}
\begin{center}
\hspace{-1.5cm}
\includegraphics[width=6.2cm]{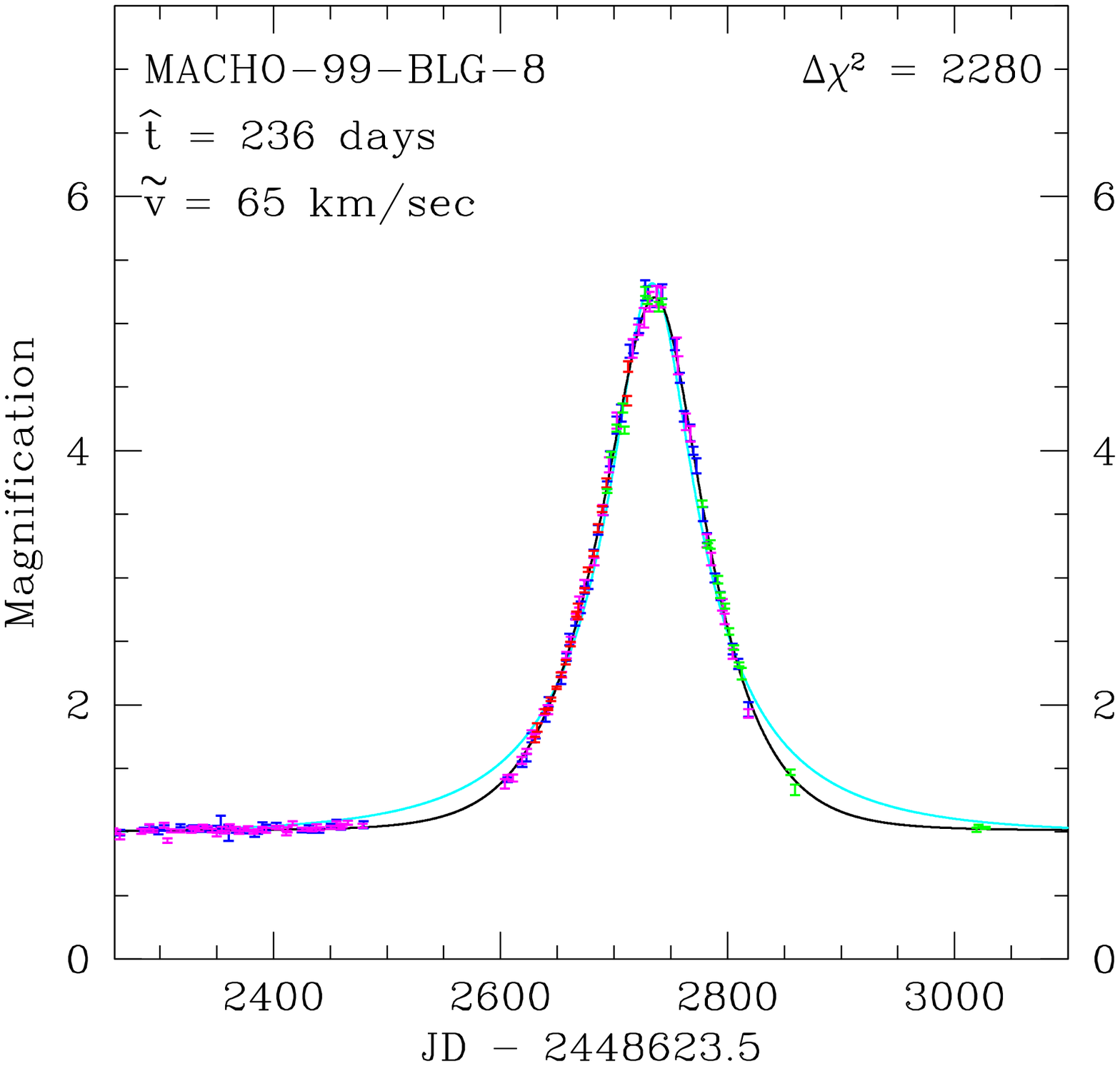}
\hspace{1.5cm}\parbox{6.cm}{
\vspace{-6cm}
\includegraphics[width=7.cm]{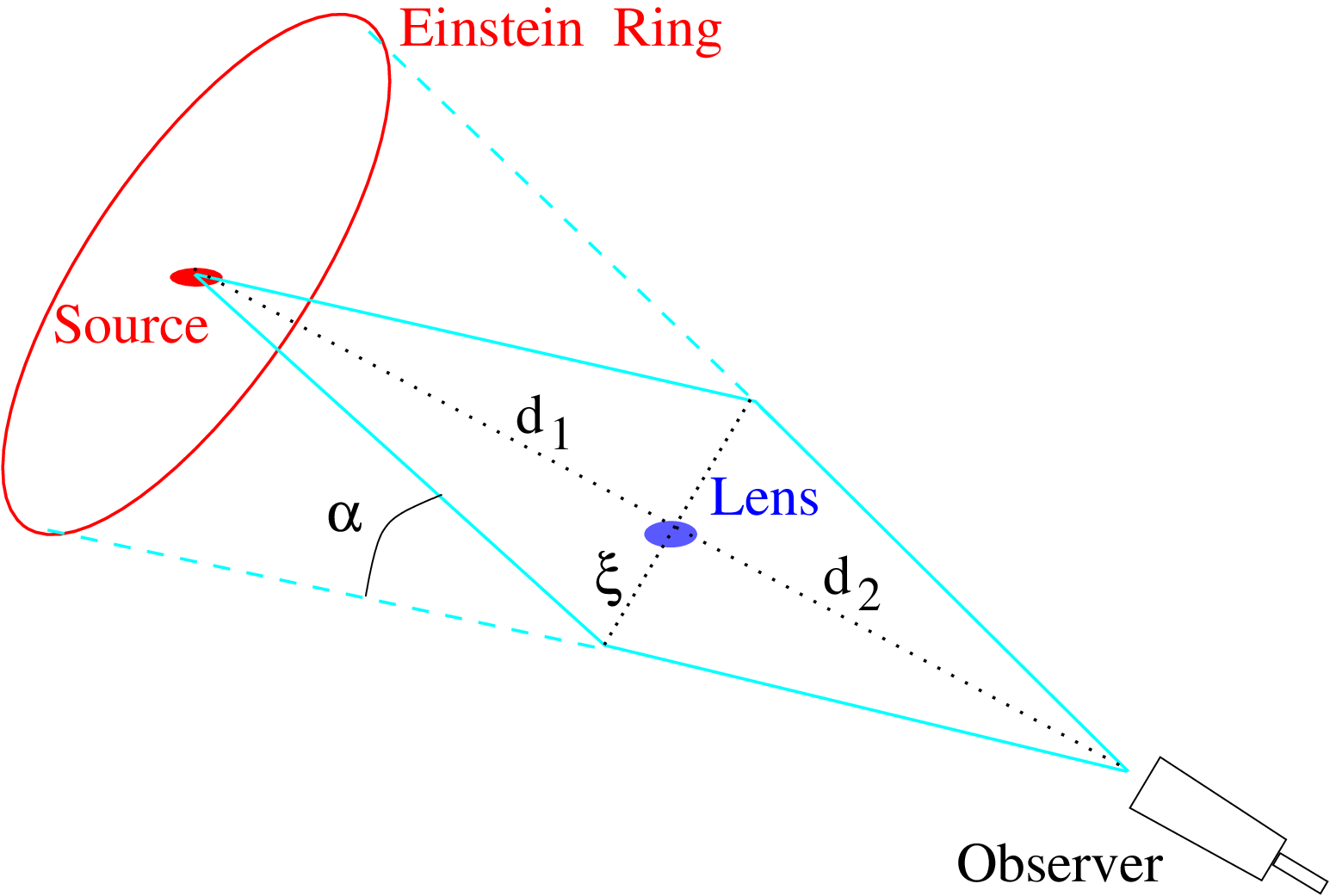}}
\caption{Left panel: one of the detected microlensing effects, see
Ref.~\cite{Bennett:2001vh}. Right panel: schematic view of gravitational
lensing by point mass.}
\label{Macho}
\end{center}
\end{figure}

Already CMBR alone tells us that there should be non-baryonic dark matter, see
Table I. BBN and CMB agree on $ \Omega_B = 0.04$, however, contribution of
stars amounts only to $\Omega_{\rm stars} = 0.005$.  There should be dark
baryons hiding somewhere. Can it be that the whole, or at least some part, of
the halo dark matter is comprised of dark baryons in the form of non-luminous
objects?  Candidates are Jupiter like planets, brown dwarfs (which are
undersized stars, too light to ignite thermonuclear reactions), or already
dead stars (white dwarfs, neutron stars, or even black holes). This class
of objects got the acronym MACHO, from MAssive Compact Halo Objects.  Special
techniques based on gravitational lensing were developed for MACHO
searches. These searches were successful, but by now it is clear that MACHOs
cannot comprise the whole dark matter, as their fraction of DM halo is
restricted to be $ < 50\%$.  Since MACHOS are the only type of dark matter
which has been detected, let us consider the issue in some more detail.
\begin{figure}
\begin{center}
\includegraphics[width=9.cm]{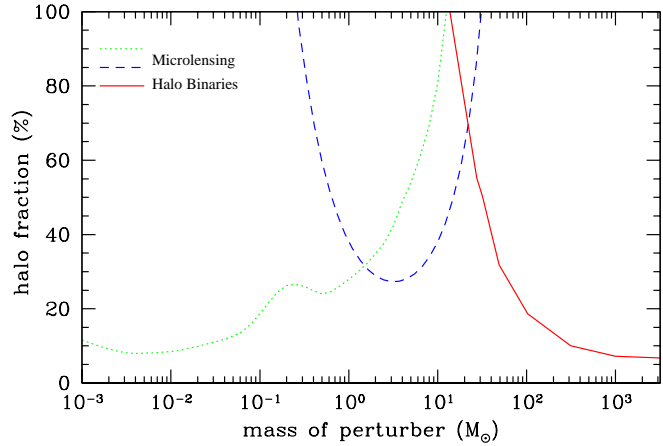}
\caption{ 95\% confidence limits on the MACHO fraction of the standard local
halo density. Green and blue lines - results of EROS \cite{Afonso:2002xq} and
MACHO \cite{Popowski:2003fm} microlensing collaborations. Red line -
constraint from the absence of distortion of distribution of binary stars in
angular separation \cite{Yoo:2003fr}.  Note that the microlensing exclusion
curve extends outside of the plotted range and up to $M \sim 10^{-7} M_\odot$,
see Ref.~\cite{Afonso:2002xq}.}
\label{End-of-MACHO}
\end{center}
\end{figure}

Consider the light deflection by a point mass $M$. If the impact parameter
$\xi$ is much larger than the Schwarzschild radius of the lens, $\xi \gg 2GM$,
then General Relativity predicts that the deflection angle of a light ray,
$\alpha$, is \be \alpha = \frac{4GM}{\xi} \; .
\label{deflection-angle}
\ee
This is twice the value obtained in Newtonian gravity.
If the lens happens to be on the line which connects the observer and a source,
the image appears as a ring with the  radius (Einstein ring radius)
\be  
r_E^{2} = 4\, G M_L\; \frac{d_1d_2}{d_1 + d_2}\; ,
\label{Einstein ring}
\ee 
see Fig.~\ref{Macho}, right panel.  If the deflector is displaced from the
line of sight by the distance $r$, then instead of the ring, an odd number of
images will appear. If the images cannot be observed separately, because the
resolution power of the telescope is not sufficient, then the only effect will
be an apparent brightening of the source, an effect known as gravitational
microlensing.  The amplification factor is
\be
A = \frac{2+u^{2}}{u\sqrt{4+u^{2}}}\; ,
\ee
where $u = {r}/{r_E}$. If the lens is moving, the distance $r$ will be
changing with time, and the image of the background star will brighten during
the closest approach to the line of sight.  If the galactic halo is filled
with MACHOs, this may happen occasionally for some of background stars.  The
typical duration of the light curve is the time it takes a MACHO to cross an
Einstein radius, $\Delta t \sim r_E/v$, where $v \sim 10^{-3}$ is typical
velocity in the halo. If the deflector mass is $1\, M_\odot$, the average
microlensing time will be 3 months, for $10^{-2}\, M_\odot$ it is 9 days, for
$10^{-4}\, M_\odot$ it is 1 day, and for $10^{-6}\, M_\odot$ it is 2 hours.

An optical depth for microlensing of the galactic halo is approximately $\tau
\sim 10^{-6}$.  Thus, if one looks simultaneously at several millions of stars
during an extended period of time, one has a good chance of seeing at least a
few of them brightened by a dark halo object. The first microlensing events
were reported in 1993. Nowadays, there are more than half a hundred registered
events. One of them is shown in Fig.~\ref{Macho}. However, derived optical
depth is not sufficient to account for all dark matter in the Galaxy halo.
95\% confidence limits on the MACHO fraction of the standard local halo
density is shown in Fig.~\ref{End-of-MACHO}.

Since MACHOs cannot account for the mass of the dark halo, non-baryonic
dark matter should be present out there.

\subsubsection{Dark Matter in Clusters of Galaxies}

Already in 1933, F. Zwicky \cite{Zwicky:1933gu} deduced the existence of dark
matter in the Coma cluster of galaxies. Nowadays, there are several ways to
estimate masses of clusters, based on the kinetic motion of member galaxies,
on X-ray data, and on gravitational lensing. These methods are different and
independent.  In the dynamical method, it is assumed that clusters are in
virial equilibrium, and the virialized mass is easily computed from the
velocity dispersion.  In X-ray imaging of hot intracluster gas, mass estimates
are obtained assuming hydrostatic equilibrium. Mass estimates based on lensing
are free of any such assumptions.  All methods give results which are
consistent with each other, and tell that the mass of the luminous matter in
clusters is much smaller than the total mass.

\begin{figure}
\begin{center}
\includegraphics[width=9.cm]{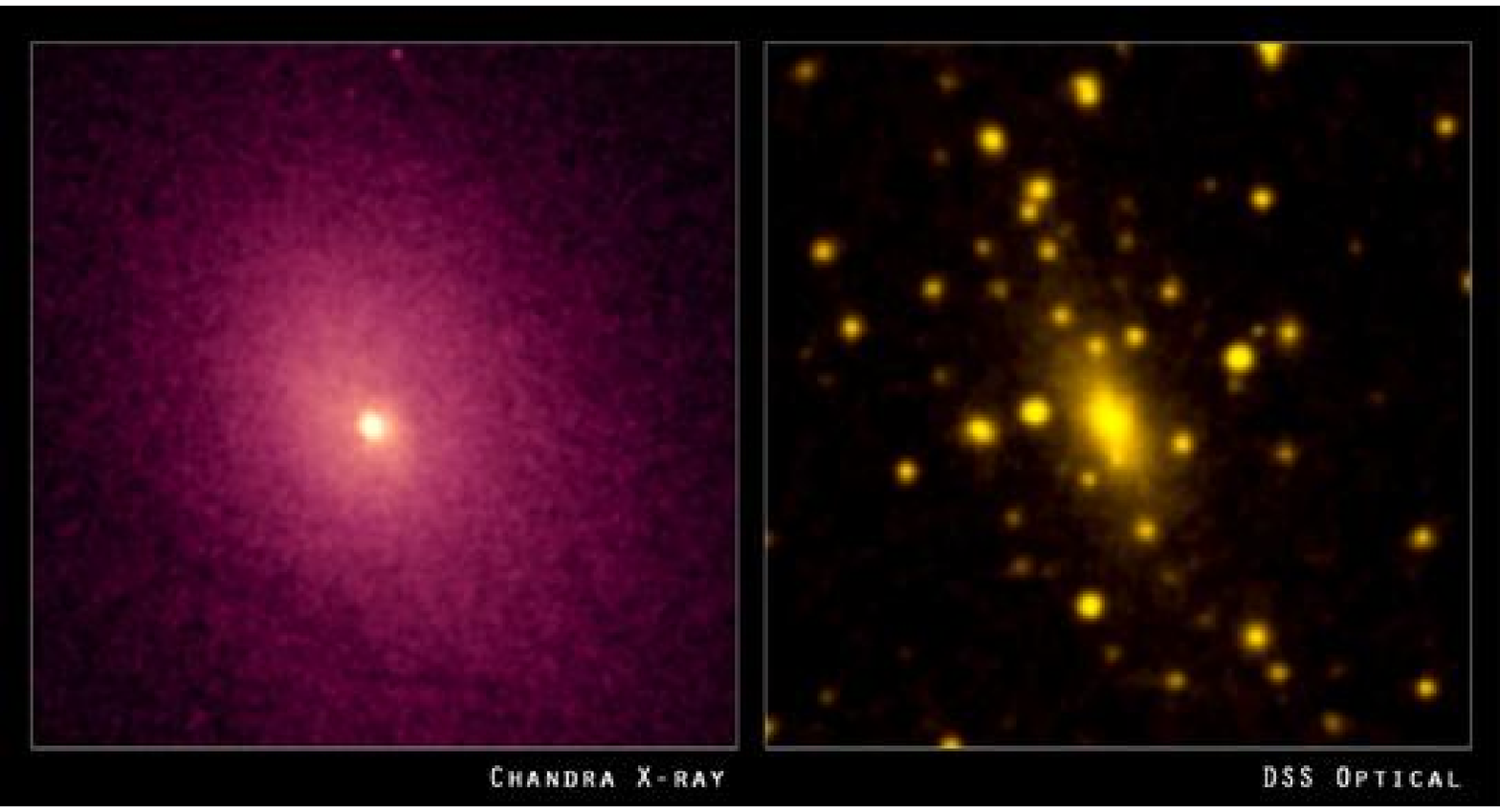}
\parbox{6.cm}{
\vspace{-5cm}
\includegraphics[width=6.cm]{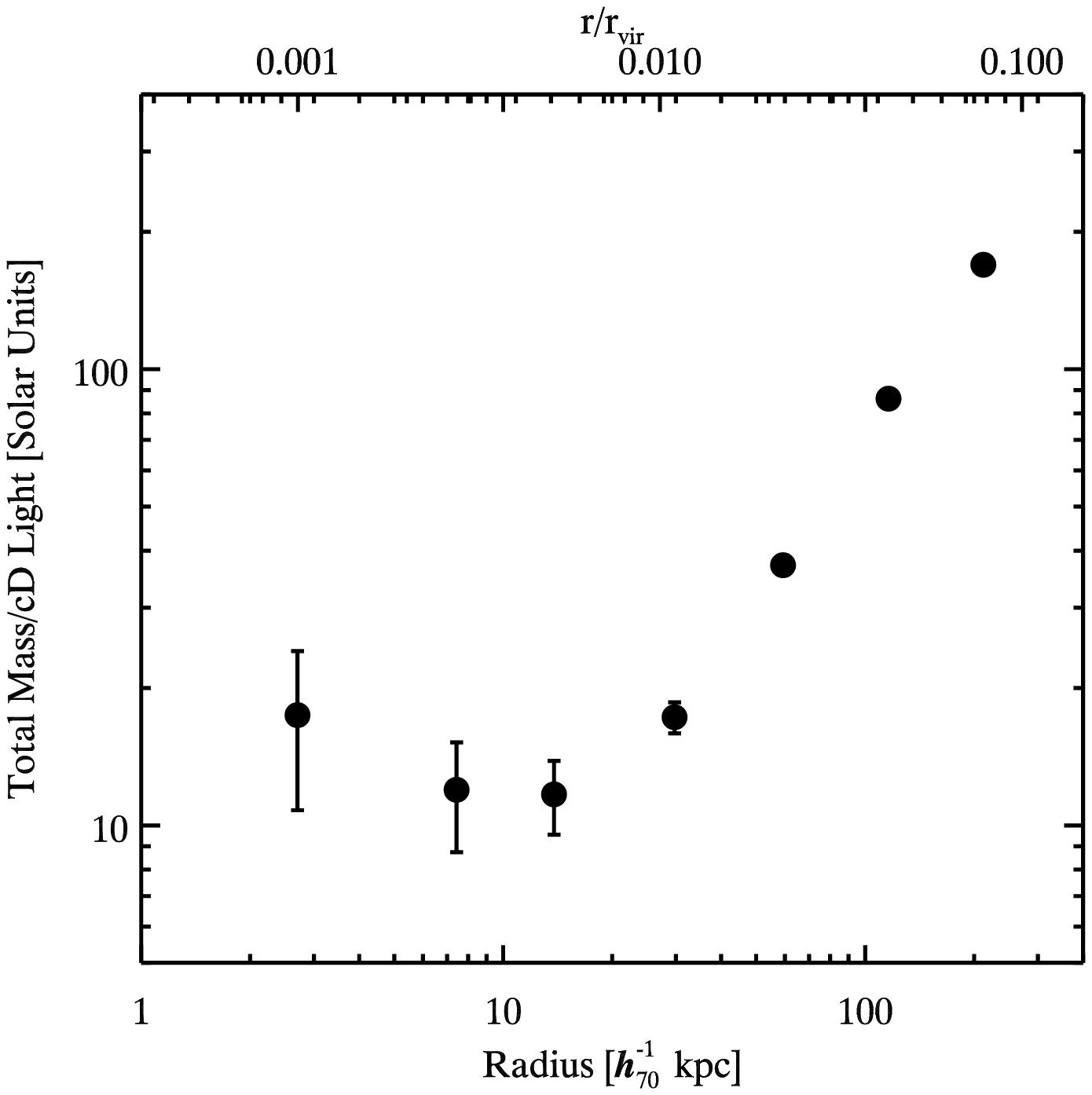}}
\caption{X-ray (left panel) and optical (middle panel) images of the Abell
2029 cluster of galaxies.  Right panel: ratio of total enclosed cluster mass
to light in A2029, from Ref.~\cite{Lewis:2002mf}.}
\label{Abell2029}
\end{center}
\end{figure}

\paragraph{Kinetic mass estimates.}
Those are based on the virial theorem,~ $\langle E_{\rm pot} \rangle +
2\langle E_{\rm kin} \rangle = 0 $.  Here $ \langle E_{\rm kin} \rangle = N
\langle m v^{2} \rangle/2$ is averaged kinetic energy of a gravitationally
bound object (e.g. cluster of N galaxies) and $ \langle E_{\rm pot} \rangle =
- N^{2} {\langle m^{2} \rangle}/2 {\langle r \rangle}$ is its averaged
potential energy. Measuring the velocity dispersion of galaxies in the
clusters and its geometrical size gives an estimate of the total mass,~ $ M
\equiv N \langle m \rangle$.
\be
M \sim \frac{2 \langle r \rangle \langle v^{2} \rangle}{G} \; .
\label{virial-mass-estimate}
\ee
The result can be expressed as mass-to-light ratio, $M/L$, using the Solar
value of this parameter. For the Coma cluster, which consists of about 1000
galaxies, Zwicky \cite{Zwicky:1933gu} has found
\be
\frac{M}{L} \sim 300\, h\, \frac{M_\odot}{L_\odot}\; .
\ee
Modern techniques end up with very much the same answer. $M/L$ ratios measured
in Solar units in central regions of galaxies range from a few to 10 in spirals
and large ellipticals. If clusters are large enough systems for their $M/L$ to
be representative of the entire Universe, one finds \cite{Carlberg:1996aq}
\be
\Omega_M \approx 0.2-0.3 \; .
\label{Omega-m-clusters}
\ee

\paragraph{Mass estimates based on X-rays.}

Mass is also traced in clusters of galaxies by hot gas which is visible in
X-rays. Assume hot gas is in thermal equilibrium in a gravitational well
created by a cluster.  Then its density distribution $\rho_g(r)$ and pressure
$P_{g}(r)$ satisfy
\be
{1 \over \rho_{g}(r)}{d  P_{g}(r) \over dr} = - {GM(\le r) \over r^2} \; .
\label{eq:hydro}
\ee
Observationally, the gas density follows from the X-ray luminosity. Gas
temperature can be measured from the shape of X-ray spectrum.  By measuring
the temperature profile of a gas, one can reconstruct the gas pressure
$P_{g}(r)$. In this way, the radial run of mass can be deduced.

For example, detailed modeling \cite{Lewis:2002mf} of Abell 2029, which is
shown in Fig.~\ref{Abell2029}, leads to the conclusion that the cluster is
dark matter dominated all the way into its core. After subtracting the
contributions of stars and hot gas into the mass budget, the density profile
of dark matter can be reconstructed. It agrees with NFW dark matter profile,
Eq.~(\ref{NFW}), $\rho \propto {1}/{x(1+x^{2})}$, where $x \equiv r/r_s$ and
$r_s = 540$ kpc. The agreement is remarkably good on all scales measured, 3 -
260 kpc. Baryons contribute $f_b \approx 14\%$ into the total mass of the
cluster. Assuming universal baryon mass fraction and $\Omega_b$ from big bang
nucleosynthesis, this also gives $\Omega_m \approx \Omega_m/f_b \approx 0.29$
for the total mass budget in the Universe, in agreement with other current
estimates.

The same methods can be employed for studies of dark matter in large
elliptical galaxies. In Ref.~\cite{Loewenstein:2002ri} the mass profile of the
elliptical galaxy NGC 4636, based on the temperature of hot interstellar gas,
was obtained for distances from 0.7 to 35 kpc. It was found that the total
mass increases as radius to the power 1.2 over this range in radii, attaining
a mass-to-light ratio of 40 solar masses per solar visual luminosity at 35
kpc. As much as 80\% of the mass within the optical half-light radius is
non-luminous in this galaxy.

\paragraph{Gravitational Lensing.}

\begin{figure}
\begin{center}
\includegraphics[width=8.5cm]{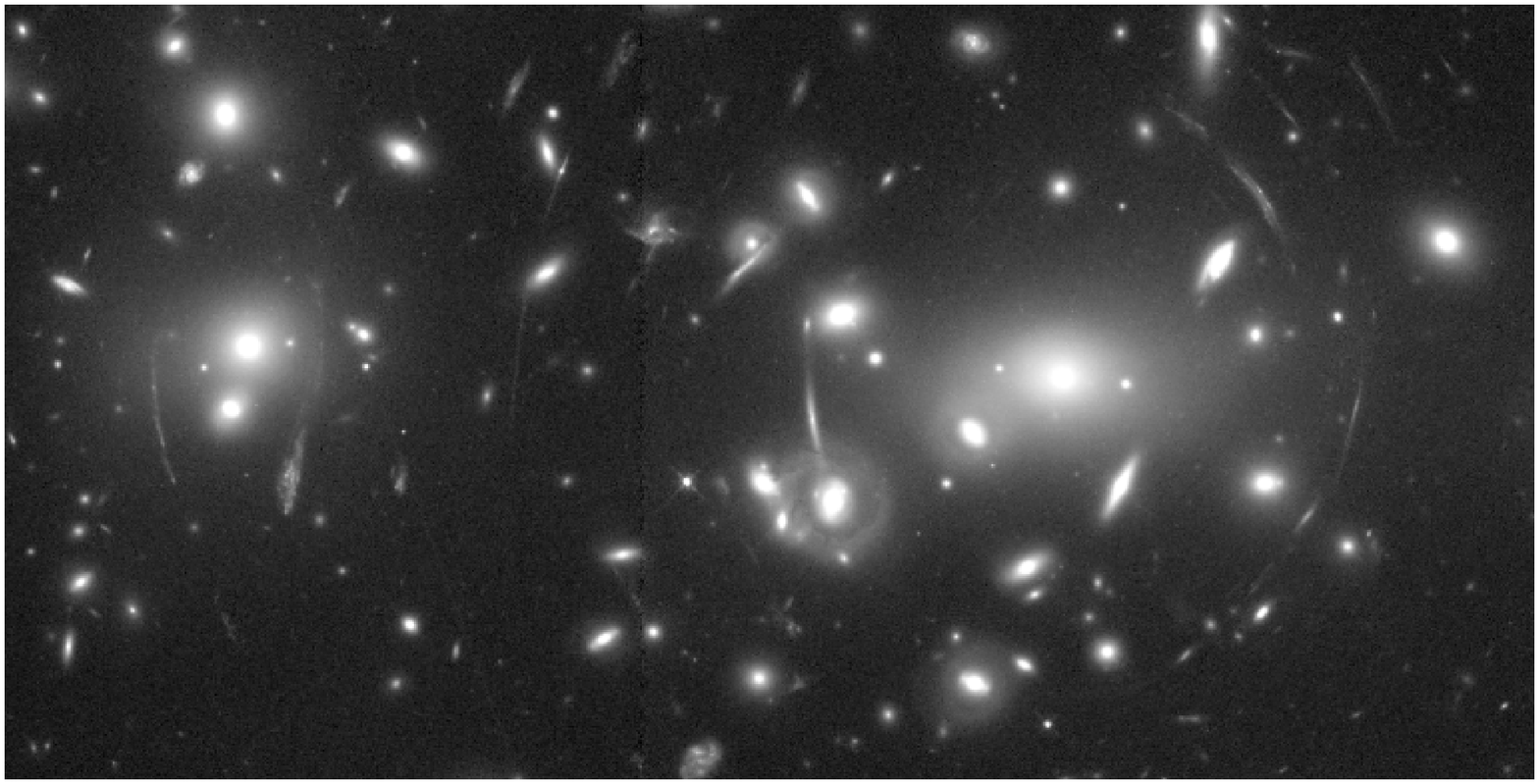}
\hspace{0.5cm}\parbox{6.cm}{
\vspace{-4cm}
\includegraphics[width=6.cm]{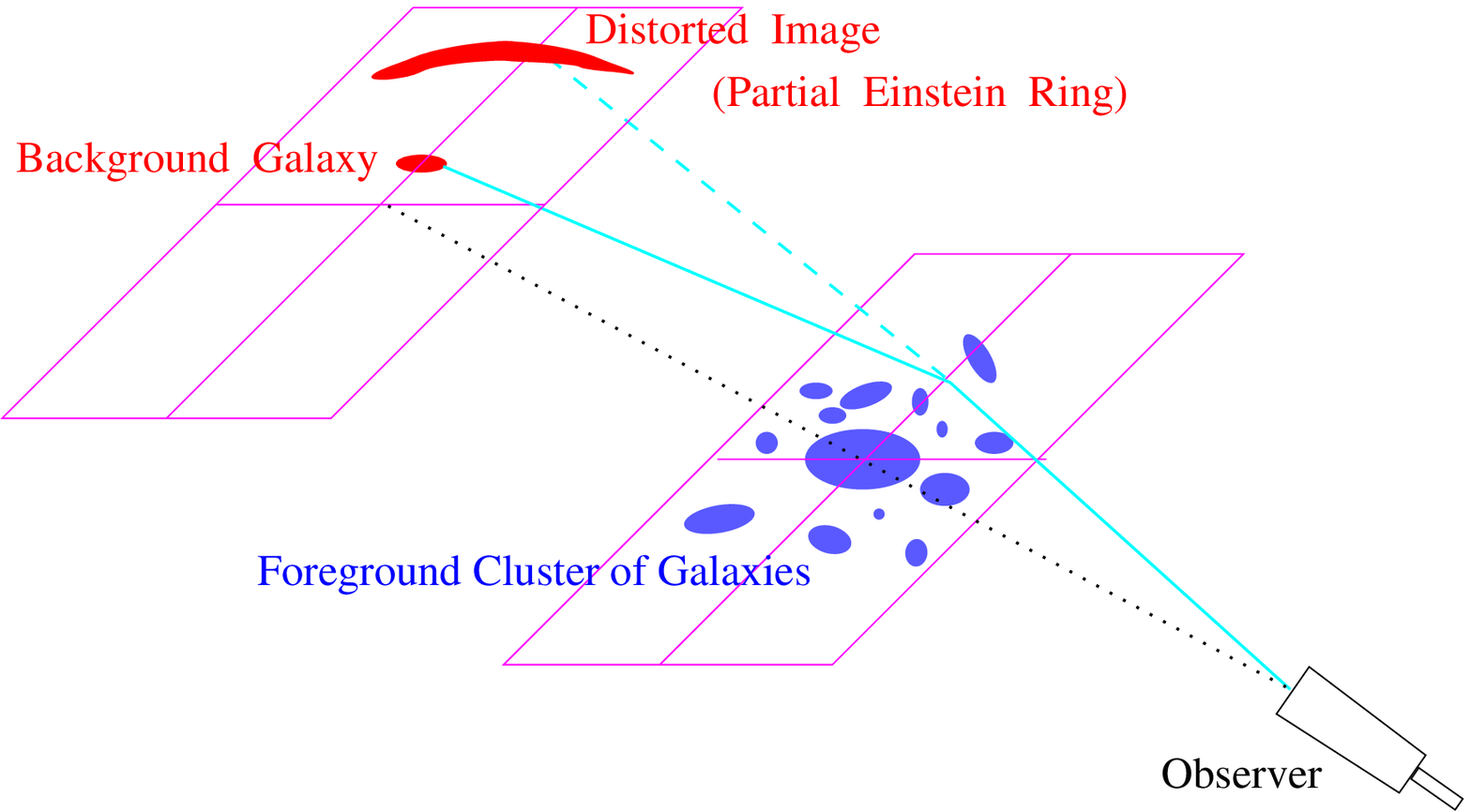}}
\caption{Left panel: an image of the cluster Abell 2218 taken with the Hubble
  space telescope (see Ref.~\cite{Kneib:1995hh}). Spectacular arcs resulting
  from strong lensing of background galaxies are clearly seen.}
\label{Abell2218}
\end{center}
\end{figure}

Gravitational lensing allows direct mass measurement without any assumptions
about the dynamical state of the cluster. The method relies on the measurement
of the distortions that lensing induces in the images of background
galaxies. As photons travel from a background galaxy to the observer, their
trajectories are bent by mass distributions, see Fig.~\ref{Abell2218}, right
panel.  Consider the deflection by a point mass $M$. For impact parameter
$\xi$ which is much larger than the Schwarzschild radius of the lens, $\xi \gg
2GM$, the deflection angle $\alpha$ is given by Eq.~(\ref{deflection-angle}).
If the gravitational field is weak, the deflection angle of an ensemble of
point masses will be the vectorial sum of the deflections due to individual
lenses.

A reconstruction of lens geometry provides a map of the mass distribution in
the deflector. For a review of the method see
e.g. Ref~\cite{Bartelmann:1999yn}.  The images of extended sources are
deformed by the gravitational field.  In some cases, the distortion is strong
enough to be recognized as arcs produced by galaxy clusters serving as a lens,
see Fig.~\ref{Abell2218}, left panel.  For the cluster A 2218, shown in this
figure, Squires et al. \cite{Squires:1996ee} compared the mass profiles
derived from weak lensing data and the X-ray emission.  The reconstructed mass
map qualitatively agrees with the optical and X-ray light distributions. A
mass-to-light ratio of $M/L = (440 \pm 80)h$ in solar units was found. Within
the error bars the radial mass profile agrees with the mass distribution
obtained from the X-ray analysis, with a slight indication that at large radii
the lensing mass is larger than the mass inferred from X-rays.  The gas to
total mass ratio was found to be $M_{\rm gas}/M_{\rm tot} = (0.04 \pm 0.02) \,
h^{-3/2}$.

\begin{figure}
\begin{center}
\includegraphics[width=7.cm]{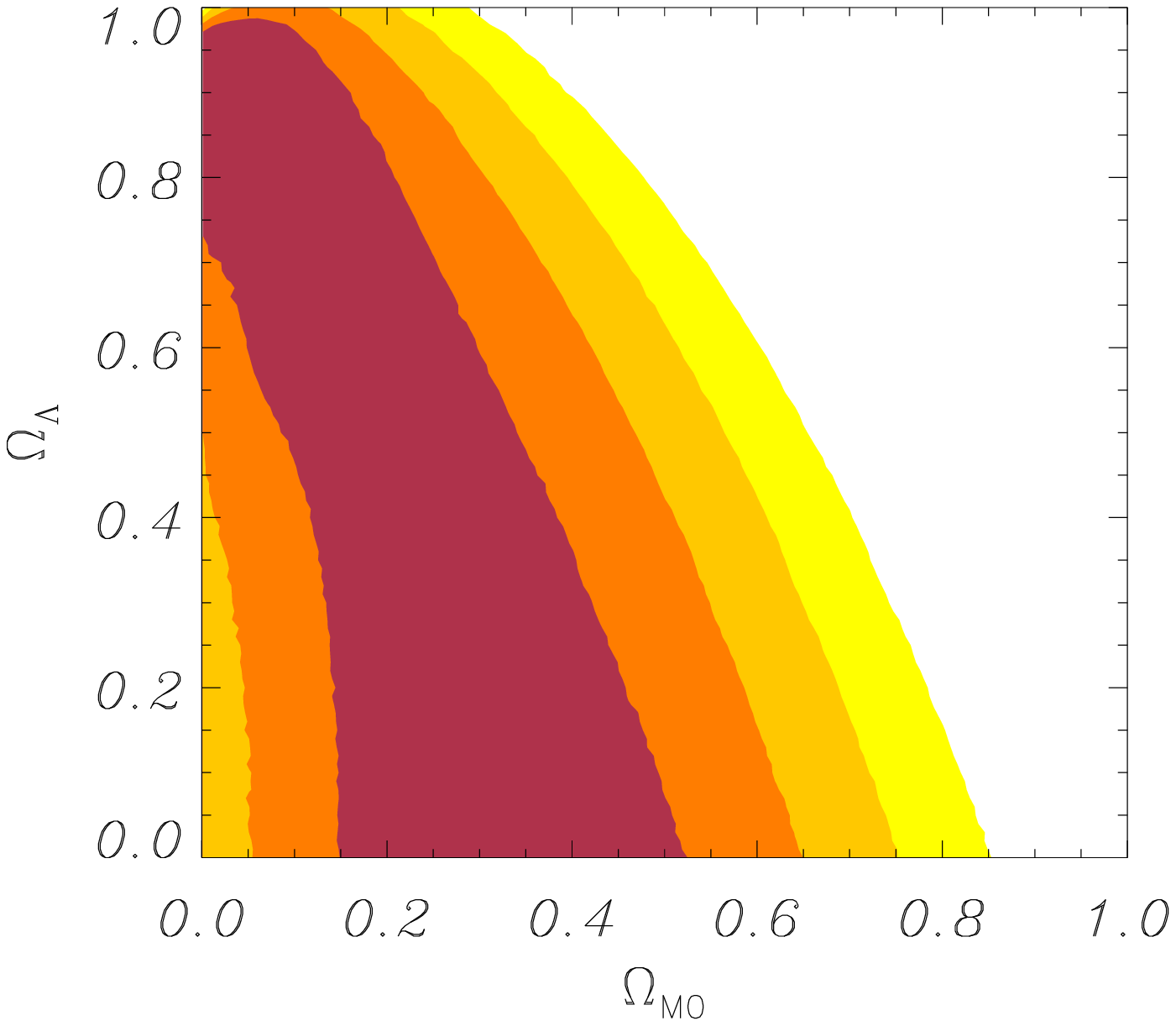}
\parbox{7.5cm}{\vspace{-5.8cm}
\includegraphics[width=7.cm]{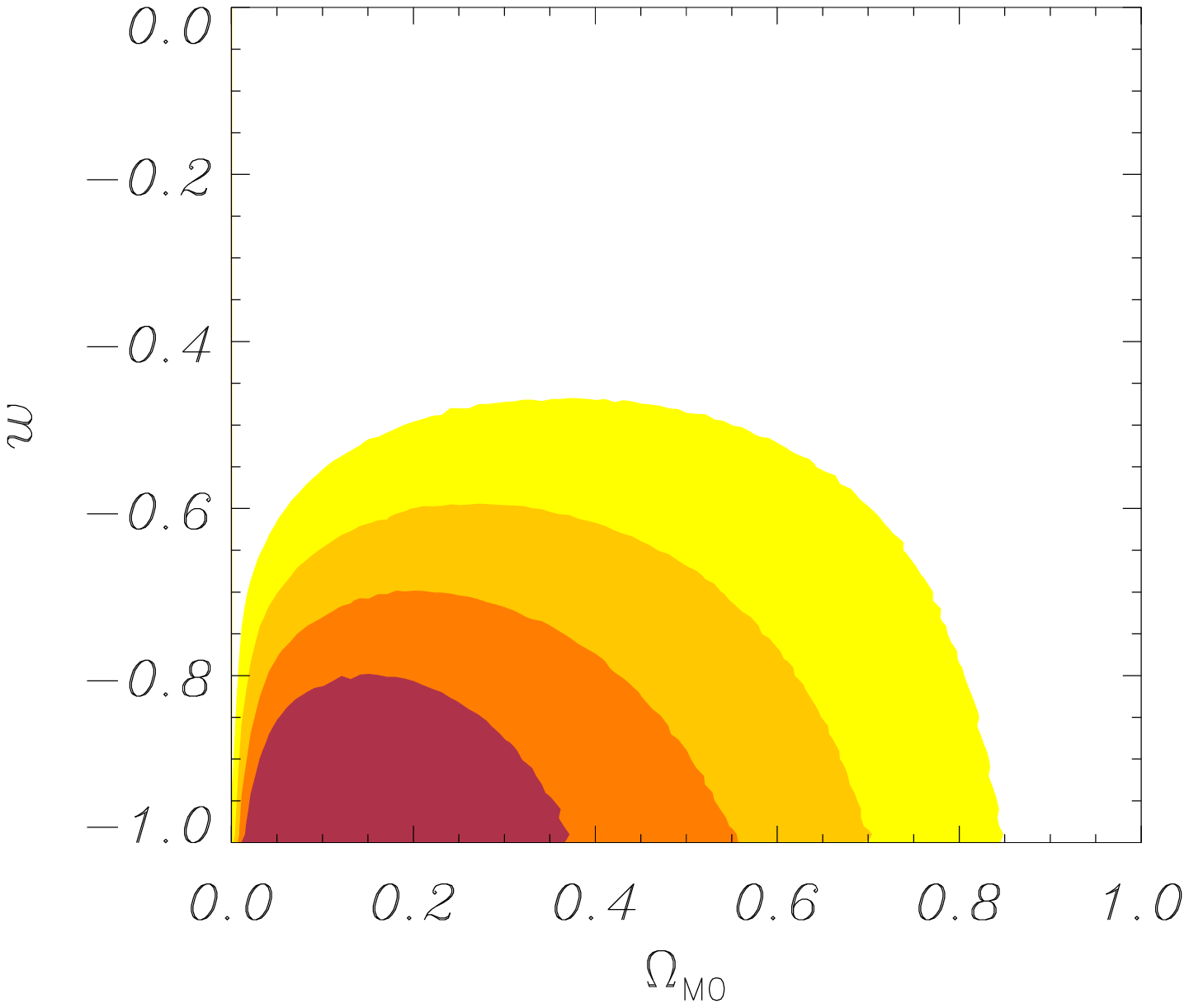}}
\caption{The constraints on cosmological parameters obtained from the study of
the lensing configuration in the Abell 2218 cluster of
galaxies ~\cite{Soucail:2004se}.}
\label{Abell2218_cosmo}
\end{center}
\end{figure}

Interestingly, the analysis of multiple images of several sources with known
(and significantly different) redshift produced by a cluster lens is sensitive
to the value of the geometrical cosmological parameters of the Universe. Study
\cite{Soucail:2004se} of the lensing configuration in the cluster Abell 2218
gives $0 < \Omega_M < 0.30$ assuming a flat Universe, and $0 < \Omega_M <
0.33$ and $w< -0.85$ assuming flat Universe with dark energy, see
Fig.~\ref{Abell2218_cosmo}. These constraints are consistent with the current
constraints derived with CMB anisotropies or supernovae studies, however this
method is a completely independent test, providing nearly orthogonal
constraints in the $(\Omega_M,\Omega_{\Lambda})$ plane.

\subsection{Structure formation and DM}

By present time the structure in the Universe (galaxies and clusters) is
formed already, the perturbations in matter $\delta \rho /\rho \agt 1$.
However, the initial perturbations were small~~ $ \delta \rho /\rho \sim
10^{-5}$.  Perturbations do not grow in the radiation dominated epoch, they
can start growing only during matter domination $ \delta \rho /\rho \sim a =
1/z$.  Moreover, baryonic plasma is tightly coupled to radiation, therefore
perturbations in baryonic matter start to grow only after recombination. For
the same reason, initial perturbations in baryons at the time of recombination
equal to fluctuations in CMBR. If baryons were to constitute the only matter
content, then perturbations in matter at present time would be equal to
\be
\frac{\delta \rho}{\rho}|_{\rm today} = z_{\rm rec}\; \frac{\delta
\rho}{\rho}|_{\rm rec} \sim 10^{-2} \; , 
\ee 
where $z_{\rm rec} \approx 1100$ is the redshift of recombination. This is one
of the strongest and simplest arguments in favour of non-baryonic dark matter.
Structure has had time to develop only because perturbations in non-baryonic
dark matter have started their growth prior to recombination. Baryonic matter
then ``catch up'' simply by falling into already existing gravitational wells.
If one aims to explain things by modification of gravity, one has to explain
not only the flat rotational curves in galaxies and the presence of dark
matter in galaxy clusters, but has also to provide the accelerated growth of
structure from recombination till present, in a consistent way.


\subsection{ Non-baryonic Dark Matter Candidates}
\label{sec:DM-models}
\begin{figure}
\begin{center}
~\hspace{-0.5cm}
\includegraphics[width=6.5cm]{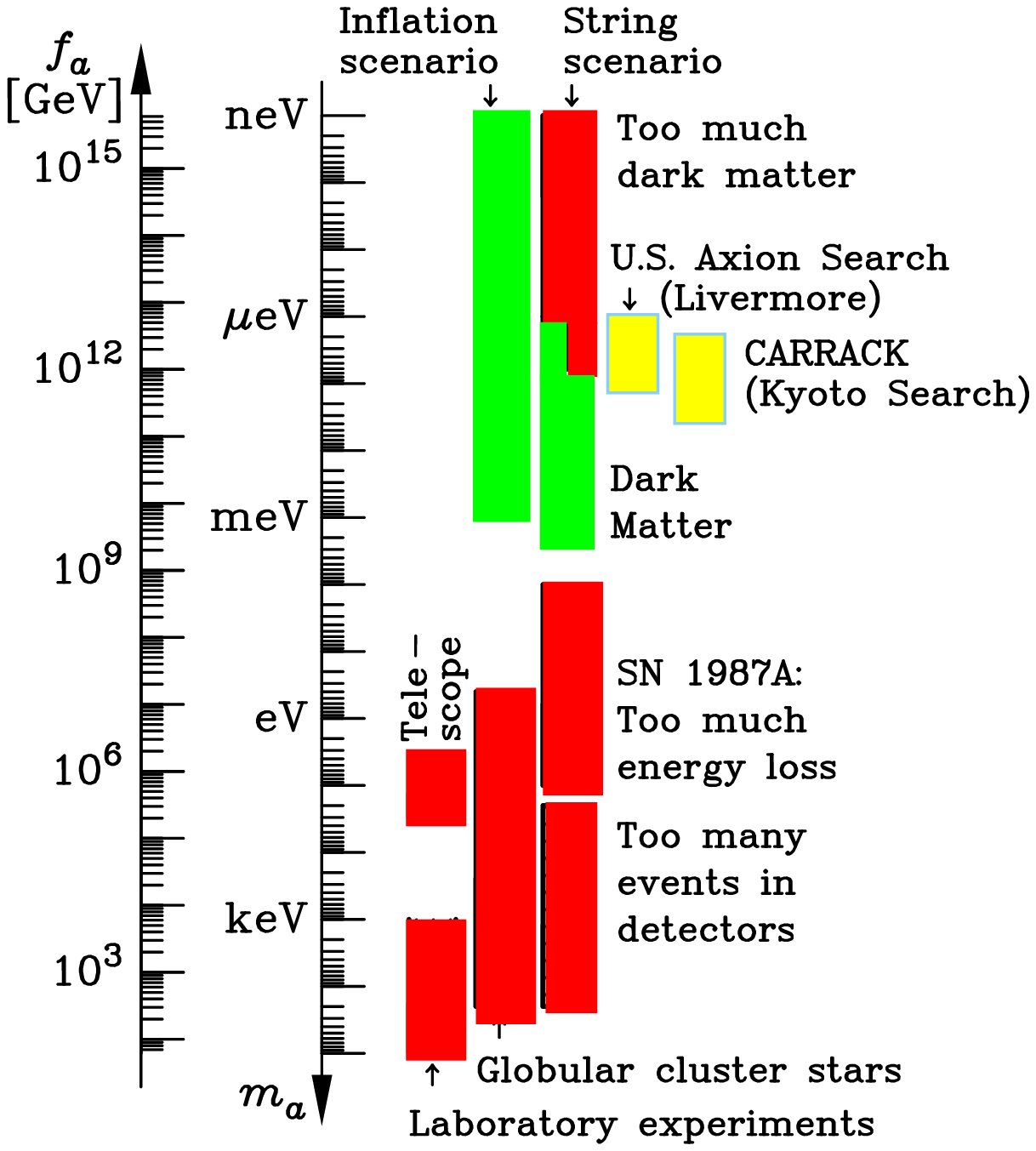}
~\hspace{1.5cm}
\includegraphics[width=7.2cm]{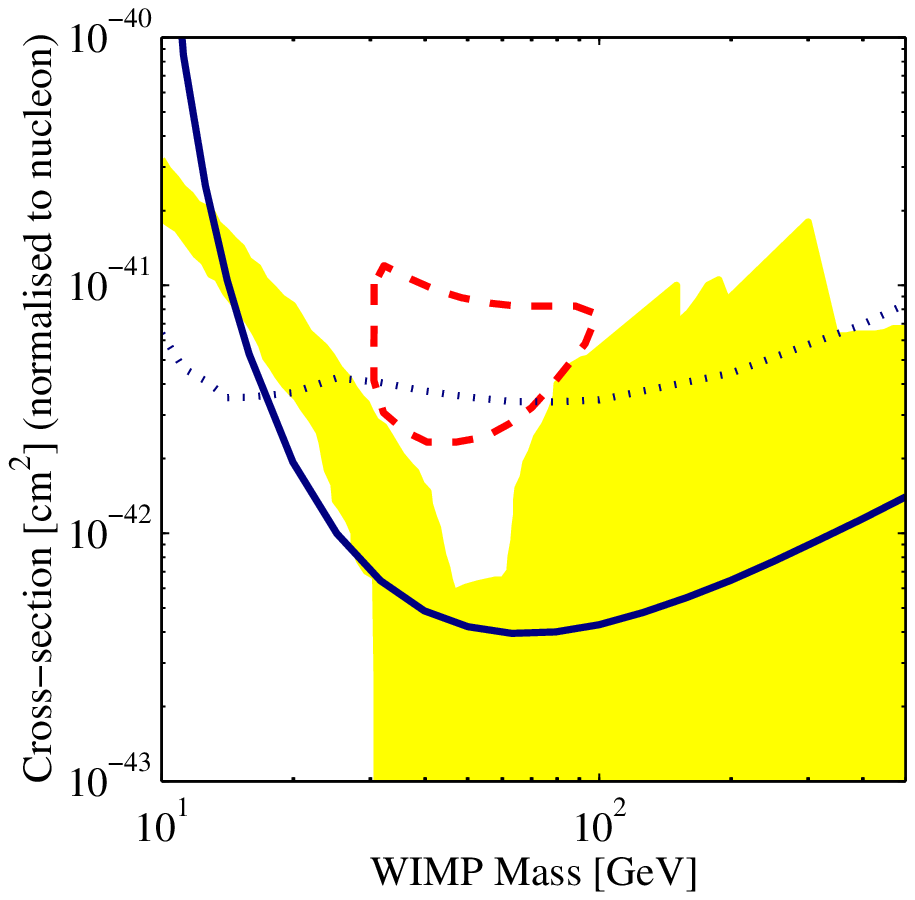}
\caption{{\bf Left panel:} The constraints on axion parameters. Red blocks -
various cosmological and astrophysical constraints; yellow blocks - exclusion
regions obtained in dedicated dark matter search experiments; green blocks -
the allowed regions in two cosmological scenarios.  {\bf Right panel:} Solid
curve - the exclusion limits on the coherent WIMP-nucleon scalar cross-section
obtained by CDMS collaboration in the direct dark matter search experiment;
the parameter space above the curve is excluded at the 90\% C.L.  These limits
are inconsistent with the DAMA 3$\sigma$ signal region \cite{Bernabei:2000qi}
(closed dashed contour) if it is due to scalar coherent WIMP interactions.
Also shown are limits from CDMS at SUF (dotted line).  The typical predictions
of supersymmetric models are shown in yellow. Adapted from
Ref.~\cite{Akerib:2004fq}.}
\label{fig:DM-limits}
\end{center}
\end{figure}

There is no lack for dark matter candidates in particle physics models.
Some of them appear naturally, and were motivated by some other reasoning, not
related to the dark matter problem.  They are the leading candidates and are
listed below.
\begin{itemize}
\item Axion. Has a mass $m \sim 10^{-5}$ eV. Appeared
  \cite{Weinberg:1978ma,Wilczek:1978pj} as a by-product of a suggested
  solution of the strong CP problem via a global $U(1$) Peccei-Quinn symmetry
  \cite{Peccei:1977hh}. The axion picks up a small mass in a way similar to
  the pion when chiral symmetry is broken. The parameters of these two
  particles are related, in particular $m_a \sim m_\pi f_\pi /f_a$, where
  $f_a$, is the axion decay constant, and determines also the strength of the
  axion coupling to other particles. There are tight astrophysical and
  cosmological bounds on $f_a$ which leave only a narrow window, $10^{10}~{\rm
  GeV} \lesssim f_a \lesssim 10^{12}~{\rm GeV} $, see
  Fig.~\ref{fig:DM-limits}, the left panel. For the review of axion physics
  and searches see e.g. Refs.~\cite{Raffelt:1990yz,Bradley:2003kg}.
\item Neutrino, $m \sim 0.1$ eV. The only dark matter candidate which is 
  known to exist. For this reason we discuss the neutrino in some more detail
  below. For the review of neutrino cosmology see
  e.g. Ref.~\cite{Dolgov:2002wy}, and for the neutrino astrophysics see
  e.g. Ref.~\cite{Raffelt:1999tx}. While the neutrino is cosmologically
  important, it cannot resolve the dark matter problem.
\item Mirror matter, $M \sim 1$ GeV. Does not belong to the list of the most
  popular candidates, but is attractive as an example of a model
  \cite{Blinnikov:1982eh,Berezhiani:2000gw,Foot:2003jt} where the approximate
  equality of baryonic and non-baryonic contributions into the energy balance
  of the Universe is attempted to be explained naturally, and not as a result
  of fine-tuning of model parameters.
\item WIMP, $m \sim 100$ GeV. The most popular candidate, a natural outcome of
  supersymmetry. The lightest supersymmetric particle (or LSP) is naturally
  stable and would have interesting cosmological abundance. Known also under
  the names Neutrallino (dark matter has to be color and electrically neutral)
  and WIMP (from Weakly Interacting Massive Particle). For recent reviews see
  e.g.  Refs.~\cite{Olive:2003iq,Ellis:2003ch}. The current status of direct
  and indirect WIMP searches is reviewed in Ref.~\cite{Gondolo:2004fg}.  The
  new limits obtained by the CDMS collaboration and not reflected in cited
  reviews are shown in Fig.~\ref{fig:DM-limits}, the right panel.
\item WIMPZILLA, $m \sim 10^{13}$ GeV. The newcomer, was initially motivated
  as a solution of the Greizen-Zatsepin-Kuzmin puzzle of Ultra-high energy
cosmic rays \cite{Berezinsky:1997hy,Kuzmin:1998cm}. The popularity was boosted
by the observation that cosmologically interesting abundance is created
naturally, just as a sole consequence of Universe expansion
\cite{Chung:1998zb,Kuzmin:1998uv}. 
\end{itemize}
Non-baryonic dark matter model building and searches makes and extensive
subject on its own.  There are many dedicated excellent reviews, I cannot list
them all, see e.g. some earlier \cite{Jungman:1996df} and the latest
\cite{Olive:2003iq,Ellis:2003ch,Gondolo:2004fg} one. For this reason, and
because of space limitations, I will not describe non-baryonic dark matter in
all its variety, instead I'll spend some time on simple and universal
relations.

\paragraph{Cosmological density of neutrino.}

Here we calculate the abundance of particles which were once in thermal
equilibrium with the rest of cosmological plasma.  Let us first consider the
case of the neutrino.

Comparing the weak interaction rate, $\Gamma \sim T^5/M_W^4$, to the expansion
rate, $H \sim T^2/M_{\rm Pl}$, one finds that neutrino are in thermal
equilibrium at temperatures $T \gg 1$ MeV and decouple from the rest of plasma
at lower temperatures. (One can do this in full detail, see
e.g. \cite{Dolgov:2002wy}.) Therefore, standard model neutrinos, which have
small masses, decouple when they are still relativistic.  The number density
of neutrino at this time is given by Eq.~(\ref{numberDensity-reativistic}).
Below this temperature, neutrinos are no longer in thermal equilibrium with
the rest of the plasma, and their temperature simply decreases as $T \propto
1/a$. However, the cosmological background of photons is heated up by the
$e^{+}e^{-}$ annihilations. Let us find a relation between $T_\nu$ and
$T_\gamma$, which will also give the relation between $n_\nu$ and $n_\gamma$.

The annihilation reaction rate is much faster than the expansion of the
universe, therefore this process is adiabatic, and entropy in comoving volume
is conserved, $g_*\; T^{3} = {\rm const}$, see Eq.(\ref
{Entropy-reativistic}). Before annihilation $g_* = g_\gamma + g_e\cdot (7/8) =
2 + 4\cdot (7/8) = 11/2$.  After annihilation $g_* = g_\gamma = 2$. Since
before annihilation $T_\nu = T_\gamma$, ~we find~~ $ T_\nu = \({4}/{11}
\)^{1/3} T_\gamma$,~~ and for one neutrino flavour we have
\be
n_\nu = \frac{3}{11}\; n_\gamma \; .
\label{nu-density}
\ee
Here we used Eq.~(\ref{numberDensity-reativistic}) and $g_\gamma = g_\nu = 2$,
since right handed neutrino do not contribute (are not excited) even if
neutrino has a small Dirac mass, see Ref.~\cite{Dolgov:2002wy}.  As a
consequence of $e^{+}e^{-}$ annihilation neutrino temperature is lower.  With
$ T_\gamma = 2.728$ K we find $T_\nu = 1.947$ K and $ n_\nu = 115\; {\rm
cm}^{-3}$.  

At temperatures larger than neutrino mass, $ T_\nu \gg m_\nu $, in the
standard model, assuming no chemical potential, we find
\be
\Omega^{}_\nu = 3\;  \( \frac{7}{8} \) \( \frac{4}{11} \)^{4/3} \; 
\Omega^{}_{\rm \gamma}
\approx 0.68\, \Omega^{}_{\rm \gamma}\; ,
\label{neutrino-Omega}
\ee
where the factor of $3$ corresponds to the number of neutrino flavours.  This
result allows us to find the epoch of equal matter and radiation
densities
\be
1+z_{\rm eq} = \frac{\Omega_M}{\Omega_\gamma + \Omega_\nu} \approx 3200 \; .
\ee

Assume that by now neutrino became non-relativistic, i.e. their masses are
larger than the present temperature. In this case, neutrino energy
density is given by $ \rho_\nu = \sum_i m_{\nu i}\, n_{\nu i}$. Since it has
to be smaller than $\Omega_m\, \rho_c $, we have the constraint
\cite{Gershtein:1966gg}
\be
\sum_i m_{\nu i} < 94 \; \Omega_m h^{2}\; {\rm eV} = 12\; {\rm eV}\; .
\label{Gershtein-Zeldovich}
\ee
For dark matter particles to boost the structure formation, their typical
velocities squared at the time of recombination should be smaller than the
depth of typical gravitational wells, $v^2 \ll 10^{-5}$. In other words, the
dark matter should be {\it cold}. This is not the case for particles as light
as those which satisfy the bound Eq. (\ref{Gershtein-Zeldovich}).  Neutrino
can make up dark matter, but it will be {\it hot} dark matter. 

\paragraph{Neutrino mass is pinned down.}

Free streaming of relativistic neutrinos suppresses the growth of fluctuations
until $\nu$ becomes nonrelativistic at $z ~\sim~ m_j/3T_0 ~\sim~ 1000\;
(m_j/{\rm eV})$. This effect of free-streaming is not seen in the data and
therefore only small corrections due to light neutrino are allowed in the
standard CDM picture. Combined CMBR and LSS analysis yields the constraint
\cite{Bennett:2003bz}
\be
\Omega_\nu h^{2} ~=~ \frac{\sum_i m_i}{93.5\; {\rm eV}} ~<~ 0.0076\; ,
\ee
which translates into the upper bound
\be
\sum_i m_i ~<~ 0.7\; {\rm eV} ~~~~~~~~(95\% \, {\rm CL})\; .
\ee
On the other hand, atmospheric neutrino oscillations provide a {lower} bound
on the heaviest neutrino mass, since $\sqrt{\delta m^{2}_{\rm atm}} ~\sim~
0.03 \; {\rm eV}$.  Combining these two limits
\be 
0.03 \; {\rm eV} ~\le~ m_{\rm heaviest} ~\le~ 0.24 \; {\rm eV}
\label{nu-mass-bound}
\ee
we see that the heaviest neutrino mass is now known to within an order of
magnitude \cite{Bhattacharyya:2003gi}.

\paragraph{Can neutrino make up a galaxy halo? }

By $z \sim 1$ the neutrino quanta satisfying the mass bound
Eq.~(\ref{nu-mass-bound}) became sufficiently non-relativistic to make their
way into gravitational wells.  The question arises, can neutrino at least make
up the dark halos and be responsible for flat rotational curves? The answer
to this question is: no.  To prove it, let us assume that neutrino does
build up a dark matter halo with a flat rotational curve 
\be
\rho^{~}_{\rm DM} = \frac{M_{Pl}^{2} v^{2}_{\rm rot}}{r^{2}} \; . 
\label{DM-halo-nu}
\ee
We can express the energy density $\rho^{~}_{\rm DM} = \rho^{~}_{\nu}$
through the integral  of phase space density over the momenta
\be
\rho^{~}_{\nu} = \frac{m_{\nu}}{(2\pi)^{3}}
\int d^{3} k\; n(k,r) \;,
\label{DM-rho-nu}
\ee
But for fermions, the phase-space density, $n(k,r) = n(E)$, should obey the
Pauli exclusion principle, $n(E) < 1$. Combining Eqs. (\ref{DM-halo-nu})
and (\ref{DM-rho-nu}), we find $m^{4} v^{3} \sim M_{Pl}^{2} v^{2}/r^{2}$, or
\be
m_\nu > 120\; {\rm eV} \( \frac{100\; {\rm km~s}^{-1}}{v_{\rm
rot}} \)^{1/4} \( \frac{1\; {\rm kpc}}{r_c} \)^{1/2}
\ee
For dwarf galaxies this constraint (the Tremaine-Gunn limit
\cite{Tremaine:1979we}) reads $m_\nu > 500$ eV, and we arrive to contradiction
with Eq. (\ref{Gershtein-Zeldovich}), which becomes even stronger when
compared with Eq. (\ref{nu-mass-bound}).

\paragraph{Cosmological density of other thermal relics.}

Assume now some weakly interacting particle has a mass larger $1$ MeV and
decouples when it is non-relativistic.  The equilibrium number density will be
Boltzmann-suppressed in this case, by the exponent $\exp (-m/T)$. The weak
interaction cross-section implies $\sigma \sim m^2/M_W^4$, if $m\ll
M_W$. Repeating calculations for abundances, one finds that in this case
$\Omega_m h^2 \approx 3\; (1\; {\rm GeV}/ m)^2$, i.e. a correct cosmological
abundance of dark matter would be achieved for $m \approx 5$ GeV.

On the other hand, if $m\gg M_W$, the annihilation crosssection becomes $\sigma
\sim 1/m^2$ and one finds $\Omega_m h^2 \approx (m /1\; {\rm TeV})^2$,
i.e. the correct cosmological abundance of dark matter is achieved for $m
\approx 300$ GeV.  Using field-theoretical unitarity and the observed density
of the Universe, it can be shown that a stable elementary particle which was
once in thermal equilibrium cannot have a mass greater than 340 TeV
\cite{Griest:1990wd}.

\paragraph{Cosmological density of non-thermal relics.}

The mass of non-thermal relics can be much larger than $O(10^2)$ TeV without
violating unitarity bound; it can also be much smaller than than $O(1)$ GeV and
dark matter still will be cold, as required by observations.

{\it 1. Axions.} Very light scalar particles, like axions, are created in a
state of coherent oscillations. This can be viewed also as a
Bose-condensate. To illustrate the general idea, let us consider a scalar
field with potential $V(\phi ) = m^2\phi^2/2$. The equations of motion for the
Fourier modes with a momentum $k$ in an expanding Universe are
\be
\ddot{\phi}_k + 3H\dot{\phi}_k + (k^2 + m^2)\phi_k = 0\; .
\label{eq-motion-scalar}
\ee
Since the term $\propto H$ can be understood as friction, amplitude of those
modes for which $9H^2 \gg(k^2 + m^2) $ (almost) does not change with time. The
oscillations of modes with a given $k$ commence when $H$ becomes sufficiently
small, $9H^2 \ll (k^2 + m^2)$. Oscillating modes behave like particles, and
their amplitude decreases with expansion. Since modes with the largest $k$
start oscillations first, they will have the smallest amplitude and the field
becomes homogeneous on a current horizon scale. Modes with all $k$ will be
already oscillating when $3H \approx m$, and will behave like cold dark matter
since then. Note that the field will be homogeneous on the horizon scale at
this time, but may be inhomogeneous on larger scales. This may lead to
formation of dense clumps, ``axion mini-clusters''
\cite{Hogan:1988mp,Kolb:1993zz} of the mass $M
\sim 10^{-12}\,M_\odot$ \cite{Kolb:1996bu}. Note also that in the case of
axions, one has to take into account the dependence of $m$ on temperature
$T$. Solving $3H(T) = m(T)$ one finds $\Omega_{\rm axion} \sim 1$ when $f_a
\sim 10^{12} {\rm GeV}$ \cite{Abbott:1983af,Dine:1983ah}. 

{\it 2. Superheavy dark matter.} Non-conformal quantum fields cannot be kept
in a vacuum in an expanding universe. This can be understood on the example of
a scalar field, Eq. (\ref{eq-motion-scalar}). In conformal time,
Eq.~(\ref{time-conformal}), and for rescaled field, $u_k \equiv \phi_k\, a$,
the mode equations take form of an oscillator equation
\be
\ddot{u}_k +  \omega_k^2u_k = 0\; ,
\label{eq-motion-scalar-conformal}
\ee
with time-dependent frequency
\be
\omega_k^2 = k^2 + a^2 m^2 - \frac{\ddot{a}}{a}(1-6\xi)\; .
\label{eq-motion-scalar-conformal}
\ee
The constant $\xi$ describes the coupling to the scalar curvature,
corresponding term in the Lagrangian is $\xi R \phi$. The case of $\xi=0$
corresponds to minimal coupling (Eq.~(\ref{eq-motion-scalar}) was written for
this case), while $\xi=1/6$ is the case of conformal coupling. Equations for
massless, conformally coupled quanta are reduced to the equation of motion in
Minkowski space-time. Particle creation does not occur in this case. For
massive particles, conformal invariance is broken and particles are created
regardless of the value of $\xi$. Let us consider the case of $\xi=1/6$ (the
general situation is considered in \cite{Kuzmin:1998kk}).  It is the particle
mass which couples the system to the background expansion and serves as the
source of particle creation in this case. Therefore, we expect that the number
of created particles in comoving volume is $\propto m^3$ and the effect is
strongest for the heaviest particles. In fact, stable particles with $m >
10^9$ GeV would overclose the Universe in the standard ``pre-inflationary''
Friedmann model \cite{Kuzmin:1998uv}. Inflation cuts the particle production
and $\Omega_{\rm SH} \sim 1$ if $m > 10^{13}$ GeV and reheating temperature is
$T \sim 10^{9}$ GeV, which is the value of reheating temperature compatible
with supergravity models \cite{Chung:1998zb,Kuzmin:1998uv,Kuzmin:1998kk}.


\section{BASICS OF INFLATION}
\label{sec:inflation}

In frameworks of ``classical'' cosmology and assuming no fine-tuning, one
concludes that a typical universe should have had Plankian size, live Plankian
time and contain just a few particles.  This conclusion is based on the
observation that Fridmann equations contain a single dimension-full parameter
$M_{\rm Pl} \sim 10^{19}$ GeV, while dimensionless parameters naturally are
expected to be of order unity. Yet, the observable Universe contains $10^{90}$
particles in it and had survived $10^{65}$ Plankian times. Where does it all
came from?  In other words, why is the Universe so big, flat ($\Omega_0
\approx 1$) and old ($t > 10^{10}$ years), homogeneous and isotropic ($\delta
T/T \sim 10^{-5}$), why does it contain so much entropy ($S>10^{90}$) and does
not contain unwanted relics like magnetic monopoles?  These puzzles of
classical cosmology were solved with the invention of Inflation
\cite{Starobinsky:1980te,Guth:1981zm,Linde:1982mu,Albrecht:1982wi,
Linde:1983gd}. All these questions are related to the initial conditions and
one can simply postulate them.  The beauty of Inflation is that it prepares
these unnatural initial conditions of Big Bang, while the pre-existing state
(which can be arbitrary to a large extent) is forgotten. Inflationary theory
came with unplanned bonuses. Not only does the Universe become clean and
homogeneous during inflation, but also the tiny perturbations necessary for
the genesis of galaxies are created with the correct magnitude and
spectrum. Below we consider the basics of inflationary cosmology.

\subsection{Big Bang puzzles and Inflationary solutions}

\paragraph{Horizon problem and the solution.}

The size of a causally connected region (horizon) scales in proportion to
time, $R_{\rm H} \propto t$. On the other hand, the physical size of a given
patch grows in proportion to the scale factor, $R_{\rm P} \propto \;a(t)
\propto t^\gamma$.  The exponent $\gamma$ depends upon the equation of state,
$\gamma = 1/2$ for radiation and $\gamma = 2/3$ for matter dominated
expansion. In any case, for the ``classical'' Friedmann Universe $\gamma < 1$
and the horizon expands faster than volume.  Take the largest visible patch
today. It follows that in the past its physical size should have been larger
than the horizon size at the time (since they are equal today) and therefore
this patch should have contained many casually disconnected regions.  For
example, as we have found in Section \ref{sec:CMBR}, the angular size of
horizon at the moment of last scattering is $\approx 2^\circ$, see
Eq.~(\ref{angSize-hor-ls}), which tells us that we observe $10^{4}$ causally
disconnected regions at the surface of last scattering. The question arises,
why is the Universe so homogeneous at large scales?

This problem can be solved if during some period of time the volume had
expanded faster than the horizon. During such a period, the whole visible
Universe can be inflated from one (``small'') causally connected region.
Clearly, this happens if $\gamma > 1$, which means $\ddot{a} > 0$.  Either of
these two conditions can be used as definition of an inflationary regime.
Using the Friedmann equation (\ref{Friedmann-Eq2}) we find that the
inflationary stage is realized when $p < - \rho /3$. In particular, if $p = -
\rho$ the energy density remains constant during expansion in accord with the
first law of thermodynamics, Eq.~(\ref{internal-energy-conservation}), and the
physical volume expands exponentially after a few Hubble expansion times,
$a(t) =\e^{Ht}$, see~ Eq.~(\ref{Friedmann-Eq1}).

\paragraph{Curvature problem and the solution.}

The Friedmann equation~(\ref{Friedmann-Eq1}) can be re-written as
\be 
k = a^{2}\(\frac{8\pi G}{3}\, \rho - H^{2}\) = 
a^{2}H^{2}\,\(\Omega -1\) = \dot{a}^{2}\,\(\Omega -1\) = {\rm const}.
\label{Friedmann-Eq1a}
\ee 
Here we immediately see the problem: during matter or radiation dominated
stages, $\dot{a}^{2}$ decreases (in general, this happens for any expansion
stage with $\ddot{a} < 0$), therefore $\blue \Omega$ is driven away from
unity.  To observe $\Omega ~\sim~ 1$ today, the observer has to live in a
universe with extreme initial fine-tuning, say at the epoch of
nucleosynthesis, when temperature was $T\sim 1$ MeV, one finds $|\Omega(t_{\rm
NS}) -1| < 10^{-15}$, and even stronger tuning is required at earlier epochs.
A possible solution is obvious: accelerated expansion $\ddot{a} > 0$ increases
$\dot{a}$ and therefore drives $\Omega(t)$ to unity.  A robust, crucial and
testable prediction of inflationary cosmology is a flat Universe, $\Omega =1$.

\paragraph{The problem of Entropy.}

As we know already, the energy of a vacuum, $p =-\rho$, stays constant despite
the expansion.  In this way, room for matter full of energy could have been
created.  The vacuum energy is converted into particles and radiation at some
later stage and, in particular, the observed huge entropy is
created. Potentially, this mechanism works for any inflationary scenario,
since the product $\rho a^{3}$ is guaranteed to grow whenever $\ddot{a} >
0$. However, the important question is whether a graceful exit out of the
inflationary stage and successful reheating is possible. In practice, this
issue has killed a number of inflationary models. Remarkably, the original
model by A. Guth \cite{Guth:1981zm} had being ruled out precisely on these
grounds \cite{Guth:1981uk}.

Inflation has to continue for a sufficiently long time for the problems of
horizon, curvature and entropy to be solved. All these give roughly the same
condition on the number of required ``e-foldings'' of inflation
\cite{Guth:1981zm} and we consider here a (simplified) derivation based on
entropy. A precise condition can be found e.g. in Ref. \cite{Liddle:1993fq}.
Multiplying the current temperature in the universe by its visible size we find
$Ta\chi_0 \sim 10^{30}$, where $\chi_0$ is the comoving size of the present
horizon. The product $Ta$ conserves (up to the change in the number of
relativistic degrees of freedom, which we neglect here) since the Universe
expansion is adiabatic after the end of Inflation, see
Eq.~(\ref{Entropy-reativistic}). Let $T_r$ denote the reheating temperature
and $\e^{N} \equiv a_f/a_i$ the number of inflationary ``e-foldings'',
where $a_f$ is the value of scale factor at the end of inflation and $a_i$ at
its beginning, respectively. We also want at least the whole visible universe
to be inflated out of a single causally connected patch, which gives
$a_i\chi_0 \sim H_i^{-1}$, where $H_i$ is the value of the Hubble parameter
during Inflation.  All this gives the condition\footnote{Strictly speaking
in this relation ${T_r}$ is not the real temperature in a state of thermal
equilibrium, but $T_r \sim \rho_r^{1/4}$, where $\rho_r$ is the energy density
at the moment when the expansion becomes dominated by relativistic particles.} 
\be 
\frac{T_r}{H_i}\; \e^N \agt 10^{30} \; .
\label{e-folds}
\ee 
In popular models of Inflation the ratio ${T_r}/{H_i}$ is within a couple
orders of magnitude from unity, and we find $N \gsim 65$.

\subsection{Models of Inflation}

Consider $T_{\mu\nu}$ for a scalar field {$\varphi$}
\begin{eqnarray}
T_{\mu\nu}=\partial_{\mu}\varphi\, \partial_{\nu}\varphi - 
g_{\mu\nu}\, {\cal L}
\label{Tmunu-scalarfield}
\end{eqnarray} 
with the Lagrangian :
\begin{eqnarray}
{\cal L}=\partial_{\mu}\varphi\, \partial^{\mu}\varphi - V(\varphi)\; .
\label{Lagrangian-scalarfield}
\end{eqnarray} 
In a state when all derivatives of {$\varphi$} are zero, the stress-energy
tensor of a scalar field simplifies to
$T_{\mu}^\nu=V(\varphi)\,\delta_{\mu}^\nu$. This corresponds to a vacuum
state. Indeed, comparing with Eq.~(\ref{Tmunu-ideal}), we find $V = \rho =
-p$. There are two basic ways to arrange {$\varphi \approx$ const} and hence
to imitate the {vacuum}-like state.

1. Consider the potential $V(\phi)$, which has a local minimum with a non-zero
energy density separated from the true ground state by a potential barrier
\cite{Guth:1981zm}.  A universe which happened to be trapped in the
meta-stable minimum will stay there for a while (since such a state can decay
only via subbarrier tunneling) and expansion of the universe will diminish all
field gradients. Then the Universe enters a vacuum state.  This model is ruled
out since the inhomogeneities created during the phase transition which
terminates the inflationary phase are too large \cite{Guth:1981uk}. However,
the model is good for illustration purposes. The frequently asked question is:
how can it be that the energy density stays constant despite the expansion?
In the model with local potential minimum the energy cannot decrease
(classically) below the local minimum value, and therefore it has to stay
constant despite the expansion.

2. A. Linde was first to realize that things work in the simplest possible
setup \cite{Linde:1983gd}.  Consider the potential
\begin{equation}
V(\phi)=\frac{1}{2}m_\phi^2\phi^2 \; .
\label{chaotic}
\end{equation}
The equation of field motion in an expanding Universe is
$
\ddot{\phi}+3H\dot{\phi}+m_\phi^2\phi=0 \; .
$
If {$H \gg m $}, the ``friction'' is too big and the field (almost) does not
move. Therefore time derivatives in $\; T_{\mu\nu}$ can be neglected, and
inflation starts (in a sufficiently homogeneous patch of the Universe). A
Hubble parameter in this case is determined by the potential energy, $H
\approx m {\phi}/{M_{\rm Pl}}$, and we see that inflation starts if the
initial field value happens to satisfy ${\phi > M_{\rm Pl}}$. During
inflationary stage the field slowly rolls down the potential hill. This motion
is very important in the theory of structure creation; inflation ends when
${\phi \sim M_{\rm Pl}}$. At this time, field oscillations start around the
potential minimum and later decay into radiation. In this way matter was
likely created in our Universe.

\subsection{Unified theory of Creation}

During Inflation and by its end the Universe was in a vacuum-like state.  We
have to figure out how this ``vacuum'' was turned into the matter we observe
around us, and how primordial fluctuations which gave rise to galaxies were
created. Fortunately, these problems can be formalized in a nice and unified
way. Basically, everything reduces to a problem of particle creation in
a time-dependent classical background. On top of every ``vacuum'' there are
fluctuations of all quantum fields which are present in a given model. This
bath of virtual quanta is indestructible, and even Inflation cannot get rid of
it. Being small, fluctuations of any field obey an oscillator equation
\begin{equation}
\ddot{u}_k ~+~ [k^2 + m^2_{\rm eff} (\eta)]\; u_k = 0 \, ,
\label{ModEq}
\end{equation}
here $u_k$ are amplitudes of fluctuating fields in Fourier space.  Effective
mass becomes time dependent through the coupling to time-dependent
background. Because $m_{\rm eff}$ is time dependent, it is not possible to
keep fluctuations in a vacuum.  If one arranges to put oscillators with
momentum $k$ into the vacuum at one time, they will not be in vacuum at a
latter time because positive and negative frequency solutions mix, see below.
Several remarks are in order.
\begin{itemize}
\item Eq.~(\ref{ModEq}) is valid for all particle species.      
\item The equation looks that simple in a conformal reference frame $ds^2 =
      a(\eta)^2\;(d\eta^2-dx^2)$. (Everywhere in this chapter a ``dot'' means
      derivative with respect to $\eta$.)
\item Of particular interest are ripples of space-time itself: curvature
      fluctuations (scalar fluctuations of the metric) and gravity waves
      (tensor fluctuations of the metric).
\item Effective mass $m_{\rm eff}$ can be non-zero even for massless fields.
      Gravitational waves give the simplest example \cite{Grishchuk:1975ny},
      with $m^2_{\rm eff} = - {\ddot{a}}/{a}$. The effective mass for curvature
      fluctuations has a similar structure $m^2_{\rm eff} = - {\ddot{z}}/{z}$,
      but with $a$ being replaced by $z \equiv a\dot{\phi}/H$, see Refs.
      \cite{Lukash:1980iv,Sasaki:1986hm,Mukhanov:1988jd,Mukhanov:1992me}.
\item For conformally coupled, but massive scalar ~$m_{\rm eff} = m_0 \,
      a(\eta)$.
\end{itemize}
Note that creation was only possible because nature is not
conformally-invariant. Otherwise, $m_{\rm eff}$ is time-independent and vacuum
remains vacuum forever.  There are two important instances of time varying
classical background in cosmology:
\begin{itemize}
\item Expansion of space-time, ~$a(\eta )$. 
\item Motion of the inflaton field, ~$\phi (\eta )$.
\end{itemize}
Both can be operational at any epoch  of creation:
\begin{itemize}
\item During inflation (superhorizon size perturbations).
\item While the inflaton oscillates (reheating).
\end{itemize}

During inflation superhorizon size perturbations of metric are created, which
give seeds for Large Scale Structure (LSS) formation and eventually lead to
the formation of galaxies, and therefore of the Solar system and all the rest
which we can see around us. During reheating matter itself is
created. Overall, there are four different situations (two sources times two
epochs). If coupling to the inflaton is not essential, the corresponding
process will be called ``pure gravitational creation'' in what follows.

There are several primary observables which can be calculated out of $u_k$ and
further used for calculation of quantities of interest. Most useful are:
\begin{itemize}
\item The particle occupation numbers, $n_k$. Integration over $d^3k$
gives the particle number density.
\item The power spectrum of field fluctuations, $P_k \equiv u_k^*
u_k$. Integration over $d^3k$ gives the field variance.
\end{itemize}
Depending on physical situation, only one or the other may have
sense. The particle number in a comoving volume is useful because it 
\begin{itemize}
\item is adiabatic invariant on sub-horizon  scales (or when $m>H$);
\item allows to calculate  abundances of various relics, e.g. dark matter.
\end{itemize}
But it has no meaning at super-horizon scales when $m < H$. The power spectrum
and/or field variance is useful because it
\begin{itemize}
\item does not evolve on  super-horizon  scales if $m<H$;  
\item allows to calculate density perturbations generated
      during inflation;
\item is crucial for dynamics of phase transitions;
\item helps to calculate back-reaction in a simple way (Hartree 
      approximation).
\end{itemize}
But $P_k$ evolves on sub-horizon  scales and when $m > H$.

Let me start with the discussion of metric perturbations.


\paragraph{Gravitational creation of metric perturbations.}

As an important and simple example, let us consider quantum fluctuations of a
real scalar field, which we denote as $\varphi$. It is appropriate to rescale
the field values by the scale factor, $\varphi \equiv \phi/a(\eta)$. This
brings the equations of motion for the field $\phi$ into a simple form of
Eq.~(\ref{ModEq}).  As usual, we decompose $\phi$ over creation and
annihilation operators $b_{\k}$ and  $b_{\k}^{\dagger}$
\be
\phi(\x,\eta) = 
\int \frac{d^3k}{(2\pi)^{3/2}}\; \left [ u_k(\eta)\, b_{\k}\, \e^{i\k\x} + 
  u^{*}_k(\eta)\, b_{\k}^{\dagger}\, \e^{-i\k\x}   \right] \; .
\label{field-fourier-decomp} 
\ee
Mode functions $u_k$ satisfy Eq.~(\ref{ModEq}). In what follows we will assume
that $\varphi$ is the inflaton field of the ``chaotic'' inflationary model,
Eq.~(\ref{chaotic}). During inflation $H\gg m$ and $H \approx {\rm const}$.
So, to start with, we can assume that $\varphi$ is a massless field on the
constant deSitter background. (The massive case can be treated similarly, but
analytical expressions are somewhat more complicated and do not change the
result in a significant way. Corrections due to change of $H$ can also be
taken into account, and we do that later for the purpose of comparison with
observations.)  With a constant Hubble parameter during inflation the solution
of Friedmann equations in conformal time is
\be
a(\eta ) = - \frac{1}{H\eta} 
\label{a-desitter-confomal} 
\ee 
and the equation for mode functions of a massless, conformally coupled to
gravity $(\xi =0)$, scalar field takes the form 
\be 
\ddot{u}_k + k^2 {u}_k -\frac{2}{\eta^2}\; {u}_k = 0 \; .
\label{ModEq-massles-conf}
\ee
Solutions which start as vacuum fluctuations in the past ($\eta \rightarrow
-\infty$) are given by
\be
{u}_k = \frac{\e^{\pm ik\eta}}{\sqrt{2k}}\left( 1 \pm
\frac{i}{k\eta}\right) \; .
\label{ModFunc-massles-solution}
\ee
Indeed, at $\eta \rightarrow -\infty$ the second term in the parentheses can
be neglected and we have the familiar mode functions of the Minkowski space
time. The wavelength of a given mode becomes equal to the horizon size (or
``crosses'' the horizon) when $k\eta = 1$. Inflation proceeds with $\eta
\rightarrow 0$, so the modes with progressively larger $k$ cross the horizon. 
After horizon crossing, when $k\eta \ll 1$, the asymptotics of mode functions
are
\be
u_k = \pm \frac{i}{\sqrt{2}k^{3/2}\eta}, {\rm ~~~~or~~~} \varphi_k 
= \frac{u_k}{a(\eta)} = \mp \frac{iH}{\sqrt{2}k^{3/2}} \; .
\label{ModFunc-massles-asimptotic}
\ee
The field variance is given by
\be
\langle 0| \phi^2(x) |0\rangle =
\int \frac{d^3k}{(2\pi)^3}\; |\varphi_k|^2 \; .
\label{variance-def}
\ee
and we find in the asymptotic (the careful reader will recognize that this is
already regularized expression with zero-point fluctuations being subtracted)
\be
\langle \varphi^2 \rangle = \frac{H^2}{(2\pi)^2} \int \frac{dk}{k} \; .
\label{variance-varphi}
\ee
Defining the power spectrum of the field fluctuations as a power per decade,
$\langle \varphi^2 \rangle  \equiv  \int  P_\varphi(k)\; {d\ln k}$, we find
\be
P_\varphi (k) = \frac{H^2}{(2\pi)^2} \; .
\label{PowerSpectrum-varphi}
\ee 

\paragraph{Curvature perturbations.}

According to Eq.~(\ref{Spatial-curvature}), the three-dimensional curvature of
space sections of constant time is inversely proportional to the scale factor
squared, $^{(3)} R \propto {a^{-2}}$. Therefore, the perturbation of spatial
curvature is proportional to ${\delta a}/{a}$, and this ratio can be evaluated
as
\be
\zeta  \equiv \frac{\delta a}{a} = H\delta t = 
H\, \frac{\delta \varphi}{\dot{\varphi}} \; .
\label{zeta-def}
\ee
This allows to relate the power spectrum of curvature perturbations
to the power spectrum of field fluctuations 
\be
P_\zeta (k) = \frac{H^{2}}{\dot{\varphi}^{2}} \;\, P_\varphi (k)\; ,
\label{Pphi-Pcurvature}
\ee 
and we find for the power spectrum of curvature perturbations
\be
P_\zeta (k) = \frac{1}{4\pi^{2}}\; \frac{H^{4}}{\dot{\varphi}^{2}}\; .
\label{PowerSpectrum-curvature}
\ee 
This very important relation describes inflationary creation of primordial
perturbations, and can be confronted with observations.  The usefulness of
curvature perturbations for this procedure can be appreciated in the following
way:

1. Consider the perturbed metric, Eq.~(\ref{metric-perturbed}). The product
$a(1-\Phi)$ for the long-wavelength perturbations can be viewed as a perturbed
scale factor, i.e. ${\delta a}/{a} = - \Phi$. Comparing this relation with
Eq.~(\ref{zeta-def}) and Eq.~(\ref{theta-phi-md}), we find for the temperature
fluctuations which are of the superhorizon size at the surface of last
scattering
\be
\frac{\delta T}{T} = \frac{2}{3}\; \zeta_k \; .
\label{power-T-zeta}
\ee

2. On superhorizon scales the curvature perturbations do not
evolve.\footnote{I should warn that this is quite a generic statement and does
holds in situations usually considered. Thus, it is forgotten
sometimes that this is not a universally true statement. To avoid possible
confusion when encountering specific complicated models, the reader should
keep this fact in mind.} This fact allows to relate directly the observed
power spectrum of temperature fluctuations to the power spectrum of curvature
fluctuations generated during inflation.

\paragraph{Tensor perturbations.}

Mode functions of gravity waves (after rescaling by $M_{\rm Pl}/\sqrt{32\pi}$)
obey the same equation as mode functions of massless minimally coupled scalar
\cite{Grishchuk:1975ny}.  Using the result Eq.~(\ref{PowerSpectrum-varphi}) we
immediately find \cite{Rubakov:1982df}
\be
P_{T}(k) = 2\; \frac{32 \pi}{M_{\rm   Pl}^{2}} \; P_\varphi(k) =  
\frac{16}{\pi}\; \frac{H^2}{M_{\rm   Pl}^{2}} \; ,
\label{tensor-perturbations}
\ee
where the factor of 2 accounts for two graviton polarizations.

\paragraph{Slow-roll approximation.}

During inflation, the field $\varphi$ rolls down the potential hill very
slowly.  A reasonable approximation to the dynamics is obtained by neglecting
$\ddot{\varphi}$ in the field equation $\ddot{\varphi}+3H\dot{\varphi} +V'
=0$. This procedure is called the slow-roll approximation
\be
\dot{\varphi} \approx -\frac{V'}{3H} \; .
\label{Slow-roll-equation}
\ee 
Field derivatives can also be neglected in the energy density of the inflaton
field, ~$\rho \approx V$
\be
H^{2} = \frac{8\pi }{3M_{\rm Pl}^2} \, V \; .
\label{Slow-roll-Hubble}
\ee 
This gives for curvature perturbations
\be
\zeta_k \equiv  P_\zeta (k)^{1/2} =
\frac{H^2}{2\pi\;\dot{\varphi}} =  \frac{4H}{M_{\rm Pl}^2}\;
\frac{V}{V'}\; .
\label{Slow-roll-curvature}
\ee

\paragraph{Normalizing to CMBR.}

As an example, let us consider the simplest model~ $V = \half m^{2}
\varphi^{2}$.~~ We have
\be
\frac{V}{V'} =  \frac{\varphi}{2}, ~~~~~~{\rm and}~~~ {H} =
\sqrt{\frac{4\pi}{3}}\frac{m\varphi}{M_{\rm Pl}}  \; .
\ee 
This gives for the curvature fluctuations
\be
\zeta_k = \sqrt{\frac{16\pi}{3}}\; \frac{m\, \varphi^{2}}{M_{\rm Pl}^3} \; .
\ee
Using the relation between curvature and temperature fluctuations,
Eq.~(\ref{power-T-zeta}), and normalizing ${\delta T}/{T}$ to the 
measured value at largest ~$l$,~ which is ${\delta T}/{T}
\sim 10^{-5}$ (see Fig.~\ref{PowerSpectrum-WMAP}) we find the restriction on
the value of the inflaton mass in this model:
\be
m \approx \frac{\delta T}{T}\; \frac{M_{\rm Pl}}{30} \approx 10^{13}
\;\; {\rm GeV} \; . 
\ee
Here I have used the fact that in this model the observable scales cross the
horizon when $\varphi \approx M_{\rm Pl}$.

\paragraph{Slow-roll parameters.}

The number of e-foldings ($a = \e^{Ht} \equiv \e^N$) of inflationary expansion
from the time when $\varphi = \varphi_i$ to the end can be found as
\be
N(\varphi_i ) = \int_{t_i}^{t_f} H(t) dt = \int \frac{H}
{\dot{\varphi}}\, d\varphi = \frac{8\pi}{M_{\rm Pl}^2} 
\int_{\varphi_e}^{\varphi_i} \frac{V}{V'}\, d\varphi \; .
\label{e-foldings-rsult}
\ee
In particular, in the model Eq.~(\ref{chaotic}) we find that the largest
observable scale had crossed the horizon ($N \sim 65$) when $\varphi_i \approx
3.5M_{\rm Pl}$. All cosmological scales which fit within the observable
universe encompass a small $\Delta \phi$ interval within $M_{\rm Pl} < \varphi
< \varphi_i$.  And inflaton potential should be sufficiently flat over this
range of $\Delta \phi$ for the inflation to proceed.  This means that
observables essentially depend on the first few derivatives of $V$ (in
addition the the potential $V(\phi_0)$ itself). From the first two derivatives
one can construct the following dimensionless combinations
\begin{eqnarray}
&& \epsilon \equiv \frac{M_{\rm Pl}^2}{16\pi}
\left(\frac{V'}{V}\right)^2 \; , \\
&& \eta \equiv \frac{M_{\rm Pl}^2}{8\pi}\frac{V''}{V}  \; ,
\end{eqnarray}
which are often called the slow-roll parameters.

The power spectra of curvature,  Eq.~(\ref{Pphi-Pcurvature}), and of tensor
perturbations, Eq.~(\ref{tensor-perturbations}), in slow-roll parameters can
be rewritten as
\be
P_{\zeta}(k) =  \frac{1}{\pi\epsilon }\frac{H^2}{M_{\rm Pl}^2}, ~~~~~~~~~~~ 
P_{T}(k) =  \frac{16}{\pi}\frac{H^2}{M_{\rm Pl}^2}\; .
\label{Power-spectra-slow-roll}
\ee
Comparing these two expressions we find
\be
\frac{P_{T}(k)}{P_\zeta(k)} = 16 \epsilon \; .
\label{consistency-relation0}
\ee

\paragraph{Primordial spectrum.}

In general, the spectra can be approximated as power law functions in $k$:
\begin{eqnarray}
&& P_\zeta(k) =  P_\zeta(k_0) \left(\frac{k}{k_0}\right)^{n_S -1}  \; ,\\
&& P_{T}(k) =  P_{T}(k_0) \left(\frac{k}{k_0}\right)^{n_T} \; .
\label{spectra-powerlow-def}
\end{eqnarray}
To the first approximation, $H$ in Eq.~(\ref{Power-spectra-slow-roll}) is
constant. Therefore, in this approximation, power spectra do not depend on $k$
and $n_S=1$, $n_T=0$. This case is called the Harrison-Zel'dovich spectrum 
\cite{Harrison:1970fb,Zeldovich:1972ij} of primordial perturbations.  
However, in reality, $H$ is changing, and in Eq~(\ref{Power-spectra-slow-roll})
for every $k$ one should take the value of $H$ at the moment when the relevant
mode crosses horizon. In slow roll parameters one then finds (see
e.g. Ref.~\cite{Lidsey:1997np} for the nice overview) 
\be
n_S = 1 + 2\eta - 6 \epsilon, ~~~~~~~~~~~  n_T = - 2 \epsilon\; .
\label{n-in-slow-roll}
\ee
We can re-write Eq.~(\ref{consistency-relation}) as a relation between the
slope of tensor perturbations and the ratio of power in tensor to curvature
modes 
\be
\frac{P_{T}(k)}{P_\zeta(k)} = -8 n_T  \; .
\label{consistency-relation}
\ee
This is called the {\it consistency relation} to which (simple) inflationary
models should obey.

Different models of inflation have different values of slow-roll parameters
$\eta$ and $\epsilon$, and therefore can be represented in the
($\eta$,$\epsilon$) parameter plane. Using the relations
Eq.~(\ref{n-in-slow-roll}) we see that this plane can be mapped into
$(n_S,n_T)$, or using also Eq.~(\ref{consistency-relation}) into the
$(n_S,r)$   parameter plane, where $r$ is the ratio of power in tensor to
scalar (curvature) perturbations. In this way, different inflationary models
can be linked to observations and constraints can be obtained.

\begin{figure}
\begin{center}
\includegraphics[width=7cm]{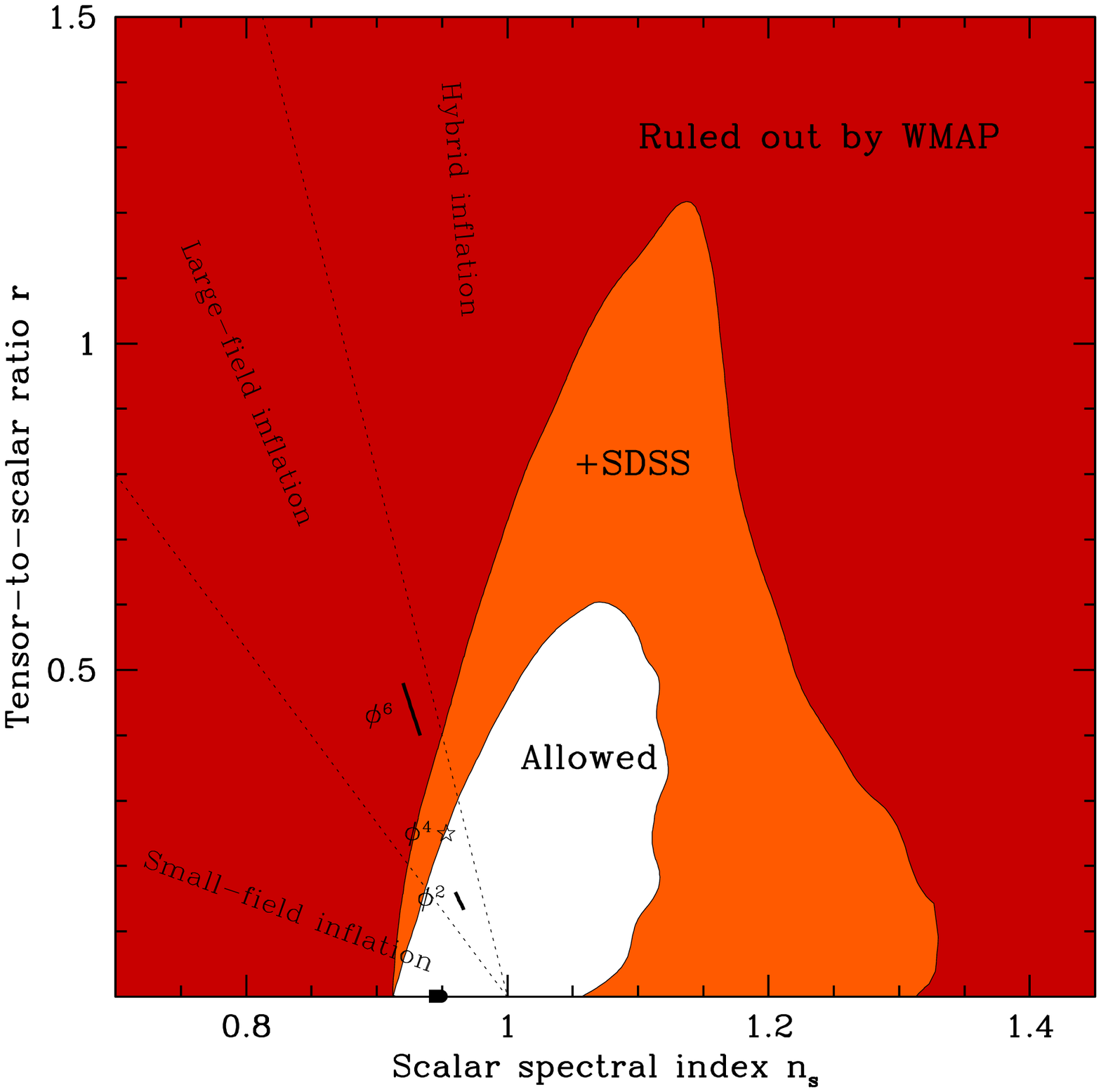}
\hspace{.7cm}\parbox{8cm}{\vspace{-6.7cm}
\includegraphics[width=7.cm]{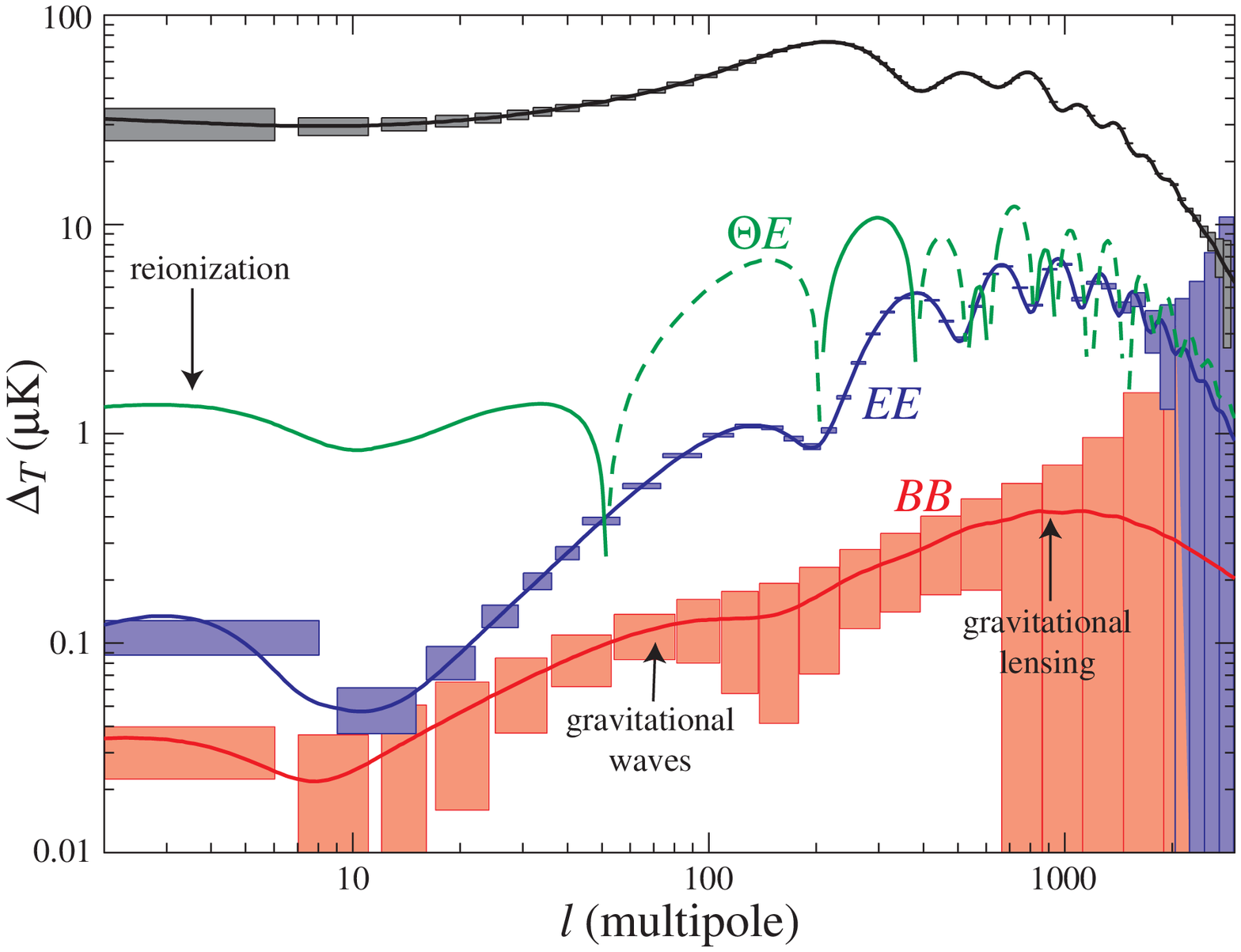}}
\caption{ {\bf Left panel:} 95\% constraints on inflationary models in the
$(n_s,r)$ plane. From Ref.~\cite{Tegmark:2003ud}. {\bf Right panel:} Forecast
for the planned ESA Planck mission.}
\label{fig:SDSS_MAP-inflation}
\end{center}
\end{figure}

The most recent constraints in the $(n_s,r)$ plane, utilizing WMAP and SDSS
data, are presented in Fig.~\ref{fig:SDSS_MAP-inflation}, the left panel. The
shaded dark red region is ruled out by WMAP alone.  The shaded light red
region is ruled out when adding SDSS information.  The two dotted lines
delimit the three classes of inflationary models known as the small-field,
large-field and hybrid models.  Some single-field models of inflation make
highly specific predictions in this plane, as indicated.  From top to bottom,
the figure shows the predictions for $V(\phi)\propto\phi^6$ (line segment;
ruled out by CMB alone), $V(\phi)\propto\phi^4$ (star; on verge of exclusion)
and $V(\phi)\propto\phi^2$ (line segment; the inflation model
Eq.~(\ref{chaotic}); still allowed).

\paragraph{Testing inflation.}

All predictions of Inflationary cosmology, which could have being tested so
far, have being confirmed. In particular, the Universe is spatially flat
(within experimental errors), $\Omega = 1.02\pm 0.02$, see Table I. The
primordial perturbations are of superhorizon size and adiabatic. The spectral
index is close to the Harrison-Zeldovich case, see
Fig.~\ref{fig:SDSS_MAP-inflation}, the left panel. Crucial test of
inflationary models would be detection of gravity waves and verification of
the consistency relation. This signatures of typical inflationary models are
within reach of future CMBR experiments, see
Fig.~\ref{fig:SDSS_MAP-inflation}, the right panel.



\section{Ultra-High Energy Cosmic Rays}
\label{sec:UHECR}

In early years, cosmic ray studies were ahead of accelerator research,
starting from the discovery of positrons, through muons, to that of pions and
strange particles. Today we are facing the situation that the puzzling saga of
cosmic rays of the highest energies may again unfold in the discovery of new
physics, now beyond the Standard Model; or it may bring to life an ``extreme''
astrophysics.

Immediately after the discovery of the relict Cosmic Microwave Background
Radiation (CMBR), Greisen, Zatsepin and Kuzmin
\cite{Greisen:1966jv,Zatsepin:1966jv} have realized that the highest energy
protons should catastrophically loose energy in photo-production of pions on
this universal background. This process limits the distance to the farthest
sources of observed rays to be roughly 100 Mpc and should lead to the cut-off
in the energy spectrum.  However, the number of events with energies beyond
the expected cut-off as measured by different installations is growing with
time \cite{Linsley:1963,Winn:1986un,Lawrence:1991cc,Glushkov:1991sn,%
Bird:1995uy,Takeda:2002at,Abu-Zayyad:2002sf}, while no nearby sources where
identified.  The findings of Greisen, Zatsepin and Kuzmin (GZK) are based on
solid fundamental physics which involve precisely measured cross-sections in a
GeV energy range (in the center of mass reference frame).  Therefore, if the
data are correct -- and it is believed they are basically
correct\footnote{While recently a disagreement in measured fluxes has emerged,
there is no reason for doubt in the reality of super-GZK events.} -- one
should either invoke new physics, or accept unconventional and uncomfortable
very ``extreme'' astrophysics.  This is the reason for the excitement and
growing interest in ultra-high energy cosmic ray research; for recent reviews
see \cite{Nagano:2000ve,Anchordoqui:2002hs,Cronin:2004ye,Torres:2004hk}.

\paragraph{Methods of detection.}

At energies below $10^{14}$ eV, the flux of cosmic rays is sufficiently high
that direct measurements using high altitude balloons or satellite experiments
are possible. Above $10^{15}$ eV, the flux is only one particle per m$^2$ per
year, which excludes direct observations on the orbit. At $10^{20}$ eV the
number is down to one particle per square kilometer per century. Here the
problem for direct measurements would be not only a vanishingly small flux,
but the enormously high energy itself. (Remember that calorimeters at modern
colliders weigh hundreds of thousands of tonnes.)  Fortunately, the major part
of our UHECR detectors is already built for us by Nature and is put, rotating,
into orbit: the Earth's atmosphere makes a perfect calorimeter.  The
atmosphere is just thick enough so that showers of secondary particles
produced by incoming cosmic rays of the highest energies, in collisions with
nuclei of air, reach their maximum intensity just above the Earth's surface.
Particles in a shower propagate with the velocity of light, forming a thin
disk perpendicular to the direction of the incident particle.  At $10^{19}$ eV
the footprint of the shower on the ground is several kilometers across.

The shower can be registered either by placing an array of particle detectors
on the earth's surface, or by measuring the Cherenkov light produced by
particles in the atmosphere, or by tracking the fluorescence light emitted
when shower particles excite nitrogen molecules of the air. Particle
detectors in a ground array can be spaced hundreds of meters apart and are
operational around the clock.  Fluorescence light telescopes see the cosmic ray
track just like a fly's eye would see the meteorite, but only moving with
the speed of light. These detectors are operational only on clear moonless
nights, but are able to measure the longitudinal shower profile and its maximum
depth directly.

With either technique, the energy and incident direction of primary particle
can be measured shower by shower.  Chemical composition also can be inferred,
but only in a statistical sense, after averaging over many showers.

1. {\it Arrival direction.} The timing of a signal in different detectors is
used to determine the direction of a shower (ground array technique).
Direction is measured with an accuracy of about $2^\circ$. The measurement is
straightforward and does not involve any uncertainties.  Inferred information
is reliable. Fluorescence light telescopes observe the whole shower track,
and in stereo mode the precision of angle determination is $0.5^\circ$.

2. {\it Energy.} Energy estimate, on the other hand, is not that
straightforward.  In fluorescent light detectors, the energy of primary
particle is derived from the observed light intensity, therefore incorrect
modeling and/or insufficiently frequent monitoring of atmosphere
transparency can be a source of errors. For the ground array detectors, the
energy estimate relies on a Monte-Carlo model of shower development and is
related to the shower density profile. Nevertheless, the currently favored
model, QGSJET \cite{Kalmykov:1993qe}, describes data well from TeV up to
highest energies and it is believed that the overall error (statistical plus
systematic) in energy determination does not exceed 30 \%.

The best would be to employ both the ground array and fluorescent light
techniques simultaneously. This should reduce systematic errors, and
this is the design of the forthcoming Pierre Auger project
\cite{Auger}.

3. {\it Chemical composition.} Chemical composition can be inferred from the
details of shower development.  For example, showers initiated by heavy nuclei
start earlier in the atmosphere and have less fluctuations compared to proton
showers. Fluorescence detectors observe shower development directly.  Using
ground array detectors, the shower depth can be extracted by measuring
e.g. the ratio of electrons to muons. At lowest energies, the chemical
composition of cosmic rays reflects primary and secondary element abundances;
for a recent review see \cite{Kampert:2000tw}; at highest energies, $E >
4\times 10^{19}$ eV, the conclusion is that less than 50\% of primary cosmic
rays can be photonic at 95\% confidence level \cite{Ave:2001xn}.

\subsection{Propagation of the ultra-high energy cosmic rays}

In this subsection, we consider the influence of different cosmological
backgrounds on the propagation of highest-energy cosmic rays.

\paragraph{Magnetic fields.}

Magnetic fields play an important role in the processes of cosmic ray
acceleration and propagation, their trajectories being bent by the
action of the Lorentz force
\begin{eqnarray}
\frac{\vec{dv}}{dt} = \frac{Ze} {E} \; \;
[\vec{v} \times \vec{B} ] \; .
\end{eqnarray} 
For a qualitative discussion, it is often sufficient to compare
a gyro-radius of the trajectory of a relativistic particle
\begin{eqnarray}
R_g = \frac{E}{Z e B}
\label{LarRad}  
\end{eqnarray} 
to other relevant length scales of the problem.  E.g., a magnetic ``trap'' can
not confine a cosmic ray if the gyro-radius exceeds the trap size. The
deflection angle $\Delta\theta$, after traversing the distance $L$ in a
homogeneous magnetic field, is proportional to $L/R_{\rm L}$.  In a chaotic
magnetic field, the deflection angle will grow as $\sqrt{L}$.  Let us estimate
a typical deflection angle, $L/R_g$, of a charged UHE particle after
traversing Galactic or extra-galactic magnetic fields.

\begin{figure}
\begin{center}
\includegraphics[width=14.cm]{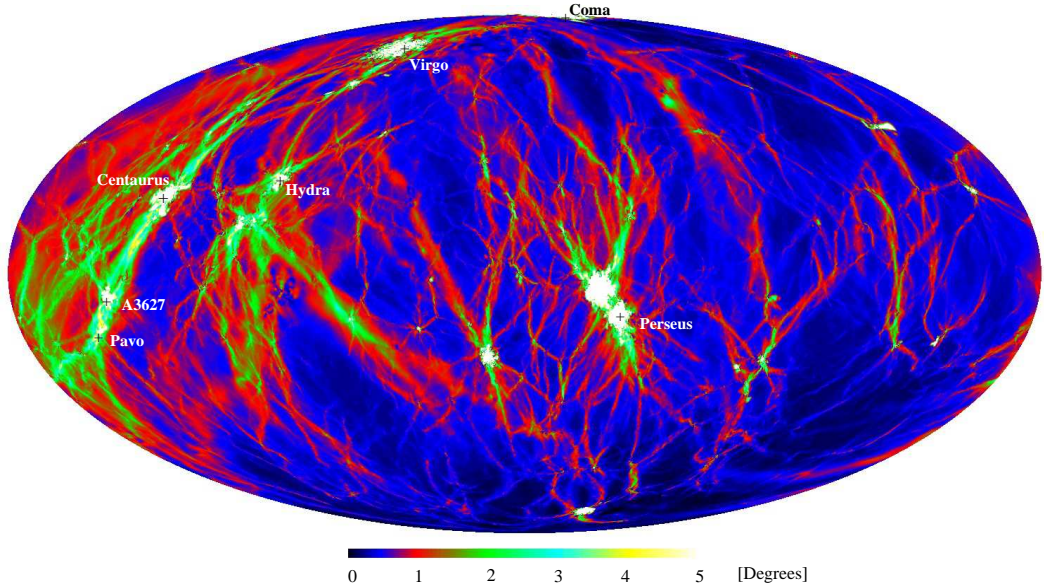}
\caption{Full sky map of deflection angles for UHECRs with energy
$4\times10^{19}$ eV after traveling 100 Mpc in an extra-galactic magnetic
field.  The coordinate system is galactic, with the galactic anti-center in
the middle of the map.  Positions of identified clusters are marked using the
locations of the corresponding halos in the simulation. The map is obtained in
a magneto-hydrodynamical simulation of cosmic structure formation that
correctly reproduces the positions and masses of known galaxy clusters in the
Local Universe. From Ref.~\cite{Dolag:2003ra} }
\label{fig:EGMF-deflections}
\end{center}
\end{figure}

1. In the Galactic magnetic field, for particles coming across the Galactic
disc, we have
\begin{eqnarray}
\frac{\Delta\theta}{Z}
\approx 2.5^\circ \;\; 
\frac{10^{20}\;{\rm eV}}{E}\;\;
\frac{B}{3 \; \mu {G}} \;\;
\frac{L}{1.5\;{\rm kpc}}\; ,
\end{eqnarray} 
where $3\, \mu G$ is magnitude of the regular magnetic field and $1.5\;{\rm
kpc}$ is the width of the disc.  We see that protons with $E >
10^{18}\mbox{~eV}$ escape our Galactic disk easily. Protons of smaller energy
are trapped and can escape the Galaxy only by diffusion and ``leaking'' from
the boundary.  Cosmic rays with $E > 10^{18}$ eV should be extra-galactic, if
protons.  Even if CRs would be all iron nuclei, at $E > 2\cdot 10^{19}$ eV
cosmic rays should be extra-galactic, otherwise strong anisotropy in the
direction of the Galactic disc would have been observed.

2. Extra-galactic magnetic fields have not yet been measured, except for the
central regions of galaxy clusters \cite{Carilli:2001hj}. However, there is an
upper bound on their strength from the (absence of) Faraday rotation of
polarized extra-galactic sources \cite{Kronberg:1994vk,Blasi:1999hu}. This
translates to the upper bound on deflections in extra-galactic magnetic fields
\begin{eqnarray}
\frac{\Delta\theta}{Z} < 2.5^{\circ} \;\;
{10^{20}\mbox{~eV} \over E}\;\; 
{B\over 10^{-9}\mbox{~G}} \;\; 
 {(L \lambda)^{1/2}\over 10\mbox{~Mpc}}\; ,
\end{eqnarray} 
where $\lambda$ is the coherence length of an extra-galactic magnetic fields
and is believed to satisfy $\lambda < 1\,$ Mpc. However, extragalactic fields
are strongly inhomogeneous, with amplitude changing by orders of magnitude
from clusters to filaments, and from filaments to voids. Deflections in
some directions, which do not cross clusters and strong filaments, may be
small, otherwise deflections can be very large. This situation cannot
realistically be described by a mean field.  

Only recently have attempts been made to simulate UHECR propagation in a
realistically structured universe~\cite{Dolag:2003ra,Sigl:2004yk}. Results of
Ref.~\cite{Dolag:2003ra} are shown in Fig.~\ref{fig:EGMF-deflections}.
Additional motivation for this simulation was to obtain, in constraint
simulations of the Local Structure, a realistic map of expected deflections,
which would reflect the positions of known clusters. Such a map can be used in
the analysis of cosmic ray arrival directions. Resulting deflections do not
exceed the resolution of UHECR experiments over most of the sky. About an
order of magnitude stronger deflections were obtained in
Ref.~\cite{Sigl:2004yk}.  There are two possible reasons for
disagreement. First, simulations of Ref.~\cite{Sigl:2004yk} were unconstrained
and therefore do not reflect our concrete local neighborhood.  Second,
variable resolution of Ref.~\cite{Dolag:2003ra} was better in cluster regions,
which is a possible reason for the larger obtained dynamical range between
fields in clusters and filaments. Since in both simulations the magnetic
fields are normalized to typical values in the core of rich clusters, their
values in the filaments will be very different, with larger fields outside
clusters in Ref.~\cite{Sigl:2004yk}. A work aimed to resolve these differences
is in progress. For now, I adopt the results of Ref.~\cite{Dolag:2003ra} and
conclude that arrival directions of UHECR should point back to the sources.
Charged particle astronomy of UHECR is, in principle, possible.

\paragraph{Interactions with cosmic radiation backgrounds.}

Ultra-high energy cosmic rays cannot propagate elastically in cosmic
backgrounds. They have enough energy to produce secondary particles even in
collisions with CMBR (important for proton primaries) or radio photons
(important for UHECR photons) or infrared radiation (important for propagation
of nuclei).  Most important is the reaction of pion photo-production for
protons (or neutrons) propagating in relic cosmic microwave background left
over from the Hot Big Bang. For the threshold energy of this reaction we find,
in the laboratory frame,
\begin{eqnarray}
E_{\rm th}({p+\gamma  \rightarrow N+\pi}) =
\frac{(m_p +m_\pi)^2 - m_p^2} {2E_\gamma 
(1-\cos \theta)} \; .
\label{GZK_thr_en}
\end{eqnarray}
Note, that in the derivation of this relation, standard Lorentz kinematic and
standard dispersion relation between particle energy and momentum, $E^2 = k^2
+ m^2$, are assumed. If any of these are violated, the threshold condition in a
laboratory frame may look different.  For the black body distribution of CMBR
photons with temperature $T = 2.7^\circ\, K$ we find
\begin{figure}
\begin{center}
\includegraphics[width=10cm]{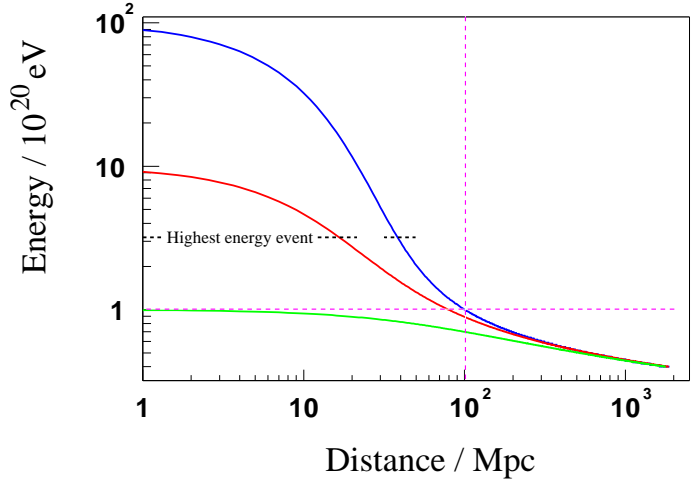}
\caption{Energy of protons as a function of the distance propagated in CMBR
  for three initial values of energy at the source, $10^{22},~10^{21}~{\rm
  and}~10^{20}$ eV respectively. }
\label{fig:UHECR-attenuation}
\end{center}
\end{figure}
\begin{eqnarray}
E_{th}
\approx 5 \times 10^{19} \;{\rm eV}\; .
\label{GZK_thr_en2}
\end{eqnarray} 
This reaction has a large cross section, being the largest
at the  $\Delta$ resonance. At half-width of the resonance
\begin{eqnarray}
\sigma \sim 300\; \mu b \approx 3 \times 10^{-28} {\rm cm}^2 \; .
\end{eqnarray} 
Density of CMB radiation is
$n \sim T^3 \sim 400$ cm$^{-3}$. This corresponds to the mean free path:
\begin{eqnarray}
L_\sigma = 
(\sigma n)^{-1} \approx 8 \times 10^{24}\; {\rm cm} \;
{\approx 2.7\; {\rm Mpc}} \;
\end{eqnarray} 
In each collision $\approx$ 20\% of energy is lost (which is the mass of a
pion).  Successive collisions rob protons of energy, which decreases
exponentially. The distance over which energy decreases by one e-fold is
called the attenuation length. At the threshold, Eq.  (\ref{GZK_thr_en2}), the
attenuation length is large, $L_A \approx 10^3$ Mpc (being determined by other
processes, see below.) With increasing energy, it rapidly decreases and at
energies above the $\Delta$ resonance for typical CMBR photons, $ E \agt
5\cdot 10^{20}\;{\rm eV}$, the attenuation length is $L_A \approx 10$ Mpc.  It
follows that the energy of protons drops below $ 10^{20}\;{\rm eV}$ after it
travels the distance of order $100$ Mpc almost independently upon initial
energy, see Fig.~(\ref{fig:UHECR-attenuation}). We conclude that
\begin{itemize}
\item[[\psp {\bf A1}\nsp ]]
{\it Protons detected with~ $ E > 10^{20}\;{\rm eV}$ 
~should have an origin within~ $ R < R_{\rm GZK} \equiv 100 \;{\rm Mpc}$}.
\end{itemize}
We will call the corresponding volume a GZK-sphere (or GZK-distance).

The reaction {$p+\gamma \rightarrow p+e^+e^-$} is sub-dominant.  While it has
a smaller threshold (by a factor of $2m_e/m_\pi \sim 10^{-2}$), it also has a
smaller cross section. But, it becomes important at sub-GZK energies.
Attenuation length for this reaction is $10^3$ Mpc -- a noticeable and
important effect.

UHE photons loose energy in $\gamma+\gamma \rightarrow e^+e^-$.  The threshold
for the reaction with CMBR photons is smaller by a factor of $2m_e^2/m_\pi m_p
\sim 10^{-5}$ compared to the GZK cutoff energy.  The cross-section decreases
fast with energy, $\sigma = \sigma_{\rm T}m_e/s^2$, where $\sqrt{s}$ is the CM
energy and $ \sigma_{\rm T} \approx 10^{-22} {\rm cm}^2 $ is the Thomson
cross-section.  Therefore, attenuation length has a minimum at the pair
production threshold.  For CMBR photons, this occurs at $E \approx 2\cdot
10^{14}\; {\rm eV}$ and $L_A \approx 10\; {\rm kpc}$. The attenuation length
increases with energy, reaching GZK distance roughly at $E \approx 10^{20}$
eV.  Photons with even larger energies are able to penetrate even larger
distances -- and this is important for many models -- but in this energy
range, the main contribution comes from poorly known radio-background, which
brings some uncertainty in the attenuation length of the highest energy
photons.

Heavy nuclei loose energy in photo-dissociation. Here, the main contribution
comes from the infra-red background which is also poorly known. 
But again, at $E \approx 10^{20}$ eV the attenuation length is comparable 
to the GZK distance \cite{Stecker:1998ib}.

\paragraph{The cut-off.}

It is easy to understand why a sharp cut-off in the spectrum of protons should
appear.  This happens because the attenuation length decreases rapidly with
increasing energy.  Assume a power law injection spectrum for UHECR,
$J_{in}(E) \propto E^{-\alpha}$, and let $n(r)$ be the density of
sources. Fluxes from individual sources decrease as $r^{-2}$, which is
compensated by volume integration, $r^{2}dr$. Therefore, the total flux
registered at energy $E$ should grow in proportion to the upper limit of
volume integration
\begin{equation}
J(E) \propto \int_0^{R(E)} n(r) dr {~\propto~ R(E)} \; ,
\label{FluxDef}
\end{equation}
if the distribution of sources, $n(r)$, does not depend on $r$.  Here,
$R(E)$ corresponds to the attenuation length, i.e. the distance
from which cosmic rays with energy $E$ can reach us. The attenuation
length of protons with $E < 5\times 10^{19}$ eV equals $10^3$ Mpc,
while the attenuation length at $E > 5 \cdot 10^{20}$ eV is only $10$
Mpc. We conclude that
\begin{itemize}
\item[[\psp {\bf A2} \nsp ]]
{\it The drop in flux by 2 orders of magnitude at GZK energy is 
expected if the distribution of sources is homogeneous .}
\end{itemize}
A word of caution is needed here. Transition in $R(E)$ from sub-GZK to
super-GZK regime is not instantaneous. Therefore, a particular value of the
drop depends upon the shape of the injection spectrum, i.e. on the value of
$\alpha$, see e.g. Refs.~\cite{Blanton:2000dr,Kalashev:2001qp,%
Berezinsky:2001wn}.

\subsection{Generation of UHECR}
\label{sec:UHECRgeneration}

The origin of cosmic rays and/or their acceleration mechanisms have been a
subject of debate for several decades. Particles can be accelerated either by
astrophysical shock waves, or by electric fields. In either case, one can
estimate the maximum energy; with optimistic assumptions, the final estimate
is the same for both mechanisms. In practice, the maximum of energy is
expected to be much lower.

1. {\it Shock acceleration.} Particles are accelerated stochastically while
bouncing between shocks. Acceleration can continue only if particles remain
confined within an accelerating region, in other words until gyro-radius, Eq.
(\ref{LarRad}), is smaller than the size of the region.  This gives
\begin{eqnarray}
E_{\rm max} = Ze B L \; .
\label{Hillas_cond}
\end{eqnarray}

2. {\it Acceleration by an electric field.} The latter can be created by a
rapidly rotating magnetized neutron star or black hole. If motion is
relativistic, the generated electric field is of the same order as the
magnetic field, and the difference in electric potentials is $\sim (B\times
L)$. This, again, reproduces Eq. (\ref{Hillas_cond}) for the maximum energy.

Known astrophysical sources with $(B\times L)$ big enough to give $E_{\rm max}
\sim 10^{20}$ eV are neutron stars, active galactic nuclei (AGN) and colliding
galaxies.

\begin{figure}
\begin{center}
\includegraphics[width=12.cm]{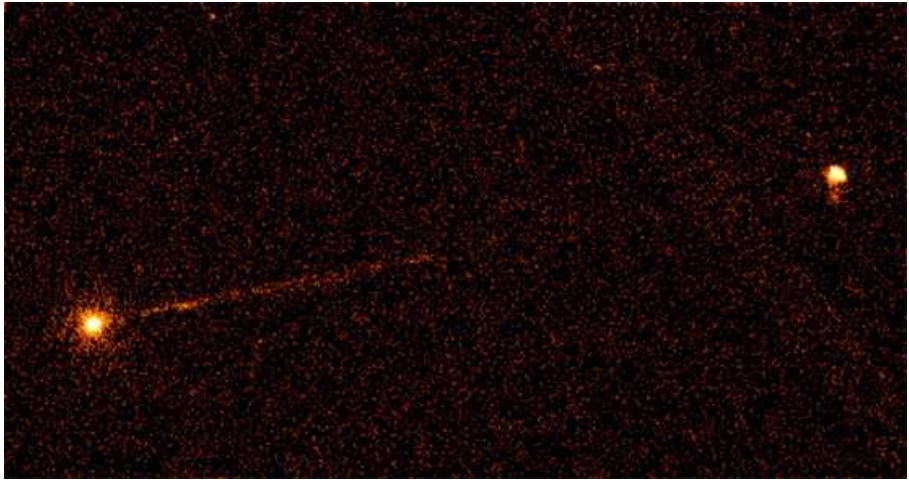}
\caption{ {\it Chandra} telescope X-ray image of the nucleus, jet, and hot
spot of Pictor A. From Ref.~\cite{Wilson:2000gp}.  }
\label{fig:pictor}
\end{center}
\end{figure}

The central engine of an AGN is believed to be a super-massive black hole
powered by matter accretion. AGNs have two jets (one of the jets may be
invisible because of the Doppler effect) of relativistic particles streaming
in opposite directions.  Interaction with the intergalactic medium terminates
this motion and at the ends of jets the radio-lobes and hot-spots are formed,
see Fig.~\ref{fig:pictor}.  The acceleration of UHECR primaries may occur
either near the black hole horizon (direct acceleration), or in hot spots of
jets and radio-lobes (shock acceleration).  The host of different AGNs is now
classified in one unified scheme, for a review see \cite{Urry:1995mg}.
Depending upon the angle between the jet axis and the line of sight we observe
different types of AGN. A typical radio galaxy, showing two strong opposite
jets, is observed at angles approximately perpendicular to the jet axis. An
AGN is classified as a quasar if the angle is smaller than $30^\circ$. If we
look along the jet axis (angle $< 10^\circ$), i.e.  directly into the barrel
of the gun, we observe a blazar.

It should be noted that not all radio-galaxies are the same.  There are
Fanaroff-Riley (FR) type I and type II galaxies (radio-loud AGNs), and Seyfert
galaxies (radio-quiet AGNs).  Both types of FR galaxies may be the sites of
UHECR acceleration, but the hot spots in FR type II galaxies are considered to
be most promising \cite{Rachen:1993pg}. It is believed that when observed
along the jet axis, FR type II galaxies make a parent population of Highly
Polarized Quasars (HPQ -- subclass of blazars), while FR type I galaxies make
a parent population of BL Lacertae objects (BL Lacs -- another subclass of
blazars).

As an example, the X-ray image of the powerful FR-II radio galaxy Pictor A
taken by {\it Chandra} observatory is shown in Fig.~\ref{fig:pictor}.  Radio
observations of jets have a long history.  Recently, {\it Chandra} started to
obtain high resolution X-ray maps of AGNs which, surprisingly, revealed very
long collimated X-ray jets. E.g. the distance from nucleus to the hot spot in
Pictor A is at least 240 kpc.  It is hard to explain such long jets as pure
leptonic, and it is possible that the population of relativistic electrons
responsible for the X-ray emission is a result of photo-pion production by UHE
protons \cite{Wilson:2000gp}.

Now, for any acceleration mechanism and independent of the actual acceleration
site (i.e. be it either the AGN's black hole or any of the hot spots) the
momentum of highest energy particles is expected to point in the direction of
the jet \cite{Tinyakov:2001nr}. In other words, if AGNs are sources of UHECR,
arrival directions of high energy cosmic rays may point back to a (subclass)
of a blazar family.  Such correlations were indeed observed
\cite{Tinyakov:2001nr,Tinyakov:2001ir,Gorbunov:2002hk,Tinyakov:2003nu}  with
BL Lacertae objects.

\subsection{UHECR spectrum.}

\begin{figure}
\begin{center}
\includegraphics[width=16.cm]{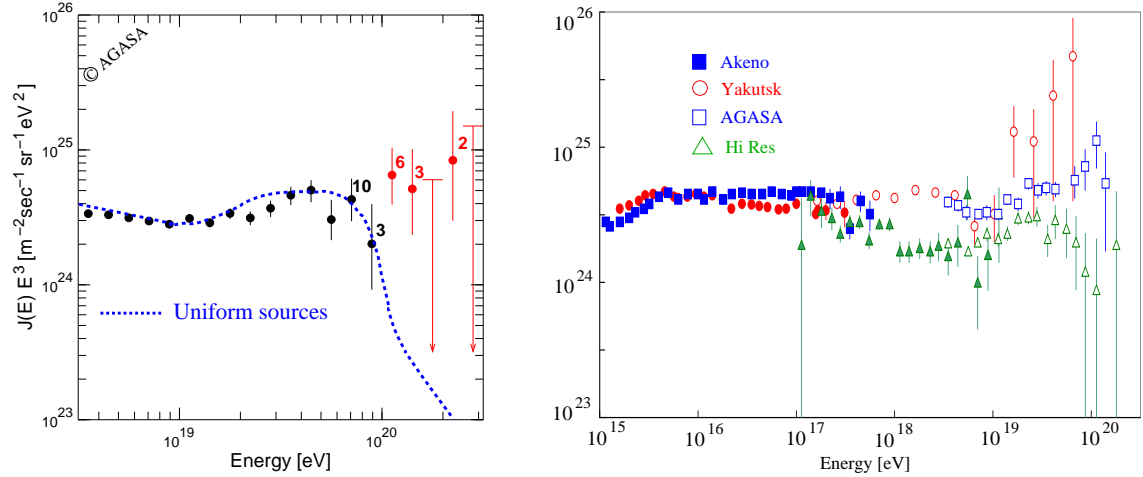}
\vspace{-2cm}
\caption{{\bf Left panel:} ~The energy spectrum of cosmic rays with a zenith
angle up to $45^\circ$ as measured by AGASA ~\cite{Takeda:2002at}. Numbers
near data points reflect the number of events in the respective energy bins.
{\bf Right panel:} ~A compilation of data from different experiments: Akeno
(filled squares), AGASA (open squares), HiRes-I and HiRes-II (open and filled
triangles), two Yakutsk sub-arrays of Cherenkov detectors (open and filled
circles).  From Ref.~\cite{Ivanov:2003iv}.}
\label{fig:AGASA_spectrum}
\end{center}
\end{figure}

The largest statistic of UHECR events has been accumulated for over 12 years
of operation by the AGASA air shower array of surface particle detectors.  The
spectrum measured by the AGASA is shown in Fig.~\ref{fig:AGASA_spectrum}, left
panel. The dotted curve represents the theoretical expectation for a
homogeneous distribution of sources and proton primaries.  This theoretical
curve exhibits the GZK cut-off at $E\approx 10^{20}$ eV. Remarkably, AGASA had
detected 11 events with higher energy and the data show no hint for cut-off.

It is hard to argue against the reality of these findings. AGASA exposure is
under control, and the only issue is the energy determination. AGASA events
have an accuracy of $\pm25\%$ in event-reconstruction resolution and $18\%$ in
systematic errors around $10^{20}$ eV \cite{Takeda:2002at}. Added in
quadrature this gives RMS error of energy determination to be $\pm 30\%$.
More importantly, the probability of an upward fluctuation to 1.5 times the
true energy is 2.8\%. There are too many super-GZK events, and with this
resolution a spectrum with GZK cutoff cannot be transformed into an excess of
post-GZK events assuming spillover \cite{Cronin:2004ye}.

Recently, the HiRes group has reported results obtained with a telescope which
measures atmospheric fluorescence light.  The energy spectrum is shown in
Fig.~\ref{fig:AGASA_spectrum} by triangles, the right panel. The spectrum is
consistent with the GZK cut-off, and there are 2 events detected with
$E>10^{20}$ eV.  Systematic error in energy measurement was estimated to be
21\%, systematic error in the aperture is not yet clear.  HiRes employs a
relatively new technique, with the following issues usually cited for
improvement: atmospheric attenuation corrections should be based on nightly
measurements and not averages, better energy calibration and aperture
calculation are called for, see e.g. \cite{Cronin:2004ye}.

The Yakutsk group uses a hybrid detection method, combining a ground array of
particle detectors with telescopes which are measuring Cherenkov light
produced by a shower in the atmosphere. A recently reported
\cite{Ivanov:2003iv} spectrum, derived from air Cherenkov light measurements,
is shown in the right panel of Fig.~\ref{fig:AGASA_spectrum} by circles. At
the low energy end it agrees well with the Akeno spectrum, and at the high
energy end it is consistent, within errors, with the AGASA spectrum. Yakutsk
and AGASA disagree significantly with HiRes at $E \sim 10^{18}$ eV were
statistical errors are negligible, which points out to some systematics. AGASA
and HiRes can be reconciled at $E<10^{20}$ by e.g. $-15\%$ and $+15\%$
respective shift of energy \cite{DeMarco:2003ig}. The discrepancy between two
experiments at $E>10^{20}$ after the shift is only $2\sigma$. However, even
after these shifts there are still 9 events with $E>10^{20}$ in the combined
data set.
 
\subsection{The Puzzle}

These measurements are regarded as a threefold puzzle, because contrary to
assertions [A1], [A2]
\begin{itemize}
\item[[\psp {\bf P1} \nsp ]] No candidate sources are found within the GZK
distance in the directions of $E > 10^{20}$ ~eV events;
\item[[\psp {\bf P2} \nsp ]] The AGASA spectrum does not exhibit the GZK
cutoff.
\end{itemize}
And finally we have the third puzzling question:
\begin{itemize}
\item[[\psp {\bf P3} \nsp ]] Which physical processes are capable of producing
events with these enormous energies ?
\end{itemize}

\paragraph{Conjectures.}

With the assumption that all three pieces of the puzzle, [P1] -- [P3], are
correct, the situation becomes desperate. There were no solutions suggested
which would not invoke new physics beyond the standard model, or very
speculative astrophysics.  In addition, all models require fine tuning, and
many do not really solve all three problems. It is not possible to consider
here all the suggested solutions. Ignoring for now the problem [P2], the
situation with [P1] and [P3] does not become easier, but we can now restrict
ourselves to the discussion of astrophysical solutions only.

Ignoring in addition [P3], the simplest suggestion is to assume very large
extra-galactic magnetic fields, which would randomize UHECR trajectories.
However, the results of Ref.~\cite{Dolag:2003ra} do not support such a
conjecture.  As we have mentioned already, a consensus regarding EGMF fields
has not yet been reached and in Ref.~\cite{Sigl:2004yk}, much stronger
extragalactic magnetic fields were advocated. Nevertheless, even in this case,
the conclusion was that the condition of global isotropy of UHECR arrival
directions requires the ``local'' value of magnetic field to be rather weak,
$B \alt 0.1~ \mu{\rm G}$, which, in turn, leads to a large number of UHECR
sources in the GZK volume, $n \agt 100$; for similar limits on the number of
sources see also
Refs. \cite{Dubovsky:2000gv,Fodor:2000yi,Blasi:2003vx,Harari:2004py}.  These
weak EGMFs of Refs.~\cite{Dolag:2003ra,Sigl:2004yk} rule out the possibility
of a single powerful radio-galaxy, which happened to be nearby
~\cite{Farrar:2000nw}, or a gamma-ray burst scenario \cite{Waxman:1995vg}, as
a potential sources of UHECR.

Another suggested astrophysical scenario was a ``dead quasars'' model
\cite{Boldt:1999ge}. This model assumes that quasars, powerful in the past,
retain the possibility to accelerate to the highest energies even after the
accretion of matter is exhausted and a quasar cease to be visible
electromagnetically. However, the process of acceleration to the highest
energies in compact sources is inevitably accompanied by a strong TeV emission
\cite{Levinson:2000nx}. Recent results obtained by several TeV telescopes, in
particular, non-observation of strong TeV sources, rule out the ``dead
quasar'' model \cite{Neronov:2004ga}. In addition, in Ref. \cite{Isola:2003jk}
it was found that known quasar remnants are typically distributed too
anisotropically to explain the isotropic ultra high energy cosmic ray flux
except in the unrealistic case where extragalactic magnetic fields of $0.1~
\mu$G extend over many Mpc.

A possibility that ultra-high-energy events are due to iron nuclei accelerated
from young, strongly magnetized neutron stars in relativistic MHD winds has
also been suggested \cite{Blasi:2000xm}.  However, with realistic parameters of
Galactic magnetic field, even iron nuclei do not propagate diffusively within
Galaxy, which disfavors this model \cite{O'Neill:2001aw}.

\paragraph{Any observational clue?}

Many quite different models were suggested for the resolution of the GZK
puzzle.  The majority of suggested models, which we have no space to consider
here, employ a new physics of one sort or another. (The reader may consult
UHECR reviews cited at the beginning of this section, but I believe that a
review which would cover all the suggested possibilities does not exist.)
Instead, let us consider the question of whether or not there is already a clue
in the data as to which model may be correct. Hints, and, in principle,
critical signatures are given by:
\begin{itemize}
\item {\it Spectral shape.}  We do not yet have enough data at the highest
energies to constrain models. Spectra below $10^{20}$ eV point to the AGN
model of UHECR origin, with protons being primaries
\cite{Berezinsky:2002nc,Berezinsky:2002vt}. 
\item {\it Chemical composition.}  Again, not enough data at the highest
energies. An analysis of Haverah Park data at lower energies shows that above
$10^{19}$ eV, less than 30\% of the primary cosmic rays can be photons or iron
nuclei at the 95\% confidence level \cite{Ave:2000nd}. In other words, at
least 70\% should be protons.
\item {\it Large-scale anisotropy.}
Gives strong signatures. Not observed, which is a hint by itself. Some
implications we had considered already, and may add that the non-observation of
anisotropy towards the Galactic center has a potential of ruling out the
model of UHECR origin based on decays of super-heavy dark matter
\cite{Dubovsky:1998pu,Berezinsky:1998rp,Kim:2003th,Kachelriess:2003rv}.
\item {\it Small-scale clustering.}  This is an observed \cite{Takeda:1999sg},
reliable feature. (Errors in angle determination are definitely small.) It is
already statistically significant.  Therefore, below I shall concentrate on
this signature.
\end{itemize}

\paragraph{Small-scale clustering.}

It was observed by different installations that arrival directions of UHECR
are too close to each other and this happens too often
\cite{Hayashida:1996bc,Takeda:1999sg,Uchihori:1999gu,Tinyakov:2001ic}.  In
particular, the AGASA collaboration has observed 6 doublets and 1 triplet of
cosmic rays with $E> 4\times 10^{19}$ within $2.5^\circ$. The chance
probability of observing just a triplet under an isotropic distribution is
only 0.9\% \cite{Takeda:1999sg}. Statistical significance of these clusters in
the AGASA data set was considered by several authors. In
Ref.~\cite{Tinyakov:2001ic}, an analysis based on the calculation of an
angular autocorrelation function was employed, and the probability $P=3\times
10^{-4}$ of chance clustering was obtained. This includes the penalty for the
choice of the energy cut, while the angular bin was chosen to be fixed at
$2.5^\circ$, which is a value previously accepted by AGASA, being consistent
with the angular resolution. In Ref.~\cite{Finley:2003ur}, this analysis was
repeated and confirmed. In addition, two more conservative estimates were
done. In the first, the penalty factor for the adjustment of the angular bin
size was added.  This returns $P= 3\times 10^{-3}$. This is a valid procedure,
but it misses prior information about the angular resolution of the
installation. In the second estimate, the bin size was kept fixed, but the
whole data set was divided in halves. The ``original data set''
\cite{Hayashida:1996bc} was used to justify the bin size of $2.5^\circ$, while
clusters in it were removed for the subsequent evaluation of statistical
significance.  This procedure returned $P=8\times 10^{-2}$. Again, this is a
valid approach too, and can be safely used with future large data
sets. However, I'd like to stress that it is {\it not} an evaluation of the
statistical significance of 6 doublets and 1 triplet. It is no wonder that a
smaller data set has reduced statistical significance. Finally, in
Ref.~\cite{:2004dx} it was found that the AGASA data set manifests a $P \sim
10^{-3}$ chance probability of clustering above background using independent
statistics of $\langle \cos \theta \rangle_{[0^\circ,\, 10^\circ]}$.  I find
this value, $P \sim 10^{-3}$, to be fair estimate of the current significance
of clustering in the AGASA data.

Note the following: if clusters are real and due to sources, the number of
events in ``physical'' clusters should be Poisson distributed. Therefore, with
the current low statistics, it is expected that roughly half of installations
should observe significant clustering, while another half should not see it
\cite{Tinyakov:2001ic}. There is no clustering in the current HiRes data
\cite{Abbasi:2004ib,:2004dx}. However, with the current statistics there is no
contradiction yet \cite{Yoshiguchi:2004np,:2004dx}.

The study of small-scale clustering is very important. If clusters are real and
not a statistical fluctuation, then UHECR should point back to sources and
UHECR astronomy is possible. Real sources should be behind the clusters and
the correlation studies make sense.  Pursuing this strategy, one should be
restricted to astrophysical sources with physical conditions potentially
suitable for particle acceleration to the highest energies.  Active galactic
nuclei (AGN) constitute a particularly attractive class of potential sources.
As we have already discussed, if AGNs are sources, those which have jets
directed along the line of sight, or blazars, should correlate with observed
UHECR events.  It is intriguing that statistically significant correlations of
UHECR with BL Lacertae objects were found \cite{Tinyakov:2001nr}.

\section{Conclusions}

Cosmology and astrophysics give us firm evidence that the standard model of
particle physics is limited. The standard model fails to explain baryogenesis,
does not contain non-baryonic dark matter and has no room for massive
neutrino. We now know that dark energy also exists, but we do not know why it
exists. There seems to be too many coincidences between numerical values of
cosmological parameters which describe the matter and energy budget.
Contributions of baryonic matter, non-baryonic matter and dark energy are
almost equal at the present epoch, while they have seemingly unrelated origin
and could differ by many orders of magnitude. Cosmology just became a
precision science and is already full of surprises; we can expect even more
exciting discoveries in the near future.

\section*{ACKNOWLEDGEMENTS}

I would like to thank the conference organizers for friendly and warm
atmosphere.   


\end{document}